\documentclass{aa}  

\usepackage{graphicx}
\usepackage{txfonts}
\begin{document}

   \title{The Extraplanar Type II Supernova ASASSN-14jb in the Nearby Edge-on Galaxy ESO 467-G051}


   \author{Nicol\'as~Meza\inst{1} \and J.~L.~Prieto\inst{2,3} \and A.~Clocchiatti\inst{1,3} \and L.~Galbany\inst{4} \and J.P.~Anderson\inst{5} \and E.~Falco\inst{6} \and C.~S.~Kochanek\inst{7,8} \and H.~Kuncarayakti\inst{9} \and J.~Brimacombe\inst{10} \and T.~W.-S.~Holoien\inst{11} \and B.~J.~Shappee\inst{12} \and K.~Z.~Stanek\inst{7,8} \and T.~A.~Thompson\inst{7,8}}

   \institute{Instituto de Astrof\'isica, Pontificia Universidad Cat\'olica de Chile, Av. Vicu\~na Mackenna 4860, 782-0436 Macul, Santiago, Chile \email{nmeza@astro.puc.cl} \and N\'ucleo de Astronom\'ia de la Facultad de Ingenier\'ia y Ciencias, Universidad Diego Portales, Av. Ej\'ercito 441, Santiago, Chile \and Millennium Institute of Astrophysics, Santiago, Chile \and PITT PACC, Department of Physics and Astronomy, University of Pittsburgh, Pittsburgh, PA 15260, USA \and European Southern Observatory, Alonso de C\'ordova 3107, Casilla 19, Santiago, Chile \and Harvard-Smithsonian Center for Astrophysics, 60 Garden St., Cambridge, MA 02138, USA \and Department of Astronomy, The Ohio State University, 140 West 18th Avenue, Columbus, OH 43210, USA \and Center for Cosmology and AstroParticle Physics (CCAPP), The Ohio State University, 191 W. Woodruff Avenue, Columbus, OH 43210, USA \and Tuorla Observatory, Department of Physics and Astronomy, University of Turku, V\"{a}is\"{a}l\"{a}ntie 20, FI-21500 Piikki\"{o}, Finland \and Coral Towers Observatory, Cairns, Queensland 4870, Australia \and The Observatories of the Carnegie Institution for Science, 813 Santa Barbara Street, Pasadena, CA 91101, USA \and Institute for Astronomy, University of Hawai’i, 2680 Woodlawn Drive, Honolulu, HI 96822, USA}

   \date{Received / Accepted}

\abstract{}{We present optical photometry and spectroscopy of the Type II supernova ASASSN-14jb, together with VLT MUSE IFU observations of its host galaxy and a nebular-phase spectrum.}{This supernova, in the nearby galaxy ESO 467-G051 ($z=0.006$), was discovered and followed-up by the All Sky Automated Survey for SuperNovae (ASAS-SN). We obtained well-sampled LCOGTN $BVgri$ and $Swift$ $w2m1w1ubv$ optical and near-UV/optical light curves and several optical spectra in the early photospheric phases. ASASSN-14jb exploded $\sim 2$~kpc above the star-forming disk of ESO 467-G051, an edge-on disk galaxy. The large projected distance from the disk and non-detection of any H~II region in a 1.4 kpc radius in projection are in conflict with the standard environment of core-collapse supernova progenitors and suggests the possible scenario that the progenitor received a kick in a binary interaction.}{We present analysis of the optical light curves and spectra, from which we derive a distance of $25\pm 1$ Mpc using state of the art empirical methods for Type II SNe, physical properties of the SN explosion ($^{56}$Ni mass, explosion energy, and ejected mass) and properties of the progenitor, namely the progenitor radius, mass and metallicity. Our analysis yields a $^{56}$Ni mass of $0.0210 \pm 0.0025$ M$_\odot$, an explosion energy of $\approx 0.25 \times 10^51$ ergs and an ejected mass of $\approx 6$ M$_\odot$. We also constrain the progenitor radius to be $R_* = 580 \pm 28$ R$_\odot$ which seems to be consistent with the  sub-Solar metallicity of $0.3 \pm 0.1$ Z$_\odot$ derived from the supernova Fe II $\lambda 5018$ line. The nebular spectrum constrains strongly the progenitor mass to be in the range 10-12 M$_{\odot}$. From {\it Spitzer} data archive we detect ASASSN-14jb $\approx 330$ days past explosion and we derive a total dust mass of $10^{-4}$ M$_\odot$ from the 3.6 $\mu$m and 4.5 $\mu$m photometry.  Using the FUV, NUV, $BVgri$,$K_s$, 3.6 $\mu$m, and 4.5 $\mu$m total magnitudes for the host galaxy, we fit stellar population synthesis models which gives an estimate of $M_* \approx 1 \times 10^9$ M$_\odot$ , an age of 3.2 Gyr, and a SFR $\approx  0.07$ M$_\odot$/yr. We also discuss the low oxygen abundance of the host galaxy derived from the MUSE data, having an average of $12+\log{(O/H)} = 8.27^{+0.16}_{-0.20}$ using the O3N2 diagnostic, with strong line methods and compare it with the supernova spectra, which is also consistent with a sub-Solar metallicity progenitor. Following recent observations of extraplanar H II regions in nearby edge-on galaxies, we derive the metallicity offset from the disk, being positive (but consistent with zero at 2 $\sigma$), suggesting enrichment from disk outflows. We finally discuss the possible scenarios for the unusual environment for ASASSN-14jb and conclude that either the in-situ star formation or runaway scenario would imply a low mass progenitor, agreeing with our estimate from the supernova nebular spectrum. Regardless of the true origin of ASASSN-14jb we show in this work that the detailed study of the environment can roughly agree with the stronger constrains of the transient observations.}{}


\maketitle

\section{Introduction}

\label{sec:intro}
Originally classified based on the absence (Type I) or presence (Type II) of hydrogen lines in their optical spectra \citep{minkowski41}, supernovae (SNe) represent the explosive ending of a star.
Decades of research have added a considerable degree of complexity to the simple scheme of Minkowski. For more detail see  \cite{Filipenko1997} or \cite{Turatto2007}.
The great diversity of core-collapse supernovae (CCSNe) is understood as the result of a rich variety of parent systems. Initial differences in mass, radius, metallicity or rotation, and evolutionary differences in mass lost to stellar winds or interacting binary companions, would result in a wide distribution of envelope masses when the progenitor stars reach the time of core collapse  \citep[e.g.,][]{heger03,kasen09,luc13,pejcha15b}. 
According to the chemical composition of the outer layers at the time of explosion the spectroscopic display will be of Type II, or Type IIb, Ib or Ic, with little or no presence of hydrogen. The latter are collectively called ``stripped envelope SNe''. Finally, according to the total mass of hydrogen in the envelope, a bona fide Type II SN will show a slower or faster rate of decline after maximum and will be named ``plateau'' (IIP) or ``linear'' \citep[IIL,][]{Barbon79}. The convention has stuck although we know now that there is a continuous distribution of decline rates \citep[e.g.,][]{Joe14b,Sanders15,pejcha15,galbany16a} and that Type IIP and IIL have similar progenitors \citep{valenti15}.
Extreme mass loss shortly before explosion may lead to the formation of a dense circumstellar medium (CSM). The interaction of SN ejecta with the CSM would produce narrow emission lines of hydrogen and the SN is named Type IIn in these cases \citep[e.g.,][]{Dopita1984,Schlegel1990,Stathakis1991,Chugai94}.

Progenitors of CCSNe have been identified in pre-explosion images \citep{Smartt09,Smartt15}, and found to be red supergiants (RSG) with Zero Age Main Sequence (ZAMS) masses between $\sim 8$ and $\sim 17$~M$_{\odot}$ for Type II SNe or objects with masses between $\sim 4$ and $\sim 10$~M$_{\odot}$ at the time of explosion after a complex interacting binary evolution for Type IIb and Ib SNe \citep{Folatelli15,Eldridge13,Folatelli16}.

 As CCSNe progenitors are massive stars with relatively short lifetimes of $\sim 4-50$~Myr, it is expected that the SNe are still associated with their birth places, namely spiral arms or H~II regions \citep{Bartunov94,Macmillan96}, although CCSNe in early-type galaxies with residual star formation have been reported in the literature \citep{hakobyan08b}.
 
 The correlation between CCSNe and star forming regions is expected to decrease toward the lower-mass progenitors and get diluted by progenitors resulting from binary evolution which can result in longer timescales before explosion \citep{zapartas17,eldridge17}. Recent work has found that stripped envelope SNe are more closely associated with star forming regions than hydrogen rich SNe, as expected from the increasing progenitor mass of the sequence $Ia \rightarrow II \rightarrow IIb/IIn \rightarrow Ib/c $ \citep[e.g.,][]{Joe12,galbany14}.
 
 Observations of the SN host metallicity, either global \citep{prieto08,Arcavi10} or local \citep{Modjaz08,Joe10,stoll13,galbany14,hanin17}, and the spatial distribution of CCSNe \cite[e.g.,][]{VanDyk1999,Petrosian2005,Mikhailova2007,kangas17}, also provide constrains on the different progenitors.
Chemical abundance studies generally associate stripped envelope SNe with higher metallicity hosts as expected from the probable increase of mass loss with metallicity \citep[e.g.,][]{Henry99,sanchez2014,SanchezM2016}.
Studies of the radial distribution of SNe find that Type Ib/c SNe are more centrally concentrated in their hosts \citep{Joe09,hakobyan09}, which is again consistent with the higher metal enrichment towards the center of galaxies.
And studies of the height distribution of SNe in edge-on disk galaxies \citep{hakobyan17c}, find that CCSNe are nearly twice as much concentrated towards the disk than thermonuclear SNe, a result consistent with the height scale of the stellar populations where their progenitors are expected to originate.

Some CCSNe, however, defy the common sense implicit in the previous description by appearing far from any identifiable birthplace. One striking example is SN~2009ip located $\sim 5$~kpc from NGC~7259, the nearest spiral galaxy \citep[e.g.,][]{Fraser13,mauerhan13,pastorello13,prieto13}.
SN~2009ip was first identified as a SNe impostor in 2009 three years before exploding as a Type IIn supernova \citep{smith14}. Late time HST data rules out the presence of star forming regions comparable to Carinae or the Orion Nebula \citep{smith16} at the explosion site. The possibility that the progenitor was a runaway star is also rejected. The peculiar velocity of the SN is smaller than 400~km/s and the high mass of the progenitor implies a lifetime shorter than the travel time from the nearest star forming site located at $\sim 1.5$~kpc.

This paper introduces another example of this sort, the Type IIP like supernova ASASSN-14jb in the edge-on disk galaxy ESO 467-G051 \citep{discovery,classification1,classification2}.
The SN exploded at 2.5~kpc from the center and 2.1~kpc above the galactic disk where no significant star forming region is detected.
Also, it is interesting that ESO 467-G051 and NGC~7259 (the host of SN~2009ip) form an interacting pair. The SNe are separated by 2.4' in the sky ($\sim 18$~kpc).

We present here photometric and spectroscopic observations of ASASSN-14jb and Integral Field Spectroscopy (IFS) of its explosion site. We perform a thorough analysis to estimate physical parameters of the SN, the progenitor star, and the parent galaxy. 
The paper is organized as follows. In Section \ref{sec:data} we describe the photometric and spectroscopic observations. Section \ref{sec:analysis} contains our comparative analysis of the SN, including the photometric and spectroscopic evolution of the ASASSN-14jb. Section \ref{sec:parameters} contains the estimates of basic physical parameters of the SN, Section \ref{sec:host} includes the analysis of the host galaxy and its H~II regions. Finally, Section \ref{sec:discusion} includes our discussion and Section \ref{sec:conclusion} our conclusions.

\section{Data} 
\label{sec:data}
\subsection{Discovery and Explosion Time}
ASASSN-14jb was discovered on UT 2014-10-19.09 (MJD = 2456949.09, $V_{\rm disc} = 16.9$~mag, \citealt{discovery}) at RA = 22:23:16.12, DEC = -28:58:30.78 (J2000.0) by the ongoing All Sky Automated Survey for SuperNovae \citep[ASAS-SN;][]{shappee14,holoien17} from the ``Cassius" 4-telescope unit at the Cerro Tololo Inter-American Observatory (CTIO), in Chile, hosted by the Las Cumbres Observatory \citep[LCOGTN;][]{brown13}. It was spectroscopically classified as a young Type II on UT 2014-10-20 \citep{classification1,classification2}. The last non-detection was reported to be on MJD $= 2456943.098$ with a 3$\sigma$ magnitude upper limit of $V>18.5$~mag, 6 days before the time of discovery. We take the midpoint between this last non-detection and  the discovery time, $t_0 = 2456946.1 \pm 3$ (MJD), as the ``explosion" time. The uncertainty is estimated as half the interval between the last non-detection and the discovery epoch.

\begin{figure}[ht!]
\includegraphics[width=\linewidth]{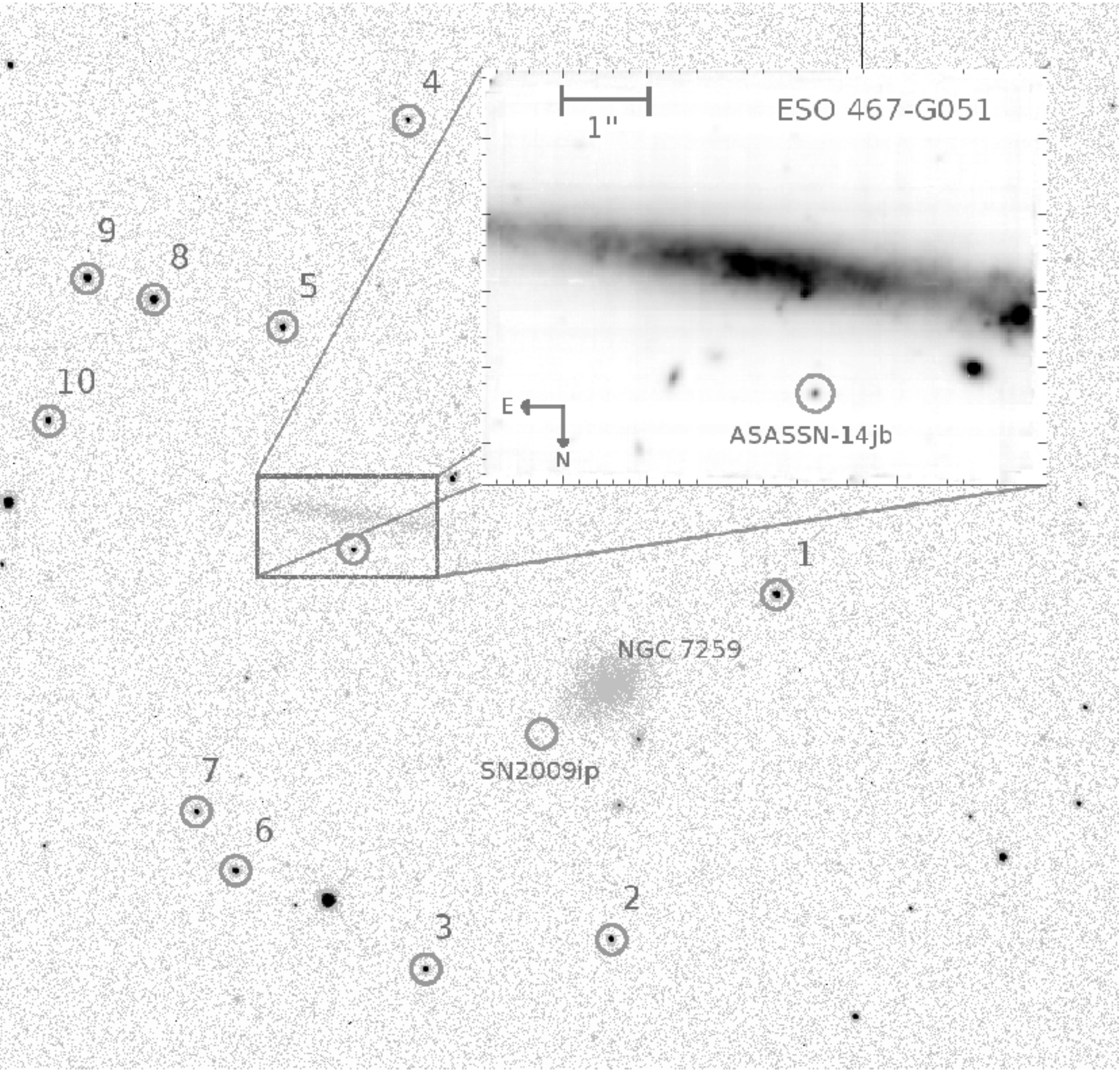}
\caption{$r$-band image of the field of ASASSN-14jb from LCOGTN. The circles show the position of ASASSN-14jb, the local photometric standards (numbered 1-10), and the nearby SN~2009ip in NGC~7259. The zoom around the position of ASASSN-14jb and its host galaxy, ESO 467-G051, shows a synthetic $r$-band image obtained from the MUSE datacube. The MUSE image field of view is of 1.5x1 arcmin.}\label{fig:field_image}
\end{figure}

\subsection{Photometry}

Optical photometric observations of ASASSN-14jb were obtained with the ASAS-SN unit ``Cassius" at CTIO in the $V$-band and the LCOGTN 1-meter telescopes at CTIO, the Siding Spring Observatory (SSO), the South African Astronomical Observatory (SAAO), and the McDonald Observatory (MDO) in the $BV$ and the Sloan Digital Sky Survey (SDSS) $gri$ filters. All the ASAS-SN images are processed in an automated pipeline using the ISIS image subtraction package \citep{alardlupton98,alard2000}, with further details given in \cite{shappee14}. Using the IRAF\footnote{IRAF is distributed by the National Optical Astronomy Observatory, which is operated by the Association of Universities for Research in Astronomy (AURA) under a cooperative agreement with the National Science Foundation.} {\em apphot} package, we performed aperture photometry on the subtracted images and then calibrated the results using the AAVSO Photometric All-Sky Survey \citep[APASS;][]{henden12}. The ASAS-SN photometry is presented in Table~\ref{tab:asassn}.

The fully reduced LCOGTN~1m images (bias/overscan subtracted and flat-fielded) were retrieved from the LCOGTN data archive. We solved the astrometry of each CCD frame using the {\em astrometry.net} software package \citep{lang10}. We obtained calibrated $BVgri$ magnitudes for local standard stars in the field from the APASS catalog. In Figure~\ref{fig:field_image} we display one of the LCOGT images showing the position of the SN, the local standards, and the nearby SN~2009ip. The data of the local standard sequence
is given in Table~\ref{tab:std}.
ASASSN-14jb was recorded with high signal-to-noise ratio (SNR) in the images of LCOGTN. This, together with the negligible background from the host galaxy at the SN site, prompted us to measure the SN flux using aperture photometry. The procedure is explained in detail in Appendix~A. 

As our observations extended only $\sim80$ days after explosion, we searched the data in public domain for images of the field which could have recorded ASASSN-14jb at later times. We found some images in the ESO  archive\footnote{\url{http://archive.eso.org/eso/eso\_archive\_main.html}}. We retrieved ESO/NTT EFOSC images in $BV$ filters that contained ASASSN-14jb's explosion site at three different epochs during 2015, obtained by the Public ESO Spectroscopic Survey of Transient Objects \citep[PESSTO;][ESO program ID 191.D-0935]{smartt15}. The CCD frames were overscan subtracted and flat-fielded with calibration frames obtained the same day using standard IRAF routines. Since the SN is fainter in these late-time images, we measured the SN brightness using PSF fitting photometry as implemented in IRAF DAOPHOT package. We calibrated the magnitudes using local standards from \citet{pastorello13}. The early LCOGTN and late ESO/NTT photometry are presented in Table~\ref{tab:allphot}.

We supplemented our ground-based optical photometry with data from the {\it Swift} and {\it Spitzer} space-born observatories.
Repeated optical and near-UV observations of ASASSN-14jb were obtained by {\it Swift}/UVOT in the passbands {\em w2,m1 w1,u,b,v} during the first $\sim30$ days after explosion. We retrieved them from the {\it Swift} Optical/Ultraviolet Supernova Archive \citep[SOUSA;][]{sousa}.
Also, the field of ASASSN-14jb and its host have been observed nine times since 2010 in the 3.6~$\mu$m and 4.5~$\mu$m bands by the {\it Spitzer} IRAC instrument \citep{fazio04}.
The observations obtained on 2015-09-13 (program ID 11053) and 2016-08-17 (program ID 12099) were taken when the SN was $\sim330$ and $\sim670$ days past explosion, respectively. We retrieved these images from the data archive\footnote{\url{http://sha.ipac.caltech.edu/applications/Spitzer/SHA/}} and also images taken on 2014-09-05 (program ID 10139) to use as templates for image subtraction.
We used {\sc HOTPANTS} \citep{becker15} for difference imaging and measured aperture photometry on the difference images. The SN is detected on 2015-09-13 and undetected on 2016-08-17. 

\begin{figure}[ht!]
\includegraphics[width=\linewidth]{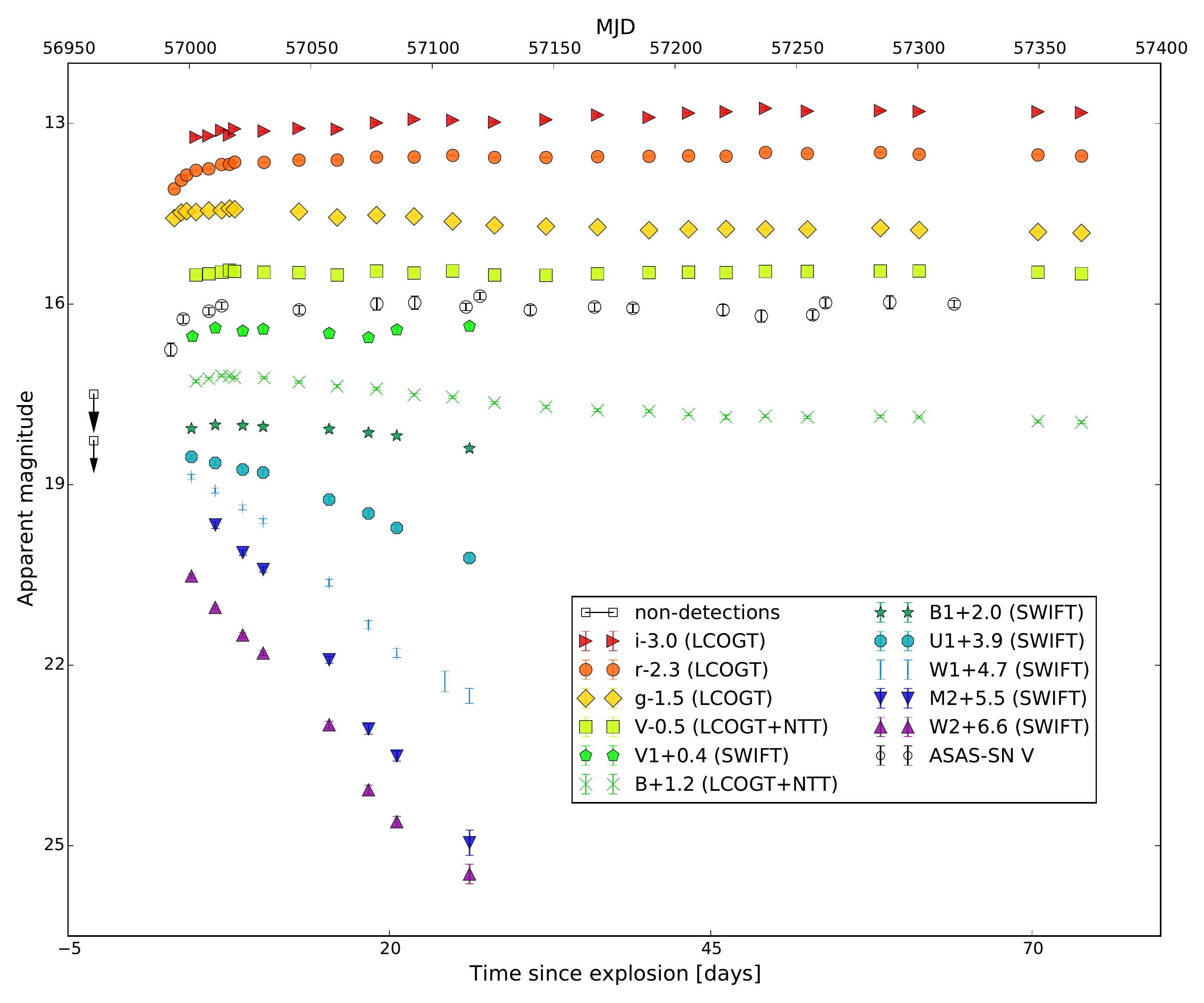}
\caption{ASAS-SN $V$, LCOGTN $BVgri$, and {\it Swift}-UVOT $w2$, $m1$, $w1$, $ubv$, light curves of ASASSN-14jb. Each band has been shifted to the value showed in the label for presentation purposes. The observed time is given in days since the estimated explosion epoch $t_0 = 56946.1$ (MJD) in the bottom x-axis and in MJD in the top axis. The black squares are the non-detection derived from the ASAS-SN $V$-band images.\label{fig:lc_early}}
\end{figure}

\begin{figure*}[ht!]
\includegraphics[width=\linewidth]{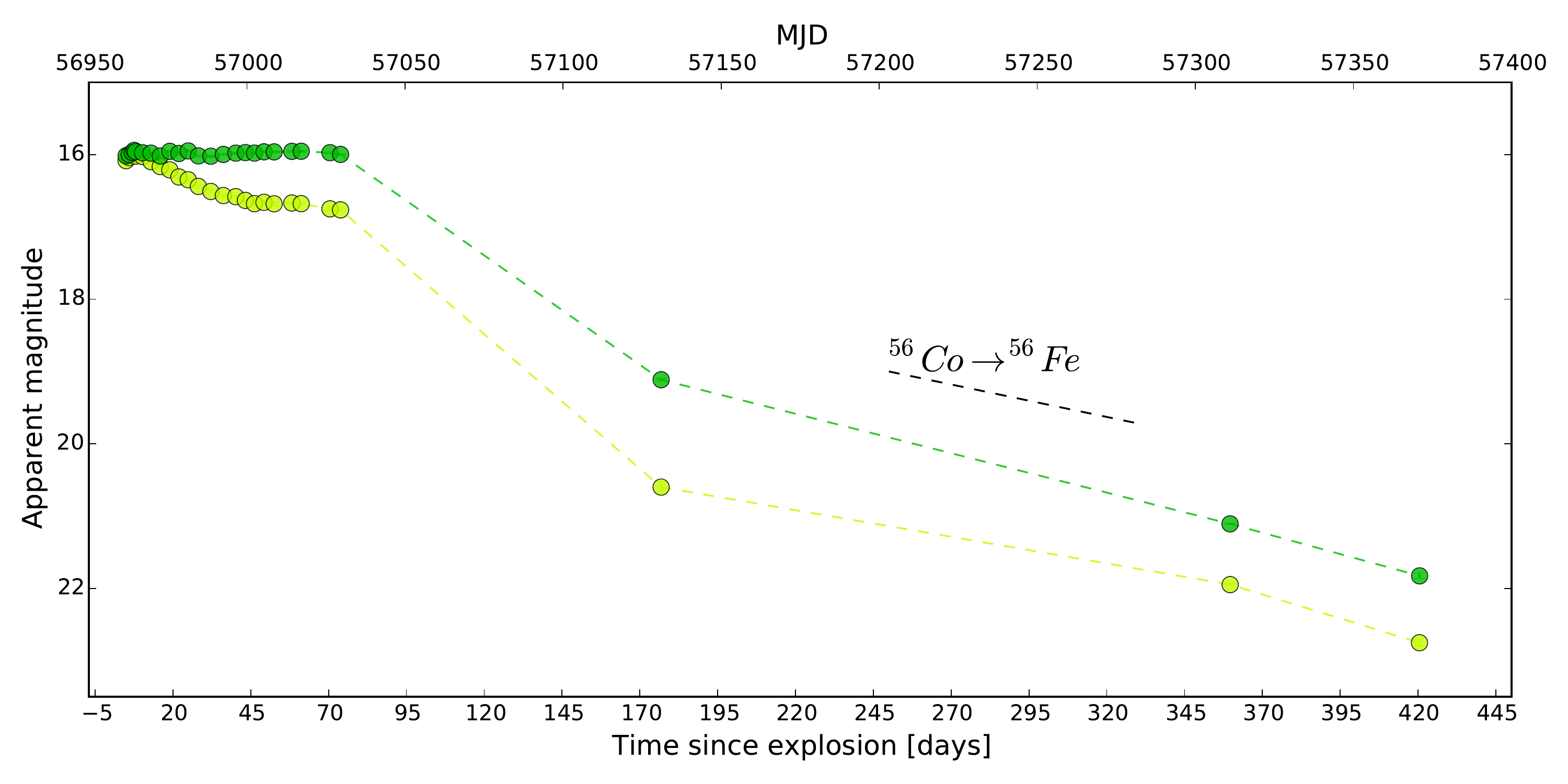}
\caption{Same as Figure \ref{fig:lc_early} for the EFOSC $B$ and $V$ late time photometry. The $^{56}Co \rightarrow ^{56}Fe$ decay slope is also showed. \label{fig:lc_late}}
\end{figure*}

Our near-UV and optical photometry is displayed in Figures~\ref{fig:lc_early} and \ref{fig:lc_late}. The first shows the near-UV and optical light curves for the first $\sim80$ days after explosion, and the second the complete light curves in $B$ and $V$ bands. The {\em Spitzer} mid-infrared photometry is discussed in \S~\ref{ss:dust}.

\subsection{Spectroscopy}

\begin{figure*}[ht!]
\centering
\includegraphics[width=\linewidth]{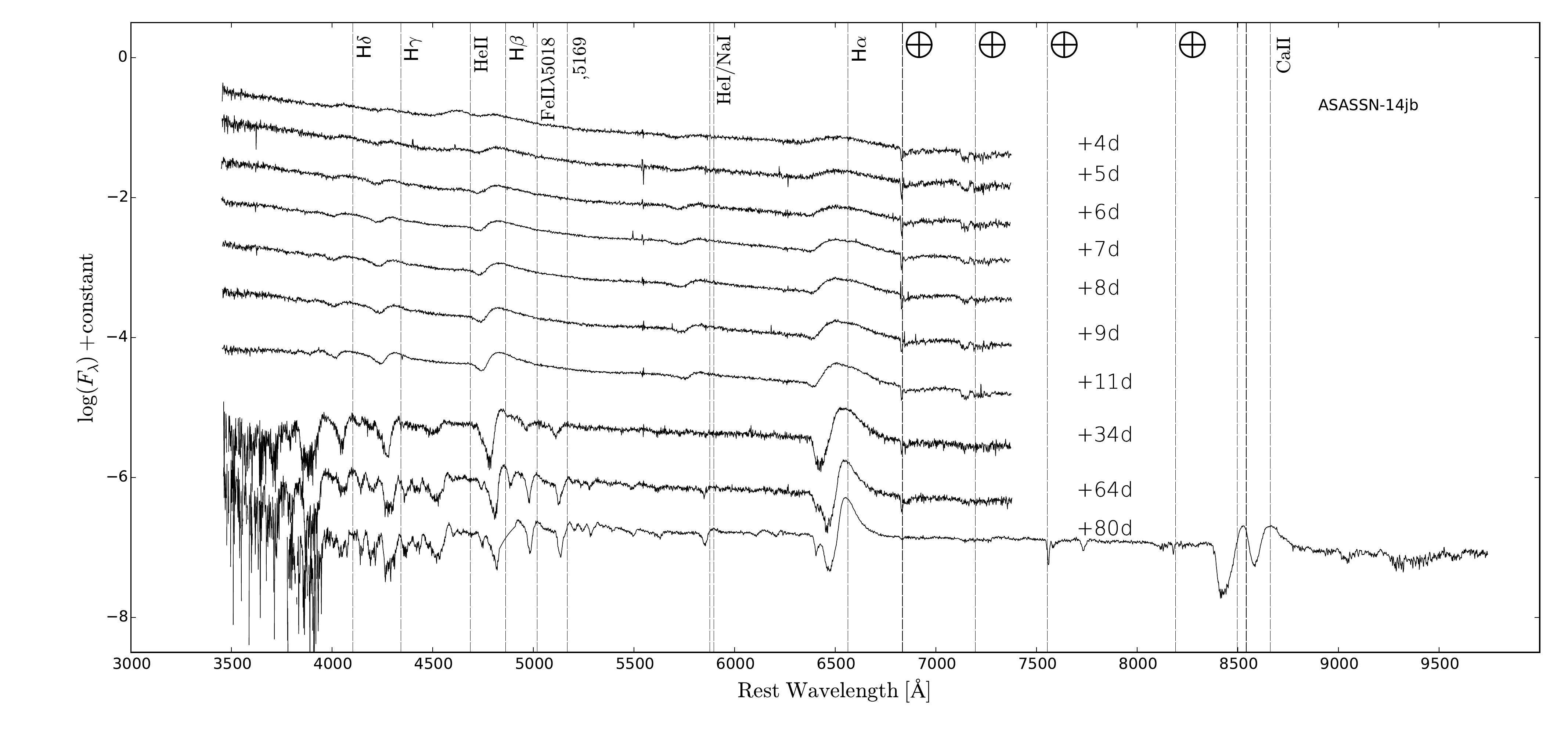}
\caption{The full spectroscopic time series, in logarithmic scale and corrected by redshift ($z = 0.006031$), for ASASSN-14jb in the photospheric phase (up to 80 days past explosion). The time next to each spectrum, in days, is measured since the estimated time of explosion $t_0 = 56946.1 \pm 3$ (MJD). The vertical lines show the wavelengths of the some of the main features in the spectra of type~II SNe. Lines with the $\bigoplus$ symbol represent the position of strong telluric absorptions. \label{fig:spectra}}
\end{figure*}

A total of ten single-slit, low-resolution optical spectra of ASASSN-14jb were obtained with the FAST spectrograph \citep{fabricant98} mounted on the Fred L. Whipple Observatory Tillinghast 1.5-m telescope ($3500-7400$~\AA, $R\sim1700$) and with the Inamori-Magellan Areal Camera and Spectrograph \citep[IMACS;][]{dressler11} mounted on the Magellan Baade 6.5m telescope at Las Campanas Observatory ($3800-9800$~\AA, $R\sim 1000$). The CCD images were reduced and the 1D spectra were extracted and calibrated using standard routines in the IRAF packages {\it ccdproc} and {\it onedspec}, respectively. The epochs of the ten spectra obtained in the photospheric phase are presented in Table~\ref{tab:spectra}. A montage of the spectra obtained in the photospheric phase is shown in Figure~\ref{fig:spectra}.  

We also observed the field of ASASSN-14jb as part of the All-weather MUse Supernova Integral field Nearby Galaxies (AMUSING; \citealt{galbany16b}) with the Multi Unit Spectroscopic Explorer (MUSE, \citealt{bacon10}) on ESO's Very Large Telescope UT4 (Yepun). MUSE is a state-of-the-art integral field spectrograph with a field of view of $1\,\mathrm{arcmin}^2$ and $0\farcs 2$ spaxels. It covers the spectral range $4800-9300$~\AA\ with a resolving power  $R \simeq 1800-3000$. Our MUSE data were obtained on 2015-11-14 and consisted of two different pointings of four dithered exposures each with an integration time of 698~sec. The sky conditions were clear, and we measure a full-with-half-maximum for the stellar point-spread function of 1\farcs{08} at 6600\,\AA. We reduced the MUSE spectroscopy with version 1.2.1 of the EO(fpipeline \citep{weilbacher14}, but also checked and corrected the astrometric zeropoint using LCOGTN images (for further detail on MUSE data reduction see \citealt{prieto16}). 

Although the MUSE observations were obtained $\sim394$~days after explosion (see Table~\ref{tab:spectra}) when ASASSN-14jb was faint, the quality and depth of the exposures allow us to easily detect the SN and extract a nebular phase spectrum. We used a circular aperture with a radius of $1\farcs 4$ to extract the spectrum at the position of the SN with QFitsView\footnote{\url{http://www.mpe.mpg.de/~ott/dpuser/qfitsview.html}}. To take care of any offset in the flux scale we estimated the $V$ magnitude of the SN at the time of the spectrum from the late time light curve ($V=21.51$) and compared it with the synthetic $V$-band magnitude obtained from the nebular spectrum ($V=21.86$). We then applied an offset of 0.35~mag (factor of 1.38 in flux) to obtain an approximate absolute flux calibration. The nebular spectrum of ASASSN-14jb is displayed in Figure~\ref{fig:Neball}.

\section{Comparative Analysis}
\label{sec:analysis}

To put ASASSN-14jb in context we assembled a comparison sample representative of the wide range of properties of Type II SNe. These include the luminous Type~IIP SN~2007pk \citep{Inserra13}, normal Type~IIP like SN~1999em \citep[e.g.,][]{hamuy01}, and the subluminous Type~II SN~2005cs \citep[e.g.,][]{pastorello09}. We chose objects with data in the public domain, good coverage of the photospheric phase, ideally including $gri$ photometry (8 SNe), and optical nebular spectra at $\sim 400$ days after explosion (6 SNe). Further details of the comparison set are given in Table~\ref{tab:comp_Table}. 

Estimates of the distance and total foreground reddening are needed
in order to compare ASASSN-14jb with the rest. We use a distance modulus of $\mu = 32.0$~mag, to be justified below.

We estimate extinction by dust in the Galaxy using the maps of \cite{schlafly11}. Taking the average in a circle with 5 arcminutes of diameter centered on the SN position we obtain $E(B-V) = 0.0154 \pm 0.001$~mag.
Extinction by foreground dust in the host galaxy is expected to be low, as the SN exploded at a significant distance from the host galaxy disk.
Consistent with this, the narrow Na~I~D doublet at the host galaxy redshift are not detected in our spectra, which is a clear sign of low reddening \citep{phillips13}. In addition, the MUSE data to be discussed below shows no significant star forming region near the explosion site. We conclude, hence, that ASASSN-14jb is affected only by Galactic extinction. Assuming a standard reddening law with a ratio of total to selective absorption $R_V=3.1$, we estimate $A_V = 0.05$~mag. 

\subsection{Photometric Evolution}

\begin{figure}[ht!]
\includegraphics[width=\linewidth]{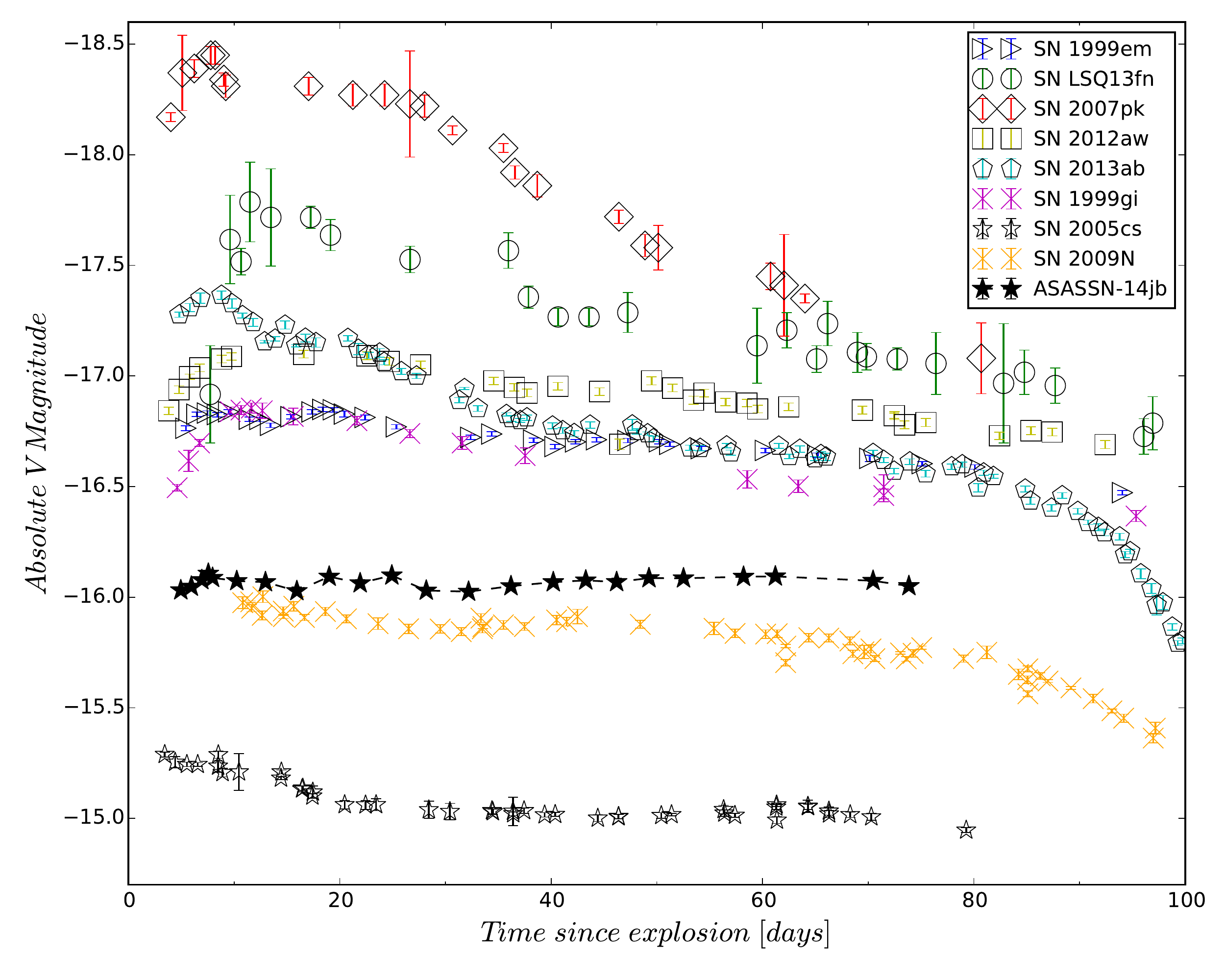}
\caption{Absolute $V$-band magnitude evolution for ASASSN-14jb (black filled stars) and the comparison sample of Type~II SNe as a function of days after their estimated time of explosion. See Table \ref{tab:comp_Table} for references of each SNe, including distance and reddening adopted.\label{fig:Vcomp}}
\end{figure}

Figure~\ref{fig:Vcomp} shows the absolute $V$-band light curve of ASASSN-14jb together with those of the comparison set. The peak at $M_{V,max} = -16.04 \pm 0.18$~mag is within the usual range of Type~IIP SNe, although lower than the average $M_{V,max} = -16.71 \pm 1.01$~mag in the sample of Type~II SNe of \citealt[][(from now onward A14)]{Joe14b}. ASASSN-14jb is dimmer than the classic, well-studied Type~II-SN~1999em by $0.80 \pm 0.18$~mag, but it shows a similar behavior in its $V$-band plateau, including the early double peak \citep{hamuy01,leonard02a}. Up to $\sim 80$ days since explosion the photometry shows the behavior typical of Type~IIP like SNe.

\begin{figure}[ht!]
\includegraphics[width=\linewidth]{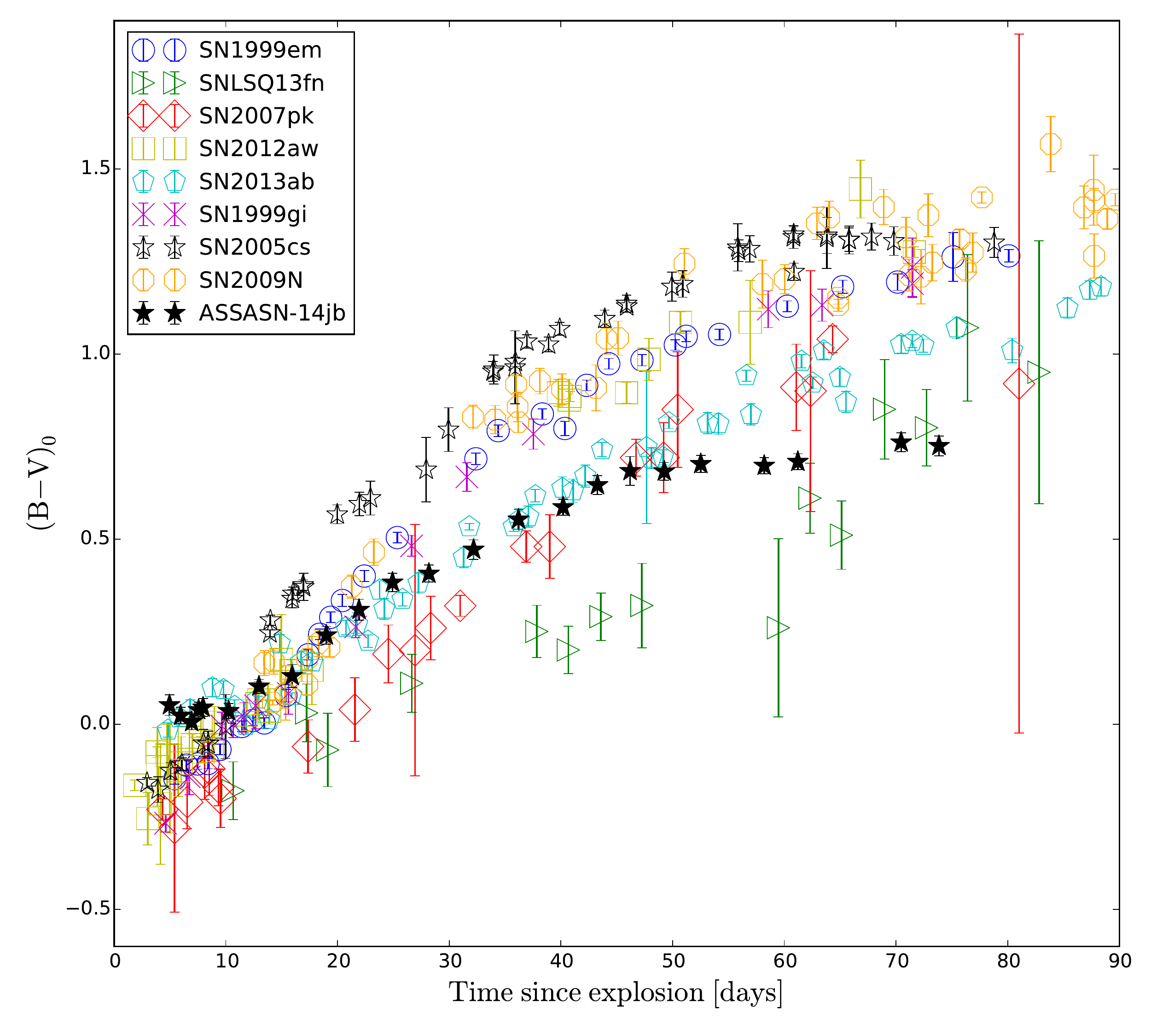}
\caption{Extinction corrected $B-V$ colors for ASASSN-14jb (black filled stars) and the comparison sample of Type~II SNe. Reddening values used are in Table \ref{tab:comp_Table}.\label{fig:BV}}
\end{figure}

\begin{figure}
\includegraphics[width=\linewidth]{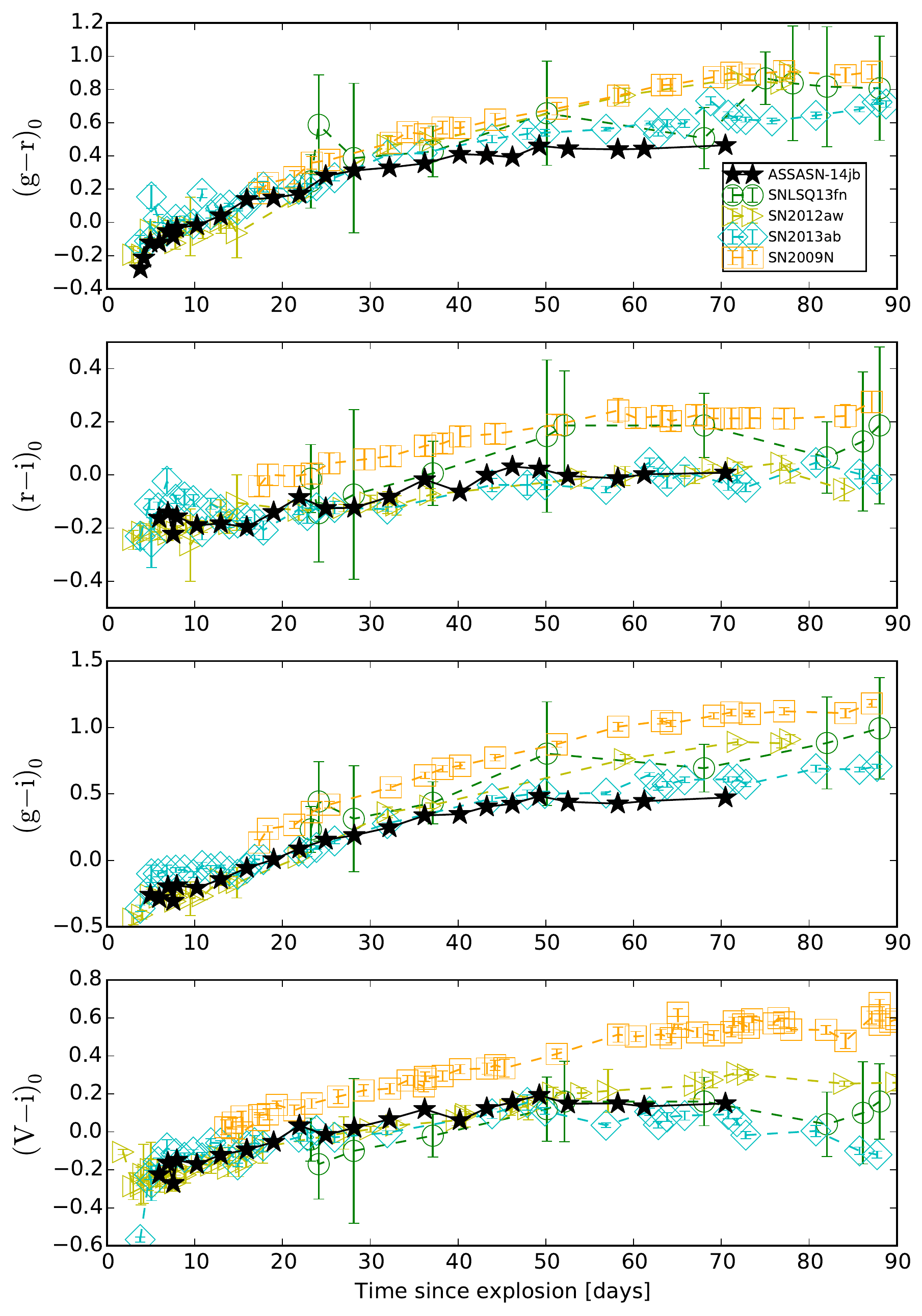}
\caption{Extinction corrected $V-i$, $g-i$, $r-i$ and $g-r$ colors for ASASSN-14jb (black filled stars) and the Type~II SNe comparison sample. Reddening values used are in Table \ref{tab:comp_Table}.\label{fig:gri}}
\end{figure}

The measured slope of the plateau ($35-70$~days after explosion) in the $V$-band for ASASSN-14jb is $s_{2,V} = -0.15 \pm 0.02$~mag per 100~days, lower than the average $V$-band plateau slope in the A14 sample ($1.27 \pm 0.93$~mag per 100~days). The slow increase in brightness of the plateau is consistent with the correlation between plateau slope and absolute $V$-band magnitude found by A14 and later work \citep[e.g.,][]{valenti16}. It is also an effect seen in bolometric light curves of CCSNe models resulting from the explosion of low mass progenitors \citep{Sukhbold2016}.

In Figures~\ref{fig:BV} and \ref{fig:gri} we present the extinction corrected color evolution of ASASSN-14jb and the comparison set. The $(B-V)_0$ color of ASASSN-14jb is bluer than average at $50$~days after explosion, being comparable only to LSQ~13fn \citep{polshaw16} and SN~2007pk \citep{Inserra13} at 80~days after explosion. The change in color from the onset of the plateau up to 80~days is $\Delta(B-V) = 0.20 \pm 0.04$~mag, low when compared with the comparison sample. The color evolution of ASASSN-14jb in $(g-r)_0$, $(g-i)_0$ and $(r-i)_0$ also indicates a comparatively bluer continuum, except in the $(V-i)_0$ color. This behavior may be due to a true difference in the temperature evolution or it may be an effect of lower line-blanketing in the blue part of the spectrum, resulting from low-metallicity of the progenitor. The similarity in the color evolution of ASASSN-14jb and LSQ~13fn, a Type~II SN that probably comes from a low-metallicity progenitor \citep{polshaw16}, might suggest that a similar physical mechanism is at play. 

\subsection{Spectroscopic Evolution in the Plateau phase}

As shown in Figure~\ref{fig:spectra}, the spectral evolution of ASASSN-14jb resembles the evolution of other hydrogen-rich Type~II SNe. At very early times the ejecta are optically thick and the spectra show a hot blue continuum with relatively weak, but broad, Balmer lines without P-Cygni absorption troughs. In the first spectrum at $\sim 4$~days after explosion, we detect broad Balmer lines in emission and the high ionization He~II $\lambda 4686$ line. The FWHM of the Balmer lines and the He~II lines in this earliest spectrum are $\sim 11000-13000$~km/s and $\sim 8000$~km/s, respectively, and all the emission peaks are clearly blueshifted by $\sim 2000-4000$~km/s. The strength of the He~II line decreases substantially in the spectrum obtained a day later, at 5~days after explosion, and the Balmer lines start to develop P-Cygni profiles. We also detect the He~I~$\lambda 5875$ line in the early spectra. At 6~days after explosion, the He~II line has weakened further and it is undetected in the spectrum obtained at 7~days after explosion. 

\begin{figure}[ht!]
\includegraphics[width=10cm]{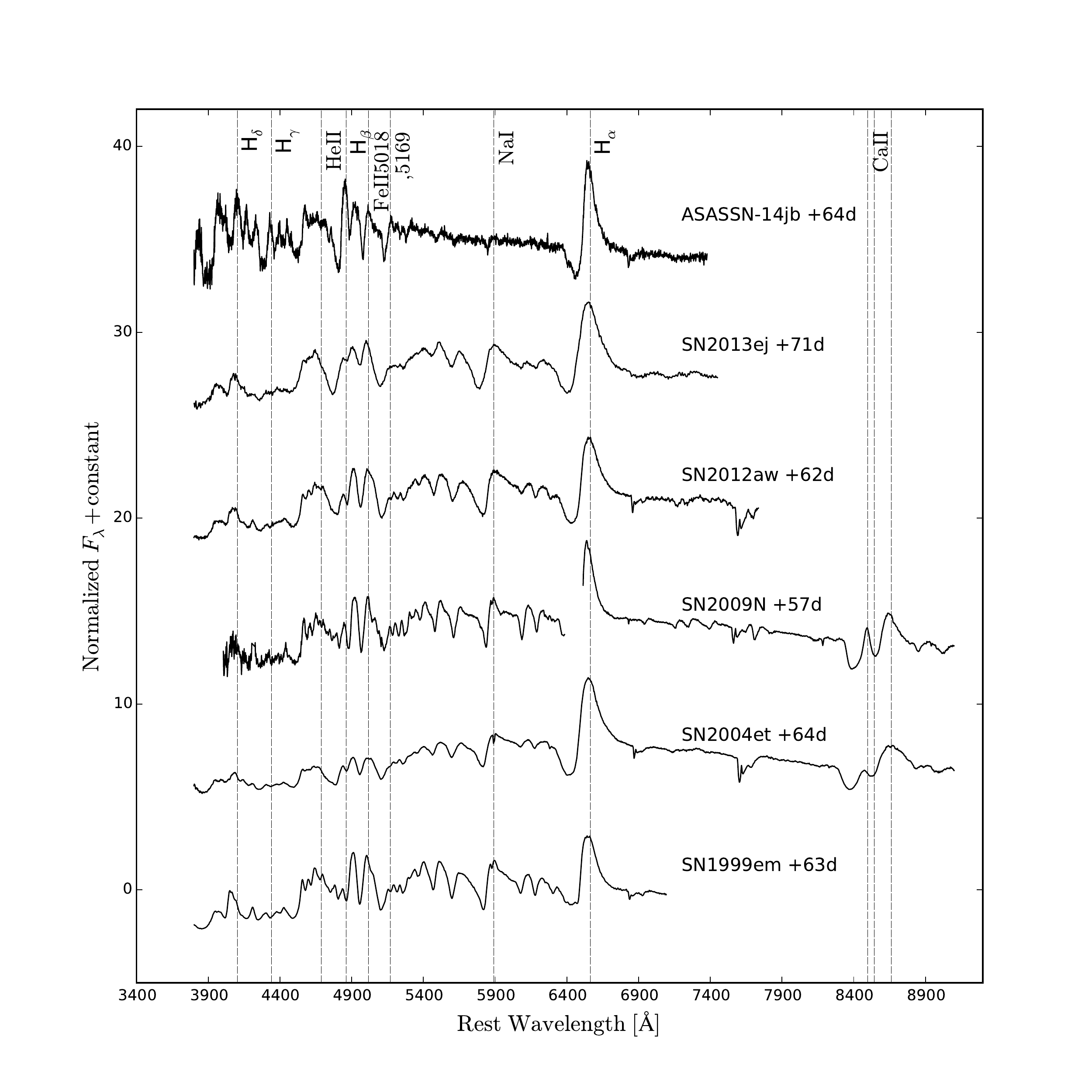}
\caption{Comparison of rest-frame photospheric phase spectra at $\approx 50$ days after explosion. The spectra are not corrected for extinction. \label{fig:speccomp}}
\end{figure}

At phases later than $\gtrsim 30$~days, the ejecta cools and the hydrogen lines develop strong P-Cygni profiles. The blueshift of the peak of the Balmer lines decreases to $\sim 1000$~km/s. This is a common feature in Type~II SNe during the photospheric epochs and it has been attributed to the shallow density profile of the ejecta compared to a typical stellar wind \citep{Luc05a,Joe14}. Lines from transitions in metals such as Fe~II, Sc~II, and Ti~II also start to appear and significantly contribute to line blanketing. The Na~I D doublet at $\lambda\lambda5890,5896$ appears now in the wavelength range where He~I~$\lambda 5875$ was detected at early times. Comparison of the photospheric phase spectra of ASASSN-14jb with those of the comparison sample at $\sim 60$ days past explosion (see Figure \ref{fig:speccomp}) shows that the absorption features of the former in the range $5300-6300$~\AA\  are relatively weak. 

\begin{figure}[h]
\includegraphics[width=9.5cm]{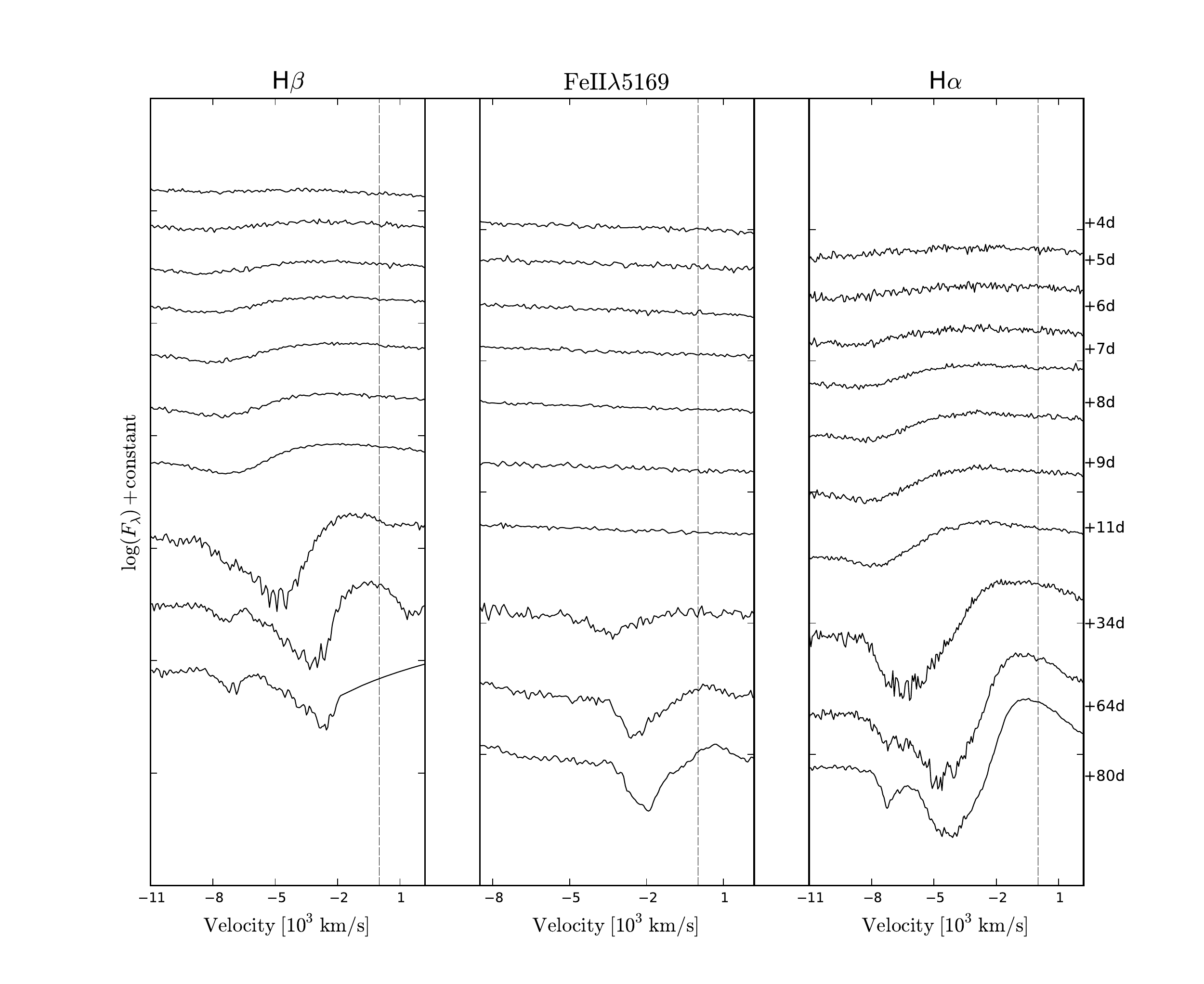}
\caption{Evolution of the H$\beta$ (left), Fe~II~$\lambda 5169$ (middle) and  H$\alpha$ (right) profiles, in velocity space, as a function of time after explosion $t_0 = 56946.1 \pm 3$ (MJD), in the plateau phase for ASASSN-14jb (up to 80 days past explosion). In each panel, a vertical dashed line indicates zero velocity.}
\label{fig:lines}
\end{figure}

In Figure~\ref{fig:lines} we present the time evolution of the H$\alpha$, H$\beta$ and Fe~II $\lambda 5169$ lines. The last two spectra we obtained in the plateau phase, at 64 and 80 days after explosion, show clear absorption features on the blue side of both the H$\alpha$ and H$\beta$ main absorption troughs.
These high velocity (HV) features are at $\sim 7000$~km/s, while the main absorption features are at $\sim 4000$~km/s. The HV absorption lines cannot be explained by Si~II~$\lambda 6355$ or Ba~II absorption features and have been observed in $59\%$ of the Type~II SNe studied by \citet{claudia17a}.

\begin{figure}[ht!]
\includegraphics[width=\linewidth]{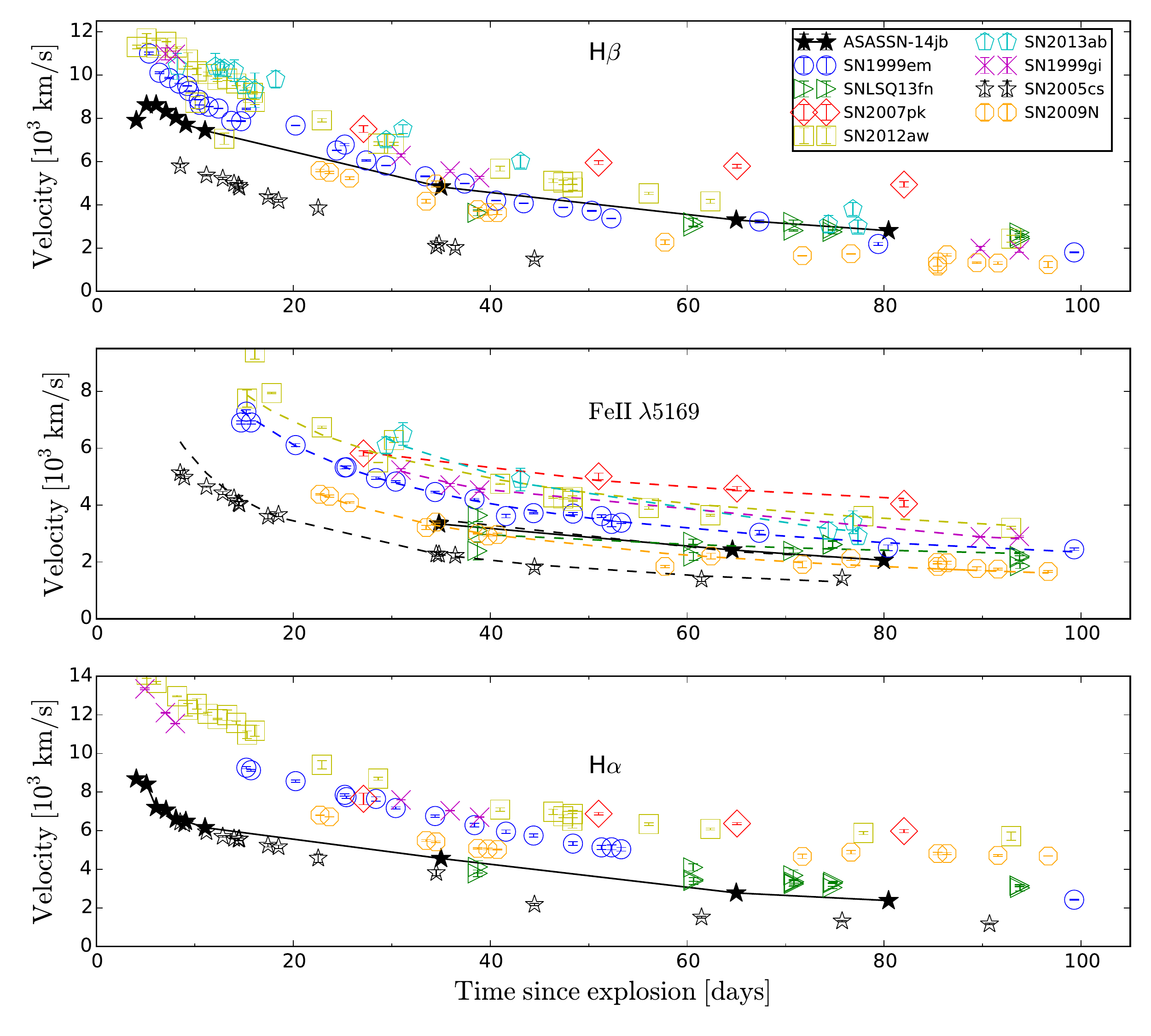}
\caption{Expansion velocities of the H$\beta$ (top), Fe~II~$\lambda 5169$ (middle), and H$\alpha$ (bottom) lines as a function of time after explosion in the plateau phase, measured from the P-cygni profile minima of the Balmer lines $\mathrm{H}{\alpha},\mathrm{H{}\beta}$ and the Fe~II $\lambda5169$ line for ASASSN-14jb (black filled stars) and the Type~II SNe comparison sample. \label{fig:comp_vels}}
\end{figure}

\subsection{Expansion velocities}

The ejecta of the Type~II SNe achieve near homologous expansion after a few days \citep{melina11} and at early times the hydrogen atmosphere is opaque and the line forming region is in the outer layers at high velocities. As the temperature drops, the opacity decreases and the photosphere recedes in mass (Lagrangian) coordinate, and appears at lower velocities.
The photosphere is tightly bound to the opacity drop caused by hydrogen recombination \citep{melina11}. In practice one estimates the photospheric velocity by measuring the blueshift of maximum absorption of an specific P-Cygni profile. This can over or under estimate the true photospheric velocity \citep{Luc05b,luc10,katy12}. Other methods used are the cross-correlation with library spectra, the comparison with detailed NLTE codes like CMFGEN or PHOENIX \citep[][]{CMFGEN,PHOENIX} or comparison with more simplified, parametrized LTE spectral modeling (e.g., SYNOW, \citealt{katy12}).

The expansion velocity measured from optical spectra is an important physical parameter directly related to the explosion energy and the plateau luminosity. These are also fundamental for distance measurements \citep{hamuy02}. Strong Balmer lines and Fe~II lines velocities were measured in our spectra.
The spectra were first normalized to a global continuum fitting a black body or a local power law. Then, when
possible, low degree polynomials were fitted to the P-Cygni profile absorptions and the minimum was taken as a proxy for the expansion velocity.
We did not apply corrections by reddening or peculiar velocities. Details of the P-cygni profile fitting are in Appendix B.
In Figure~\ref{fig:comp_vels} we show the expansion velocities of ASASSN-14jb together with those of the comparison set. For LSQ~13fn  we only present the velocities after 30 days due to the low signal to noise of the spectra. The velocities of SN~2013ab are directly taken from \cite{bose15}. The measured velocities are also collected in  Table~\ref{tab:vel} together with the average velocity of our comparison set and those of \cite{claudia17b}.

To compare with well known correlations \citep{hamuy03,faran14a}, we interpolated the velocities of our sample at 50 days after explosion using a Monte-Carlo approach. For each SN we generated a re-sampling of each velocity using the measured errors and following a non-linear least squares procedure, using the curve fit function from the NumPy library, we fit a power law model each time. For this procedure we select only velocities after 20 days, as early velocities do not follow the same decay rate as in the plateau. We then take the average and standard deviation of the re-samplings as the 50 days velocity and error, respectively. For all the lines measured the velocities for ASASSN-14jb are slightly under average. The Fe II $\lambda5169$ velocity interpolated at 50 days is measured to be $2774 \pm 69$ km/s, compared to the average of $3442 \pm 977$ km/s of our comparison sample. Our sample average velocity is in agreement with the value obtained with a larger sample in \cite{claudia17b}, of  $3537 \pm 851$ km/s at 53 days after explosion. The photospheric velocities inferred are in agreement with the velocity-luminosity correlations (see Section 3.2 below). 
 
As is common in Type II SNe distance measurements \citep{hamuy02,tomato17}, we fitted a power law to interpolate the velocities. The Fe~II $\lambda 5169$ logarithmic decay presents an average of $-0.55 \pm 0.16$ in our sample, compared to the value obtained for ASASSN-14jb, of $-0.62 \pm 1.80$. We observe a strong anti-correlation between the velocity decay slope and the $V$-band absolute magnitude at 50 days. This is expected as there is an internal correlation between both parameters of the power law \citep{tomato17} and the velocity scale correlates with the luminosity \citep{hamuy02}. 

In conclusion, ASASSN-14jb presents expansion velocities consistent, both from the scale and velocity decay, with below average luminosity Type II SNe.
 
 \begin{figure*}[ht!]
\includegraphics[width=\linewidth]{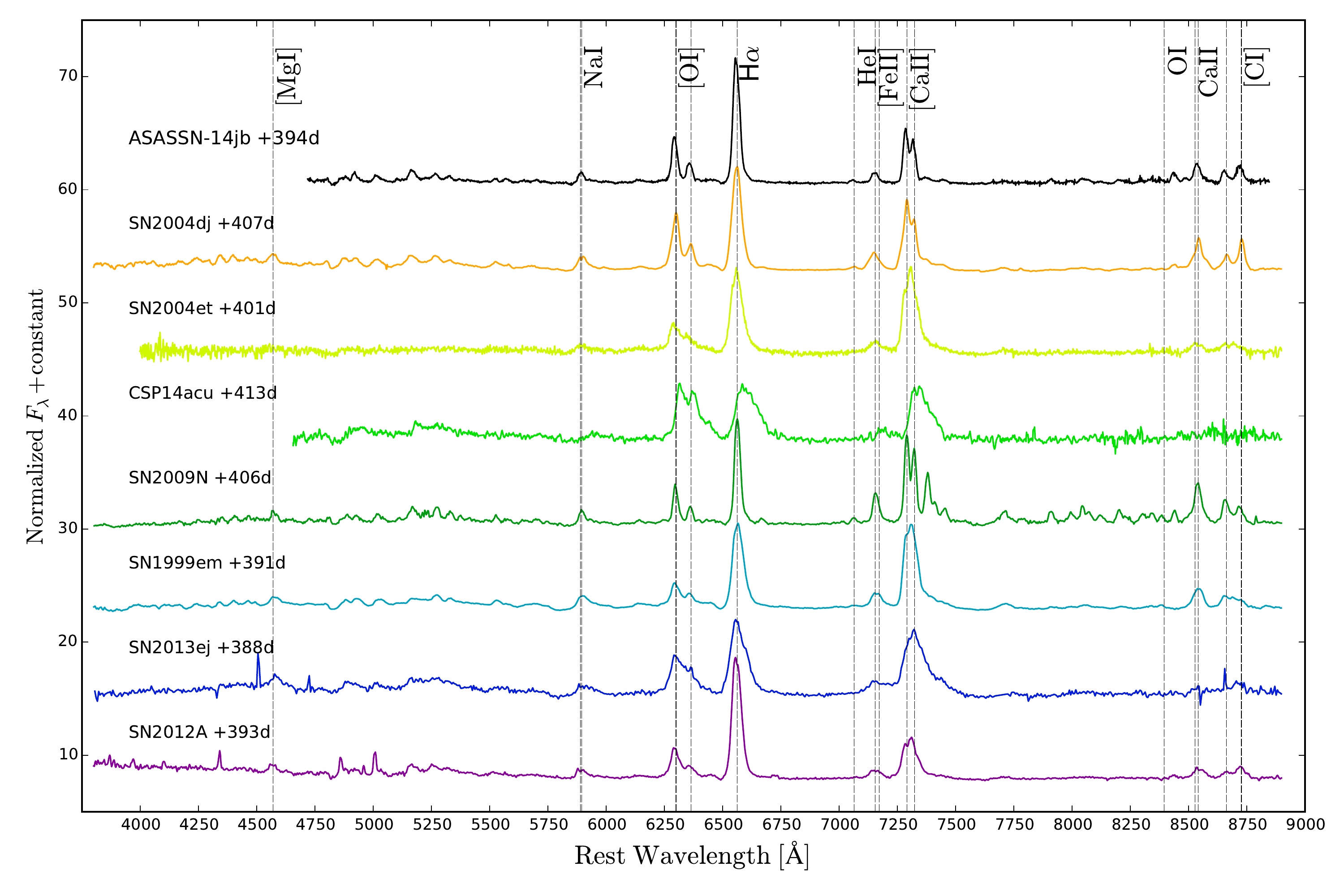}
\caption{Comparison of nebular spectra of observed Type II SNe at $\approx 400$ days past explosion. \label{fig:Neball}}
\end{figure*}

\subsection{Nebular Spectrum}

The nebular spectrum of ASASSN-14jb is shown in Figure \ref{fig:Neball} together with a sample of Type II SNe.
It shows the typical emission lines of Type IIP like SNe at nebular phases: H$\alpha$, [Ca II] $\lambda\lambda7291,7324$, [O I] $\lambda\lambda6300,6364$, Ca II $\lambda\lambda\lambda8498,8542,8662$, [Fe II] $\lambda\lambda7155,7172$, with a boxy profile, Na I $\lambda\lambda5890,5896$ and [O I] $\lambda5577$, in order of observed intensity \citep{silverman17}.
Also, a strong [C I] $\lambda8727$ line is present, which has been detected in other Type IIP like SNe such as SN~2004dj \citep{silverman17}. A weak He~I line at $\lambda 7065$ is observed, but with no He I counterpart at $\lambda 6678$. The clear separation observed in different multiplets like [OI], [CaII] and Ca II + [CI] is a sign of a low explosion energy for ASASSN-14jb.

The blueshift with respect to rest-frame of $\sim 200$~km/s observed in the peaks of the strongest lines is particularly interesting. It cannot be caused by the same physical effect as in the plateau because in this optically-thin phase the outer density profile is less relevant to the emission \citep{Joe14}. Other hypotheses to explain this blueshift may be: 1) a shift in velocity relative to the host due to dynamical effects or an intrinsically high-velocity of the progenitor star; 2) dust production; 3) Some lines are still optically thick at 400 days past explosion; or 4) an asymmetry in the explosion itself. Notable is the double peak observed in $H\alpha$ which is not observed in other lines, although this is also seen in the nebular spectra of other Type IIP like SNe \citep{elmhamdi03_99em,Chugai2005,Utrobin2009}.  

\section{Distance and Physical parameters of ASSASN-14jb}
\label{sec:parameters}
Progenitors of Type IIP like SNe have been constrained by pre-explosion images to be RSGs with ZAMS masses of  $\approx 8-17$ M$_{\odot}$ \citep{Smartt09,Smartt15}. The discrepancy between these initial mass constraints and the predictions from evolutionary codes \citep{heger03,ekstrom12} has been called the ``RSG problem".
The data set we have compiled allow us to constrain the basic physical parameters of the progenitor of ASSASN-14jb,
compare them with the expected values for SN Type IIP like progenitors, and contribute to this discussion. But doing so requires a good estimate of the distance to the SN.

\subsection{Distance}

We measure the distance towards ASSASN-14jb using some of the most recent empirical methods: The Photospheric Magnitude Method \citep{Osmar14}, the Standard Candle Method \citep{hamuy02}, and the Photometric Color Method, \citep{tomato17}. Further details of our hypothesis and calculations are given in Appendix C. 

\begin{figure}[ht!]
\includegraphics[width=9.5cm]{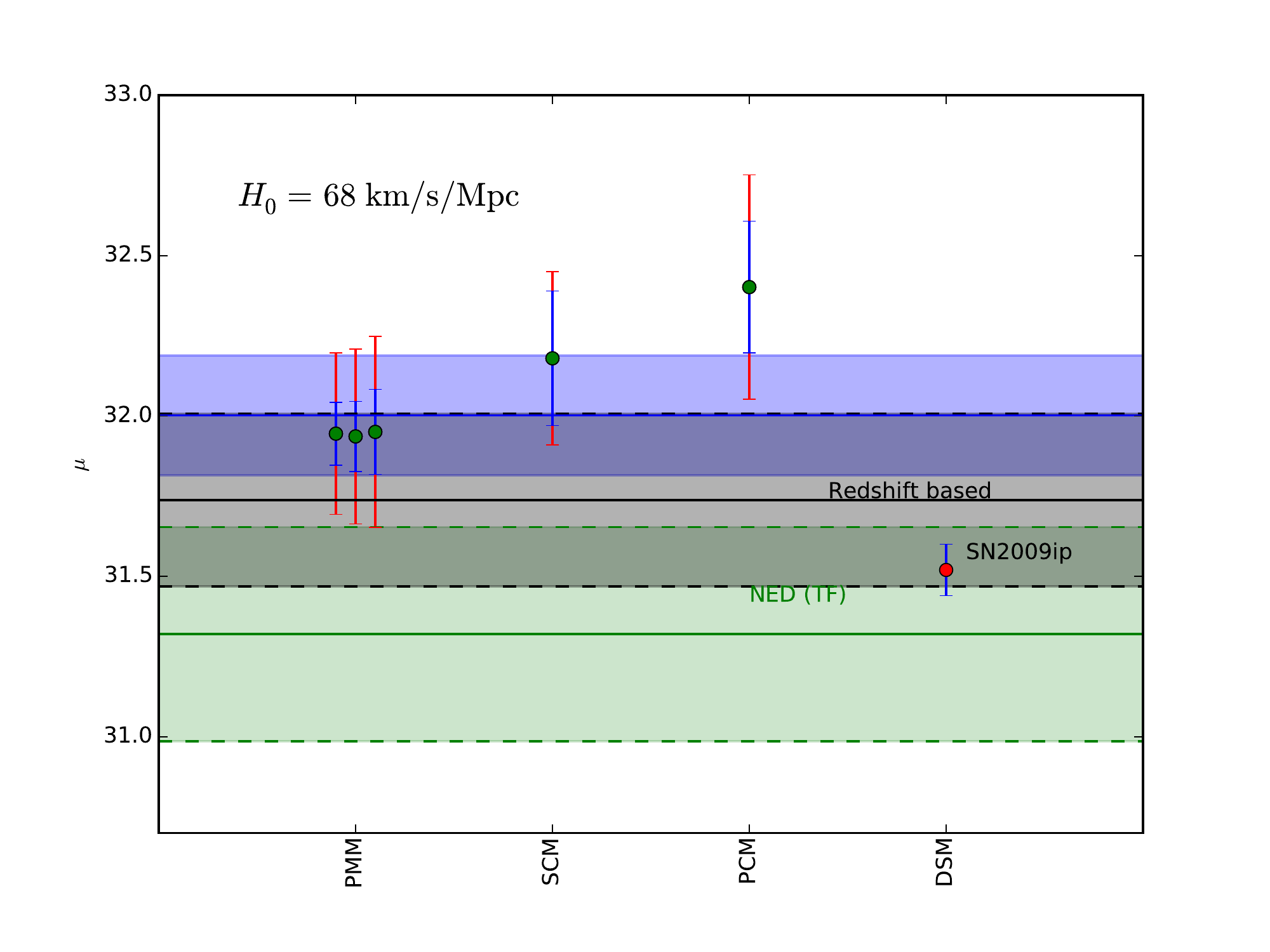}
\caption{Distance modulus estimates for ASASSN-14jb using the PMM, SCM and PCM methods (green dots). The black horizontal line is the distance modulus from the host galaxy redshift and a Hubble law with $H_0 = 68$ km/s/Mpc, and the grey area shows a dispersion of $200$ km/s as a reference error for the CMB frame redshift. The thick green line is the average from the Tully-Fisher measurements obtained from the NED database and the green area represent the standard deviation of the measurements, all normalized to $H_0 = 68$ km/s/Mpc. The thick blue line is the adopted distance modulus in this work, $\mu = 32.00\pm 0.18$, obtained as the weighted average of the PMM, SCM and PCM methods and the standard deviation is showed as a blue fill.  We also show as a reference the distance measured to SN 2009ip in NGC 7259 \citep{potashov}.\label{fig:distances}}
\end{figure}

The three different estimates of distance are shown in Figure \ref{fig:distances}. They are consistent with each other within each method's intrinsic dispersion. They are also consistent with the distance obtained from the Virgo-inflow corrected recession velocity of the host galaxy and the Hubble law, assuming $H_0 = 68\, \rm km/s/Mpc$. Given this, we simply take as the final distance the weighted average, considering the statistical errors only, of $\mu = 32.00 \pm 0.18$~mag ($D =25.1$~Mpc).

\subsection{Temperature Evolution and the Progenitor Radius}

The early emission, a few days after explosion, for a Type II SN comes from the adiabatic cooling of the outer envelope and depends mainly on the pre-supernova radius \citep{rabinak11,sapir2017}. Observationally, the pre-supernova radius has been constrained from early UV and optical photometry, using semi-analytic models \citep{rabinak11,nakar10}, on individual Type II SNe \citep{gezari10,valenti14,bose15,huang16,huang18,tartaglia18}, and on larger samples to make statistic studies \citep{santiago15,gall15,rubin16}. 

We build the spectral energy distribution (SED) for ASASSN-14jb at early times, from $\sim$4 to 26~days after explosion to derive an estimate of the progenitor radius. Using the {\it Swift/UVOT} near-UV and optical photometry, we build the SED of ASASSN-14jb at each epoch, after correcting for extinction assuming an standard Galactic extinction law with total to selective extinction ratio $R_V = 3.1$ \citep{cardelli}. We fit a blackbody function to the SED and obtain the photospheric temperature and radius evolution, assuming the distance to the host galaxy obtained in Section 3.2. 

\begin{figure}[ht!]
\includegraphics[width=9.5cm]{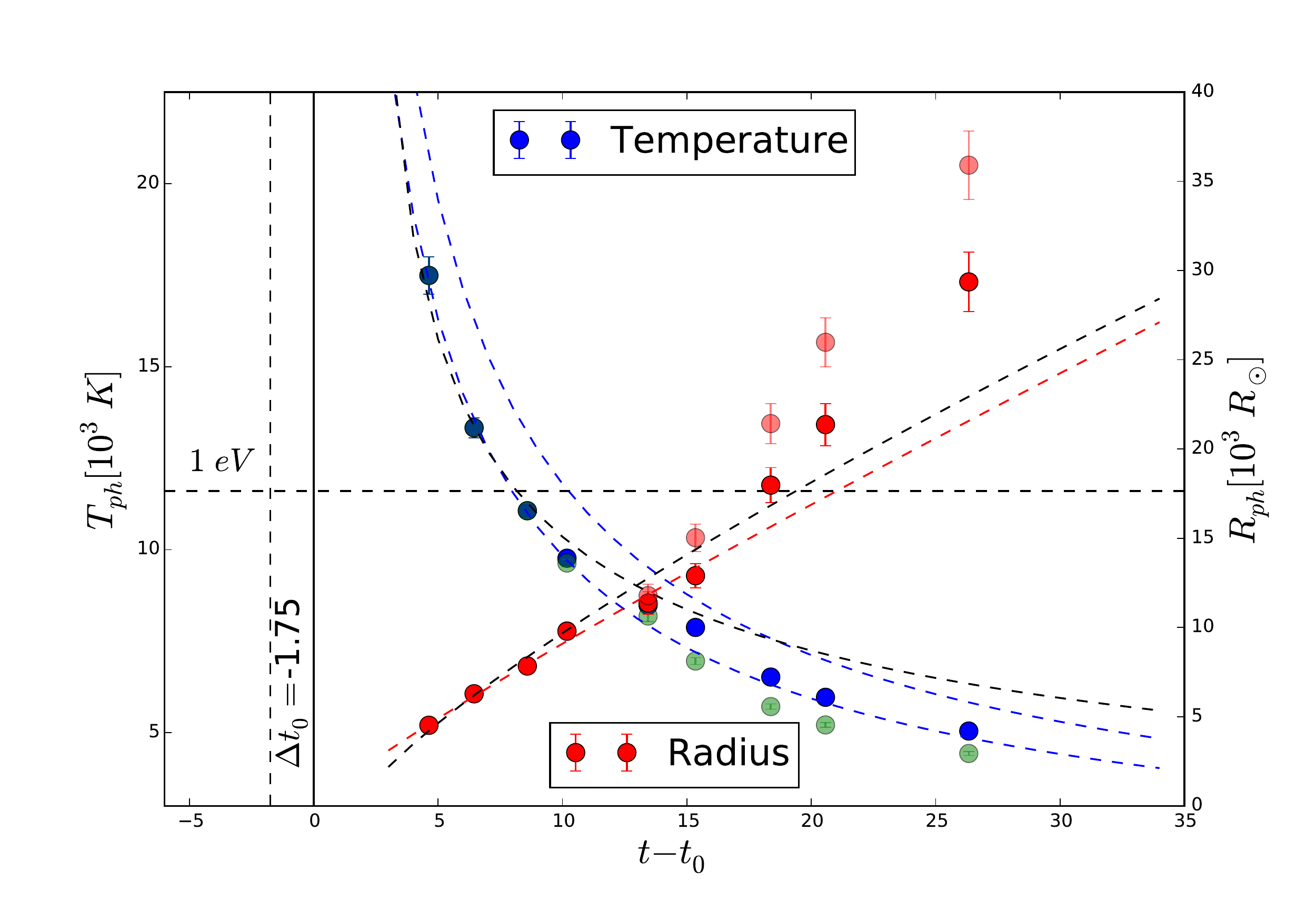}
\caption{Black-body temperature (blue dots) and radius evolution (red dots) as inferred from the SED of $Swift$ data. The SED was corrected by a reddening value of $E(B-V) = 0.0154 \pm 0.001$~mag. The horizontal line indicates the 1 eV level, where recombination is expected to become relevant afterwards. The vertical black solid line indicates the assumed explosion epoch at MJD $56946.1 \pm 3$ and the vertical dashed lines is at the assumed offset ($\Delta t_0 = -1.75$) in the explosion epoch, for the optimal cooling model from RW11. The red dashed lines shows a power-law fit to the photospheric radii (only using the data points with $1-1.2\times T_{ph}< 1 eV$) while the blues dashed lines shows the same for the Temperatures plus a line with a factor of 1.2 (i.e., the lines shows a fit to $1-1.2 \times T_{ph}$). The black curved dashed lines are the fits assuming the RW11 for $T_{ph}$ and $R_{ph}$.\label{fig:temp}}
\end{figure}

The results are presented in Figure~\ref{fig:temp}. The observed photospheric temperatures are in the range 5-21$\times 10^3\,K$ and the radii are in the range 2-18$\times 10^3$~R$_{\odot}$. We observe a change in the evolution $\sim 10$~days after explosion. The observed temperature logarithmic decay increases from $-0.726 \pm 0.055 $ to $-0.850 \pm 0.019$ and the photospheric radius growth rate increases from $ 0.890 \pm 0.012 $ to $  1.45 \pm  0.06 $.

\citealt[][hereafter RW11,]{rabinak11} developed analytic expressions for the early evolution of the photospheric radius $R_{ph}$ and temperature $T_{ph}$, expected to be valid before hydrogen recombination is significant. The expressions depend on progenitor properties at the time of explosion such as radius $R_{*}$, opacity $\kappa$, and density profile. They also depend on the explosion energy $E$ and ejecta mass $M_{ej}$.
\begin{eqnarray}
\frac{R_{ph}}{[cm]} &=& 3.3 f_{\rho}^{-0.062} E_{51}^{0.41}M_{ej}^{-0.31}\kappa_{0.34}^{0.093} t_5^{0.81} \times 10^{14}  \label{eq:Rph}\\
\frac{T_{ph}}{[eV]} &=& 1.6 f_{\rho}^{-0.037}
E_{51}^{0.0027}M_{ej}^{-0.054}R_{*,13}^{1/4}\kappa_{0.34}^{-0.28} t_5^{-0.45} \label{eq:Tph}
\end{eqnarray}
where $0.079 < f_{\rho} < 0.13 $ for RSGs, depends on the progenitor density profile, $M_{ej}$ is the ejected mass in solar masses, $E_{51}$ is the explosion energy in units of $10^{51}$~erg, $\kappa_{34}$ is the opacity normalized by 0.34, $R_{*,13}$ is the progenitor radius in units of $10^{13}$~cm and $t_5$ is the time since shock breakout in units of $ 10^5$ seconds.

It is well known that the emission from Type II SNe differs from a pure black body \citep{eastman96,Luc05b}. When the atmosphere is highly ionized, the electron scattering opacity dominates and the photon thermalization layer differs from the photosphere, the former being underneath at higher temperatures. Because of this the observed temperature (or color temperature $T_c$, as it is usually called) can differ significantly from the photospheric temperature $T_{ph}$. This effect, together with the increasing line blanketing and departures from the plane parallel atmosphere approximation, can make the observed SED significantly different from a black body. This effect can be corrected using a theoretically estimated {\em dilution factor} $\zeta$, so that the emergent flux is defined as, 
\begin{equation}
\label{eq:zeta}
F_{\lambda} = \theta^2\ \pi B_{\lambda}(T_c) = \zeta_{\lambda}^2 \left(\frac{R_{ph}}{d_L}\right)^2\pi B_{\lambda}(T_c)
\end{equation}
where
$d_L$ is the luminosity distance and $B_{\lambda}(T_c)$ is the Planck function \citep{eastman96,Luc05b}.
The dilution factor is in general wavelength dependent and directly affects the angular diameter distance estimate to a supernova. However, as we take the luminosity distance from empirical methods independent of \ref{eq:zeta}, the dilution factor only affects the photospheric radius estimation and not the temperature. %
Note, however, that in the RW11 scheme $\zeta := 1/f_T^2 \approx 0.694$, which translates in a factor of $1/\zeta \approx 1.44$ increase in the photospheric radius (see equation 38 in RW11). In our case we choose a representative value of $\zeta_{\lambda} = 0.5$ for this high temperature regime, as obtained in \cite{Luc05b,pejcha15}.

To test the effect of the deviations from a black body, we used a semi-empirical approach. We first selected the 15~$M_\odot$ progenitor mass supernova explosion model from \cite{luc13} that best fits the observed near-UV colors (named m15z8m8). Then, at each epoch we fit the SED using,
\begin{equation}
F_{\lambda}^{data} = \theta^2 \left(\pi B_{\lambda}(T_c)\right)^w \times \left(\frac{F_{\lambda}^{model}}{max(F_{\lambda}^{model})}\right)^{(1-w)}
\end{equation}
a model built from the weighted geometric mean of a black body and the normalized explosion model. The normalization ensures that the model's intrinsic brightness does not affect the fitted parameters. In principle, the weighting parameter $w$ is unconstrained but to ensure a smooth correction we fixed its value to be between $0.6-0.9$ and selected the model with the minimum residuals.
Figure~\ref{fig:temp} show our results for both approaches. We see that both temperature and radius evolution become smoother when including the models, probably because the hybrid approach accounts better for line blanketing.
Since the available models start from $\approx 10$~days past explosion, all the SEDs before that time were fitted with a single black body.
The derived slope in the temperature changes to $-0.703 \pm  0.017$  after 10 days. The photospheric radius follows roughly a linear expansion with a logarithmic slope of $1.167 \pm 0.068$ during the same epochs.

Now we apply the RW11 models using the modified black body. We fitted the temperature evolution assuming the fiducial values of $E_{51}=1$, $\kappa_{34}=1$, $f_{\rho} = 0.1$, and use only temperatures greater than $\sim 1$~eV because RW11 inform that their models are more reliable in this range. The radius we obtain is $ 985\pm 49$~R$_{\odot}$.
Noting the impact of the time since explosion in the result, we varied the shock-breakout (SBO) epoch and look for the one that minimizes the $\chi^2$ of the fit. We then obtain $R = 580 \pm 28$~R$_{\odot}$, for a SBO occurring $1.75$~days earlier than our initial estimate (reduced $\chi^2= 3.62$).

Inverting Equation \ref{eq:Tph} is another form of understanding possible variations of $R_*$:
\begin{equation}
R_* \approx 1.52 \times f_{\rho}^{0.148} \left[\frac{T_{col}(t)}{(f_T/1.2)}\right]^4 t_5^{1.8} \times 10^{12} [cm]. \label{eq:R*}
\end{equation}
where $f_T = T_{col}/T_{ph}$ is the ratio of the color temperature and true photospheric temperature.
Following RW11 we take $f_T = 1.2$ as a good approximation at early times
\footnote{We note that this equation is a corrected version of Equation 33 in RW11. The original one is $R_{*} = 0.7 \times f_{\rho}^{0.1} \left[T_{col}/(f_T/1.2)\right]^4 t_5^{1.9} \times 10^{12} [cm]$, which is inconsistent with Equation~13 of the same paper (our Equation~\ref{eq:Rph}).}.

The values obtained using Equation~\ref{eq:R*} with early time observations are consistent with $R_* \approx 400-1000$~R$_{\odot}$. As expected, the radius decreases as the actual decay of the photospheric temperatures is faster than the model, and the $f_T$ varies due to the changing opacity of the ejecta.

We choose as our best estimate of the progenitor radius, the RW11 fit with an SBO offset of $-1.75$~days, which gives an estimate of  $R_* = 580 \pm 28$~R$_{\odot}$. This value is consistent with the color evolution of the mildly sub-solar (Z = 0.4
Z$_\odot$) and compact ($R_* =  611$~R$_{\odot}$) RSG progenitor (m15z8m3) used in \cite{luc13}.
As the photospheric radius growth strongly depends on the explosion energy to ejecta mass ratio (equation~\ref{eq:Rph}) we can estimate the ejecta mass as a function of the explosion energy.
Fixing $\kappa_{34}=1$,$f_{\rho} = 0.1$ as before
we obtain $M_{ej} \approx 36.0 \times E_{exp}^{1.32}$.

\subsection{Bolometric Light Curve and the $^{56}$Ni Mass}

\begin{figure}[ht!]
\includegraphics[width=9.5cm]{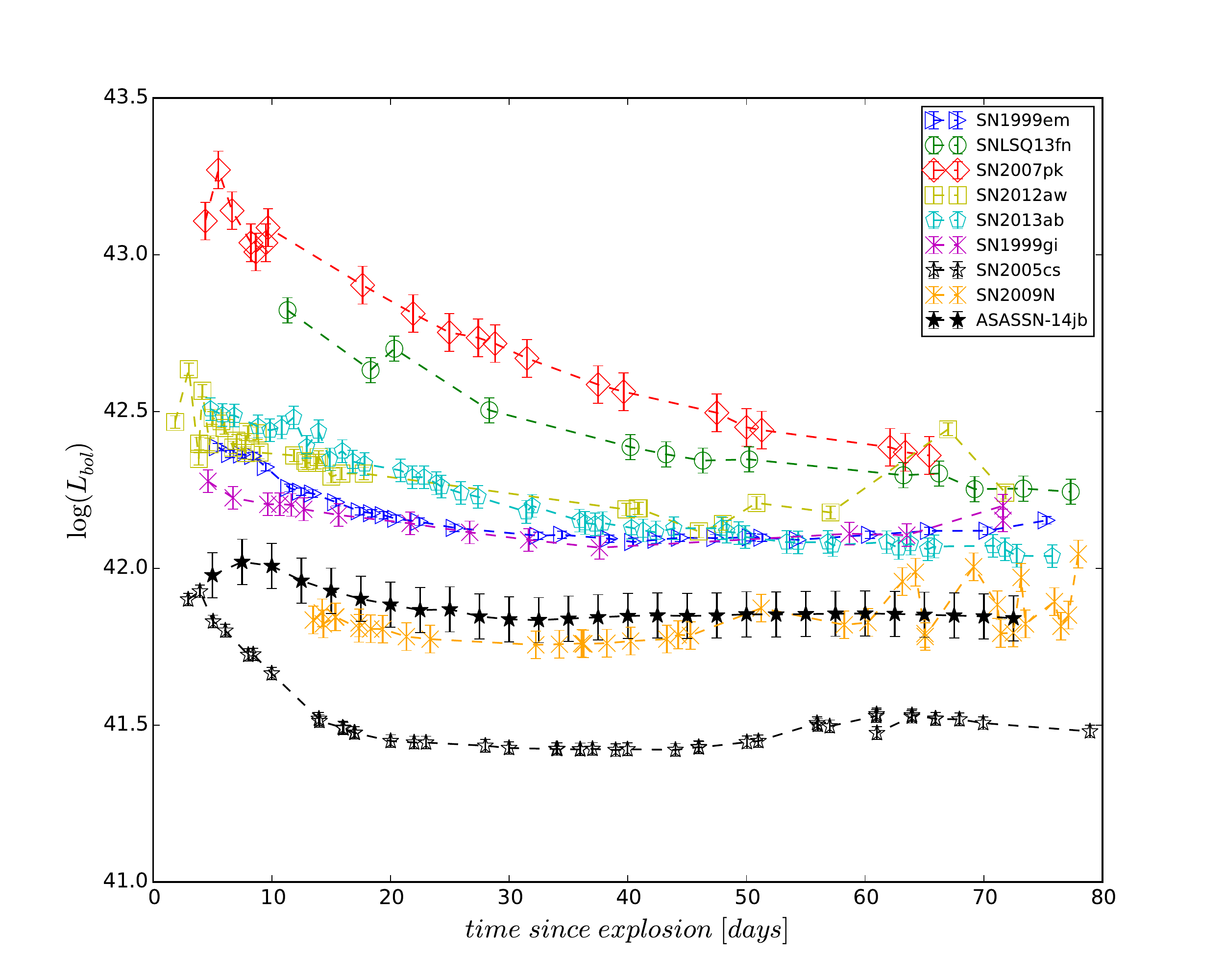}
\caption{Comparison of bolometric light curves using the bolometric corrections of \cite{pejcha15}. Each SN was corrected for extinction. See Table \ref{tab:comp_Table} for references of the comparison sample.  \label{fig:bolall}}
\end{figure}

We constructed a bolometric light curve of ASASSN-14jb, by applying the semi-analytic bolometric corrections from \cite{pejcha15} for the extinction corrected $B-V$ colors. Figure~\ref{fig:bolall} shows the result, together with the bolometric light curves of the comparison sample.
The bolometric luminosity of ASSASN-14jb $\sim 50$~days after explosion is $\log{(L/L_{\odot})} = 8.4 \pm 0.18$, which is under-luminous compared to SN~1999em, SN~2012aw or SN~2013ab.

\begin{figure}[ht!]
\includegraphics[width=9.5cm]{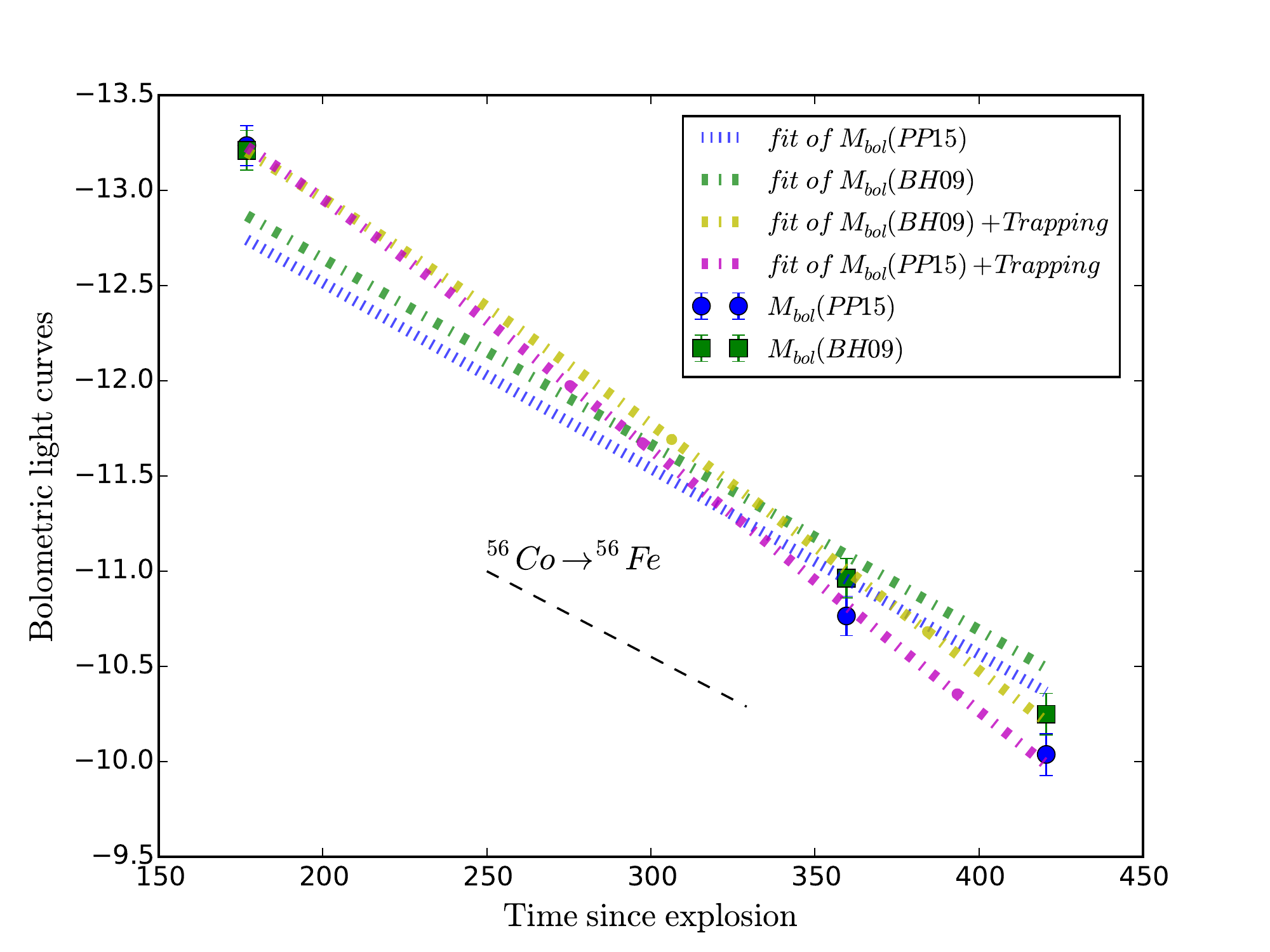}
\caption{Bolometric light curves in the radioactive tail derived from the $B$ and $V$-band photometry from EFOSC imaging, using the bolometric corrections of \cite{pejcha15b} (PP15, blue dots) and \cite{melina09} (MH09, green squares). The decay is faster than the expected from full trapping of $^{56}$Co decay (showed as a dashed black line). The non-full trapping model from \cite{clocchi97} gives a $^{56}$Ni Mass of $0.0210 \pm 0.0025 \ M_{\odot}$ for the \cite{pejcha15b} bolometric corrections. The time scale for this fit is $T_0 = 319 \pm 33$ days, which corresponds to a ejecta mass of 12 $M_{\odot}$ for an explosion energy of 1 foe.\label{fig:Ni56}}
\end{figure}

The late-time bolometric light curve can be used to estimate the $^{56}$Ni and total masses ejected in the explosion.
We assume a model of partial trapping of gamma rays \citep{clocchi97}, where the energy released by Cobalt decay is scaled by the transparency factor $I(t) = (1-\exp{(-(T_0/t)^2)})$,
and the characteristic time is given by $T_0 = C(\eta,\kappa_{\gamma})ME^{-0.5}$.
We assumed a fixed density slope $\eta = 10$ and constant
gamma-ray opacity $\kappa_{\gamma} = 0.06$ (which involves assuming an electron fraction per nucleon of $Y_e = 0.03$).
The late-time light curve and fitted model are plotted in Figure~\ref{fig:Ni56}.
The best fit provides $T_0 =  319 \pm 33$~days, which would correspond to an ejecta mass of $ 12 \pm 1 M_{\odot}$ for an explosion energy of $10^{51}$~erg (1 foe).
The $^{56}$Ni mass results $0.0210 \pm 0.0025$~M$_{\odot}$, which is slightly lower than the median of 0.031~M$_{\odot}$ corresponding to the Type~II SNe sample studied by \citet{muller17}. 

Late-time light curves that decay faster than Cobalt are expected in SNe with low mass ejecta, as in the case of stripped enveloped SNe and some fast declining Type~II SNe \citep[e.g., SN~2014G, SN~2013hj and ASASSN-15nx,][]{bose16,terreran16,bose18}.
Given the low energetics shown in, for example, the expansion velocities, the ejected mass estimated above is consistent with the low mass range of Type~IIP SNe progenitors.

The ejected mass estimated with the late-time light curve can be compared with that obtained in the previous section using the early light curve.
The scaling in this case is $M_{ej} \approx 12.0 \times E_{exp}^{0.5}$~M$_{\odot}$ which implies an ``agreement" pair ($M_{ej},E_{exp}$) at (6 M$_{\odot}$,0.25 foe).

\citealt{Joe14} discusses the observations of higher deviation from the expected full trapping decay for the brightest Type II SNe in their sample (in the $V$-band) and they suggest that this would be explained by the expected more diluted (i.e., less massive ejecta or more extended) hydrogen associated with faster declining SNe.
In our case, the ejecta mass is slightly below what would be needed for a relatively low energy explosion of $\approx$ 0.25 foe to have a high gamma ray opacity ($\tau_{\gamma} = (T_0/t)^2 \geq 1$) at $\approx 400$ days.\\

We can also estimate the ejected $^{56}$Ni mass using its correlation with the FWHM of H${\alpha}$ \citep{maguire12}.
A value of $\rm FWHM \simeq 35$~\AA\ gives us a $^{56}$Ni mass of $0.0118 \pm 0.0012$~M$_{\odot}$. This is lower than, but consistent with, the estimate obtained from the late-time bolometric light curve.

The low nickel mass $\log{(M_{56Ni})} \approx -1.5$ agrees with the explosion parameters according to the classic explosion models for Type II SNe \citep{Popov1993,Litvinova1985} and observed correlations \citep{hamuy03}.

\subsection{Nebular Spectra and the Progenitor Mass}

In the nebular phase, spectra of Type~IIP like SNe show prominent H$\alpha$, O~I, Ca~II and Fe~II emission lines. The [O I] $\lambda\lambda6300,6364$ is used to estimate the Oxygen core mass and therefore the main sequence mass of the progenitor star \citep[e.g.,][]{uomoto86,fransson89,elmhamdi03_99em,jerkstrand12,hanin15}. 
We compare our spectra with models from \cite{luc13}, \cite{jerkstrand14} and \cite{Jerkstrand18}. 
\cite[][hereafter D13]{luc13}, in particular, presented radiative transfer models for Type~II SNe using a grid of 15~M$_{\odot}$ stellar models from MESA STAR, with varying mixing length, metallicity, and rotation. We choose the nebular spectra in their ``m15" model, which reach up to 400 days past explosion (see Table~1 in D13 for details). \cite[][hereafter J14]{jerkstrand14}, calculated models for 12, 15, 19 and 25~M$_{\odot}$ stars evolved using the KEPLER code, assuming Solar metallicity and no rotation. We also included the recent model of 9~M$_{\odot}$ from \cite[][hereby J18]{Jerkstrand18}.

\begin{figure}[ht!]
\includegraphics[width=9.5cm]{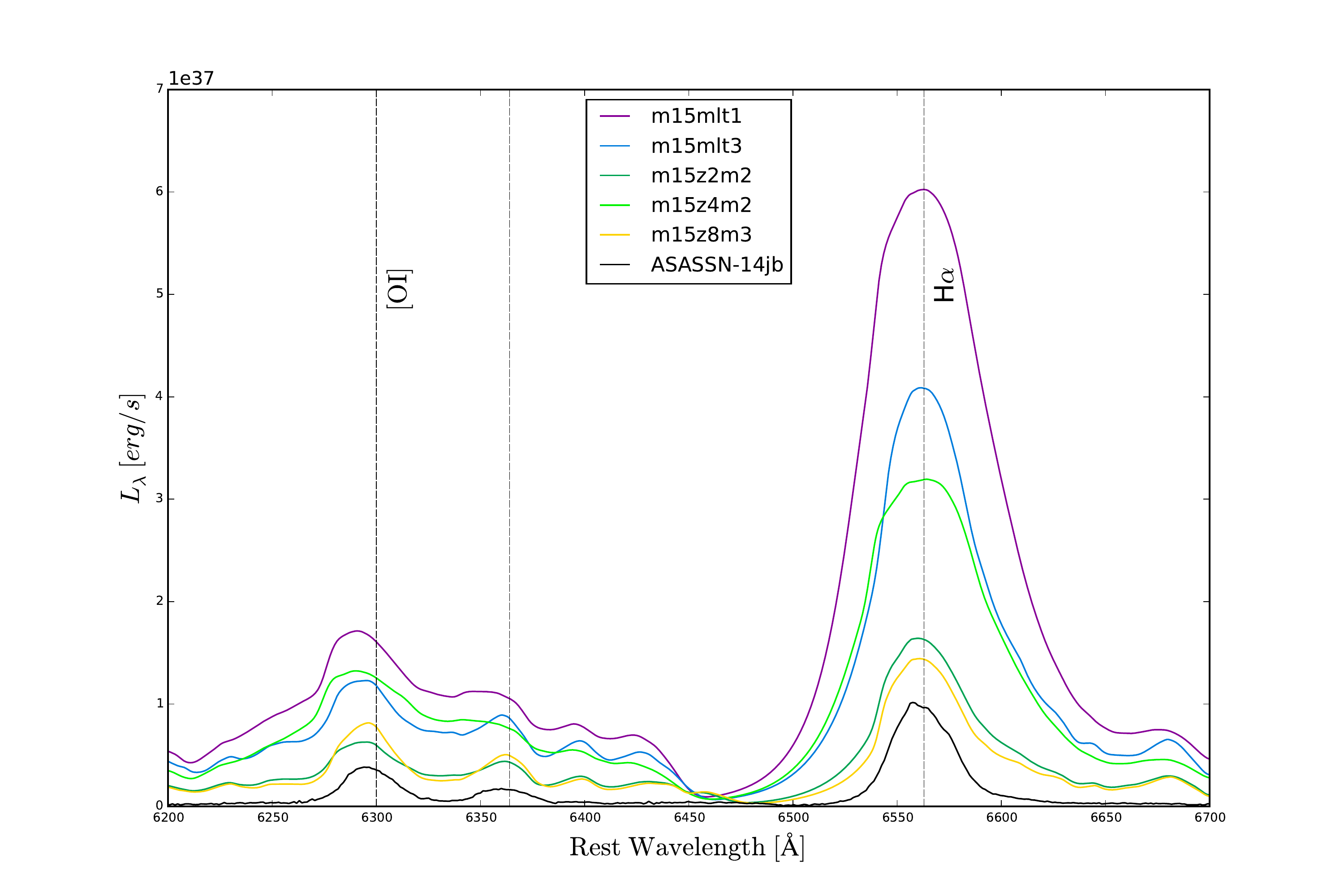}
\caption{Comparison of the nebular spectra of ASASSN-14jb with the NLTE models of \cite{luc13} at $\approx 400$ days past explosion. The vertical dashed lines shows the rest wavelength of H$\alpha$ and the [O I]  $\lambda\lambda$6300, 6364 doublet. \label{fig:Neb_Luc}}
\end{figure}

\begin{figure}[ht!]
\includegraphics[width=9.5cm]{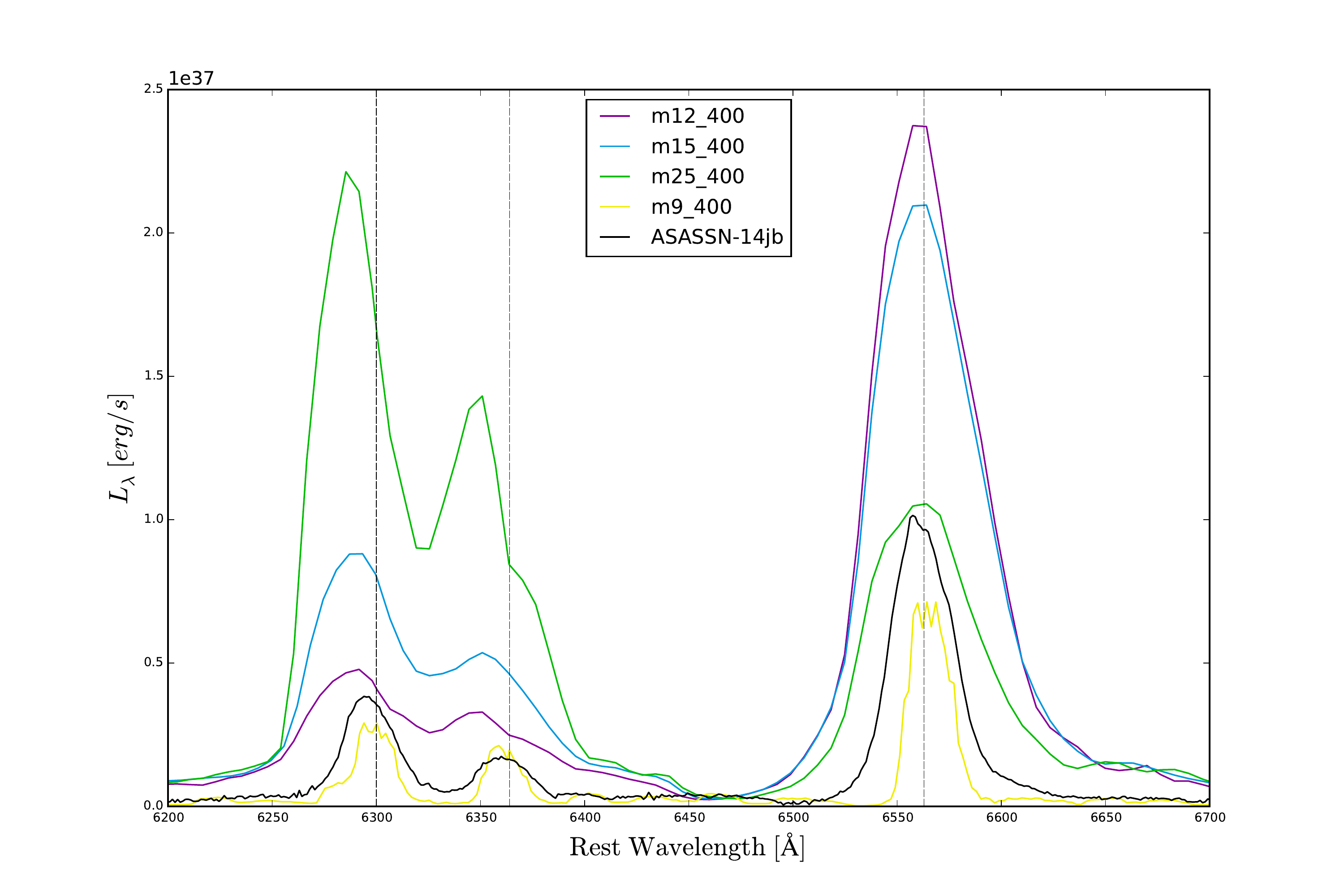}
\caption{Comparison of the nebular spectra of ASASSN-14jb with the NLTE models of \cite{jerkstrand14,Jerkstrand18} at $\approx 400$ days past explosion. The vertical dashed lines shows the rest wavelength of H$\alpha$ and the [O I]  $\lambda\lambda$6300, 6364 doublet.\label{fig:Neb_Jerk}}
\end{figure}

In Figure~\ref{fig:Neb_Luc} and \ref{fig:Neb_Jerk} we compare the data and the models for the [O I] $\lambda\lambda6300,6364$ and H$\alpha$ lines.
In the case of D13 models m15z2m2 and m15z8m3 provide the best fit. They correspond to a progenitor metallicities of $Z = 0.02$ and $0.008$, oxygen mass of $M_O = 0.325$ and $0.507$~M$_{\odot}$, and nickel masses of $M(^{56}{\rm Ni}) = 0.05$ and $0.081$~M$_{\odot}$, respectively (for details see Table~1 in D13).
In the case of the J14 and J18 models the 9 M$_{\odot}$ and 12 M$_{\odot}$ are closest to the data, although all the models over-predict the H$\alpha$ emission line.
A problem with this direct comparison is that the explosion models presented in D13, J14 and J18 produce higher $^{56}$Ni masses than measured for ASASSN-14jb.
If the spectra are normalized by the H$\alpha$ emission to approximately take care of this difference, the J14 models of 15 M~$_{\odot}$ and 9 M$_{\odot}$ become closest to the data. On the other hand, the D13 models reveal a high degeneracy given their differences in stellar evolution.

Other approach we can use to alleviate the impact in the spectra of the $^{56}$Ni mass discrepancy between ASASSN-14jb and the models, is to normalize the flux in a given line by the $^{56}$Ni decay power. Doing this for the [O I] $\lambda\lambda6300,6364$ line, we obtain a ratio of $\approx 0.04$. This leads to a surprising better consistency with the higher mass range of J14 models ($15-19$~M$_{\odot}$), but again to a better fit of the 9 M$_{\odot}$. According to \cite{Jerkstrand18} this results both from the oxygen shell in the 9 M~$_{\odot}$ model lying closer to the $^{56}$Ni and a significant contribution of Fe I to the [O I] $\lambda\lambda6300,6364$ doublet.

As a final estimate we calculate an upper limit to the emitting oxygen mass using the analytic formula provided by J14,
\begin{equation}
max (M_{\mathrm{O \ I}}) = \frac{L_{\mathrm{[O \ I]}}/\beta_{\mathrm{[O \ I]}}}{9.7\times{10^{41}}} \times{\exp\left(\frac{22720}{T}\right)},
\end{equation}
where $L_{\mathrm{[O \ I]}}$ is the luminosity of the [O I]$\lambda\lambda 6300,6364$ doublet in erg/s, $\beta_{\mathrm{[O \ I]}}$ is its Sobolev escape probability, and $T$ the equilibrium temperature in K.
A temperature in the range 3900--4300 K was estimated from the ratio between the lines [O I] $\lambda\lambda 6300,6364$ and [O I] $\lambda 5577$\footnote{Assuming $\frac{\beta_{6300,6364}}{\beta_{5577}} = 0.5-1$ as in J14 and estimating the fluxes by fitting a skewed Gaussian to the line profiles.}. The resulting values of $max (M_{\mathrm{O \ I}})$  fall in the range 0.09--0.18~M$_{\odot}$, which is  $20-40\%$ above that of the 9 M$_{\odot}$ model of J18 and 1.5--3 times lower than that of the 14 M$_{\odot}$ model of J14.

We conclude that the nebular spectra confidently points at a low mass progenitor for CCSNe standards, in the range $\sim10-12$~M$_{\odot}$. This is supported by surveys of theoretical explosions spanning a wide grid of progenitor properties \citep[e.g.,][]{Sukhbold2016}, which show that models above $\sim 12.5 $M$_\odot$ are much more efficient at producing oxygen than below that ZAMS mass. 

\subsection{Warm Dust in the Ejecta} \label{ss:dust}

ASASSN-14jb is clearly detected in the 2015-09-13 {\em Spitzer} images, with $m_{3.6} = 17.85 \pm 0.03$~mag and $m_{4.5}= 16.38 \pm 0.03$~mag (magnitudes in the AB system), but undetected on 2016-08-17.
The implied absolute magnitudes ($M_{3.6} \simeq -14.1$, $M_{4.5}\simeq -15.6$) and red color ($m_{3.6}-m_{4.5} \simeq 1.5$~mag) at $\sim 333$ days after explosion are consistent with warm dust formation in other Type~II SNe \citep[e.g.,][]{prieto12,szalai18}.
We used a black body function with dust emissivity $Q_\lambda\propto \lambda^{-1}$ to fit the mid-infrared SED and found $L \sim 1.1\times 10^6$~L$_{\odot}$, $T \sim 444$~K, and $R\sim 839$~AU.
Using Equation~1 in \citet{prieto09} (from \citealt{Dwek1983}), we obtain a total dust mass of $M_d \sim 10^{-4}$~M$_{\odot}$.
These estimates are consistent with models of newly formed dust in the SN ejecta and observations for other Type II SNe at a similar epoch after explosion \citep[e.g.,][]{Sarangi2015}.

\subsection{Progenitor Metallicity}
\label{subsec:Fe_abundance}
\begin{figure}[ht!]
\includegraphics[width=9.5cm]{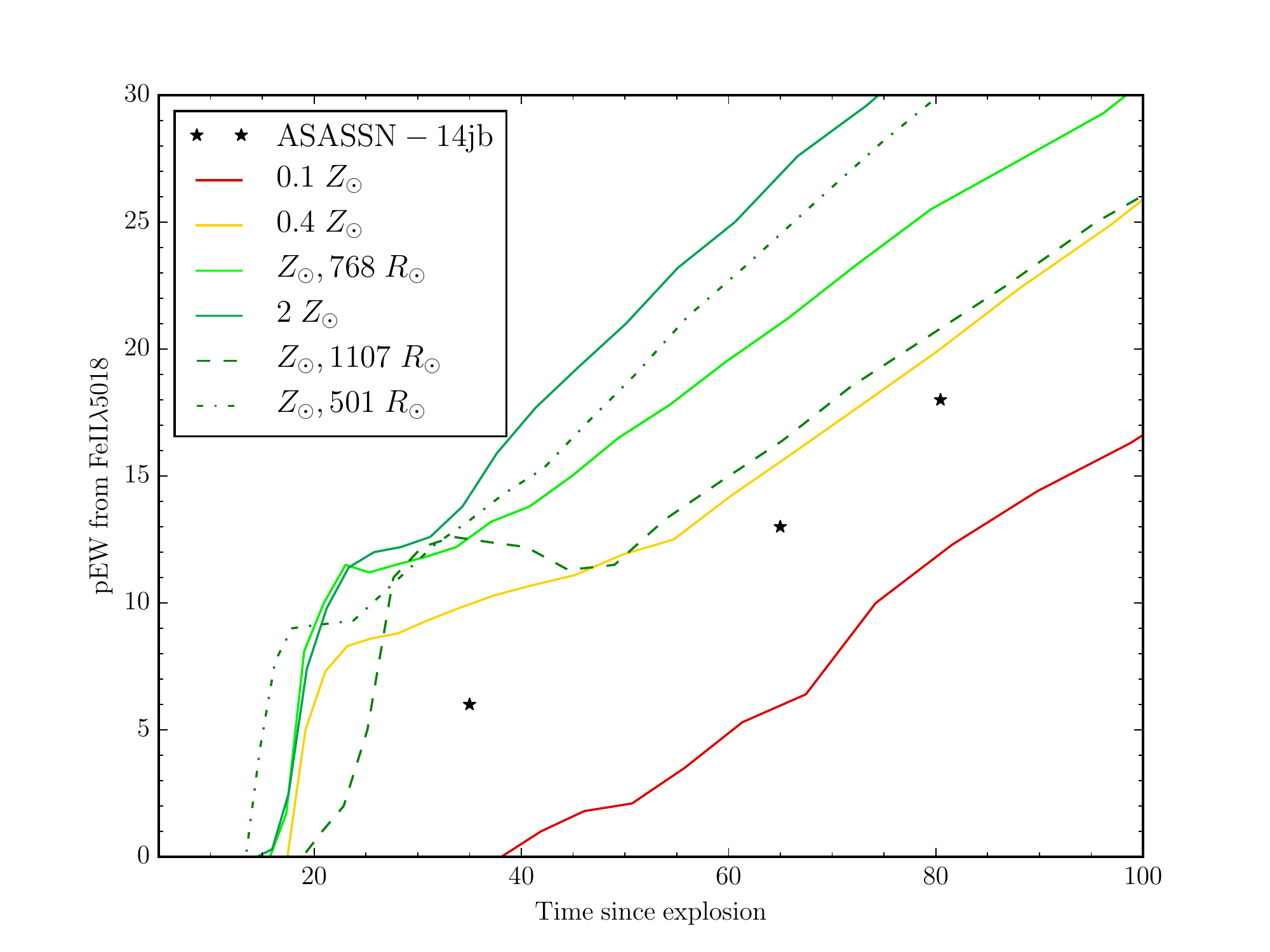}
\caption{Measurements of the pseudo-equivalent width of the Fe II $\lambda5018$ line for ASASSN-14jb (black stars), with the correspondent epoch since explosion $t_0 = 56946.1 \pm 3$ (MJD). The solid lines show \cite{luc13} models for progenitors with different metallicities and the dashed line model progenitors with Solar metallicities and different initial radii. \label{fig:pEW}}
\end{figure}

The pseudo-equivalent widths (pEWs) of the Fe II~$\lambda 5018$ line during the photospheric epoch of Type IIP SNe are a proxy for the progenitor metallicity. This has been shown both by models (D14) and correlation of observed pWEs with metallicities of H~II regions in the host galaxies \citep{Joe16}.
We measured the pEW evolution during the plateau phase of ASSASN-14jb and compare it with those of D13 models. The result is shown in Figure~\ref{fig:pEW}.
The models span four progenitor metallicities (2 $Z_{\odot}$, 1 $Z_{\odot}$, 0.4 $Z_{\odot}$ and 0.1 $Z_{\odot}$) and four different mixing-length scales for the Solar composition model, which greatly change the progenitor radius.
We see that the observed pEWs of the Fe~II line are most consistent with the curve for 0.4 $Z_{\odot}$, but the Solar metallicity model for the progenitor with largest radius is also close enough.
If we linearly interpolate the models and, assume that the dispersion given by different radius at Solar metallicity is a fair estimate of the uncertainty, we obtain Z = $0.3  \pm 0.1 Z_{\odot}$ for ASASSN-14jb.

\section{The Host Galaxy of ASASSN-14jb}
\label{sec:host}
Statistical constrains on properties of the progenitor stars can be also establish through the study of the galactic environment of the SNe \citep[e.g.,][]{stoll13,Joe15,galbany16d}.
Relevant to our discussion are recent integral field spectroscopy observations (IFS) which have provided metallicities and ages of nearby stellar populations \citep{hanin13,galbany16b,galbany18,hanin17}. 

Of particular interest are the studies of H~II regions, which rapidly fade away after the massive stars which provide the ionizing photons explode as SNe \citep[e.g.,][]{Pols1998}. Since typical lifetimes are up to a few tens of Myr they trace well the ongoing star formation. Typical sizes of giant H~II regions are in the hundreds hundreds parsecs \citep{sanchez2012}. Our MUSE observations of ASASSN-14jb were done with a seeing of 1\farcs{08} in the $R$-band, providing a resolution of $\sim 130$ pc at the distance of the host. 

We built 2D $H\alpha$ emission maps of ESO 467-G051 from the MUSE datacube following the procedures of \citealt{galbany16b}.
In brief, we fitted the spectrum of each spaxel using a combination of the stellar continuum modeled with the stellar population synthesis code STARLIGHT \citep{cidfernandes09}, with Gaussian profiles at the central wavelengths of the emission lines detected with SNR greater than 3, and we corrected the line ratios using the Galactic reddening map of  \cite{schlafly11} and an estimate of the internal extinction in ESO 467-G051 using the Balmer decrement. In addition, we used the H~II-Explorer code \citep{sanchez2012} to detect and analyze H~II regions in the datacube.
 
\subsection{Overall Properties of ESO 467-G051} 

The host galaxy of ASASSN-14jb, ESO 467-G051, is an edge-on, Scd galaxy \citep{devac91}. Our observations, together with extensive data in the literature, make it possible to measure integrated fluxes from the UV up to the IR.

To do so, we (1) used co-added LCGOTN images with good seeing and measured the total $BVgri$ magnitudes using a large elliptical aperture in {\em ds9}; (2) used a near-infrared $K_s$-band image of the field from the ESO data archive\footnote{Image obtained with SOFI at the NTT by the PESSTO program (ID 191.D-0935) with SOFI mounted on the ESO/NTT} to measure the magnitude of the host within an elliptical aperture calibrated to 2MASS \citep{2mass} stars in the field; (3) obtained archival ultraviolet magnitudes from GALEX \citep{galex} and mid-infrared magnitudes from {\it Spitzer}'s S4G survey \citep{sheth10}. The integrated apparent magnitudes are presented in Table~\ref{tab:hostmags}. Using our distance, the absolute magnitudes of ESO 467-G051, corrected by Galactic extinction, are $M_B\approx -17.6$~mag, $M_V \approx -18.0$~mag, and $M_{K_s}\approx -19.1$~mag.

\begin{figure}[ht!]
\includegraphics[width=\linewidth]{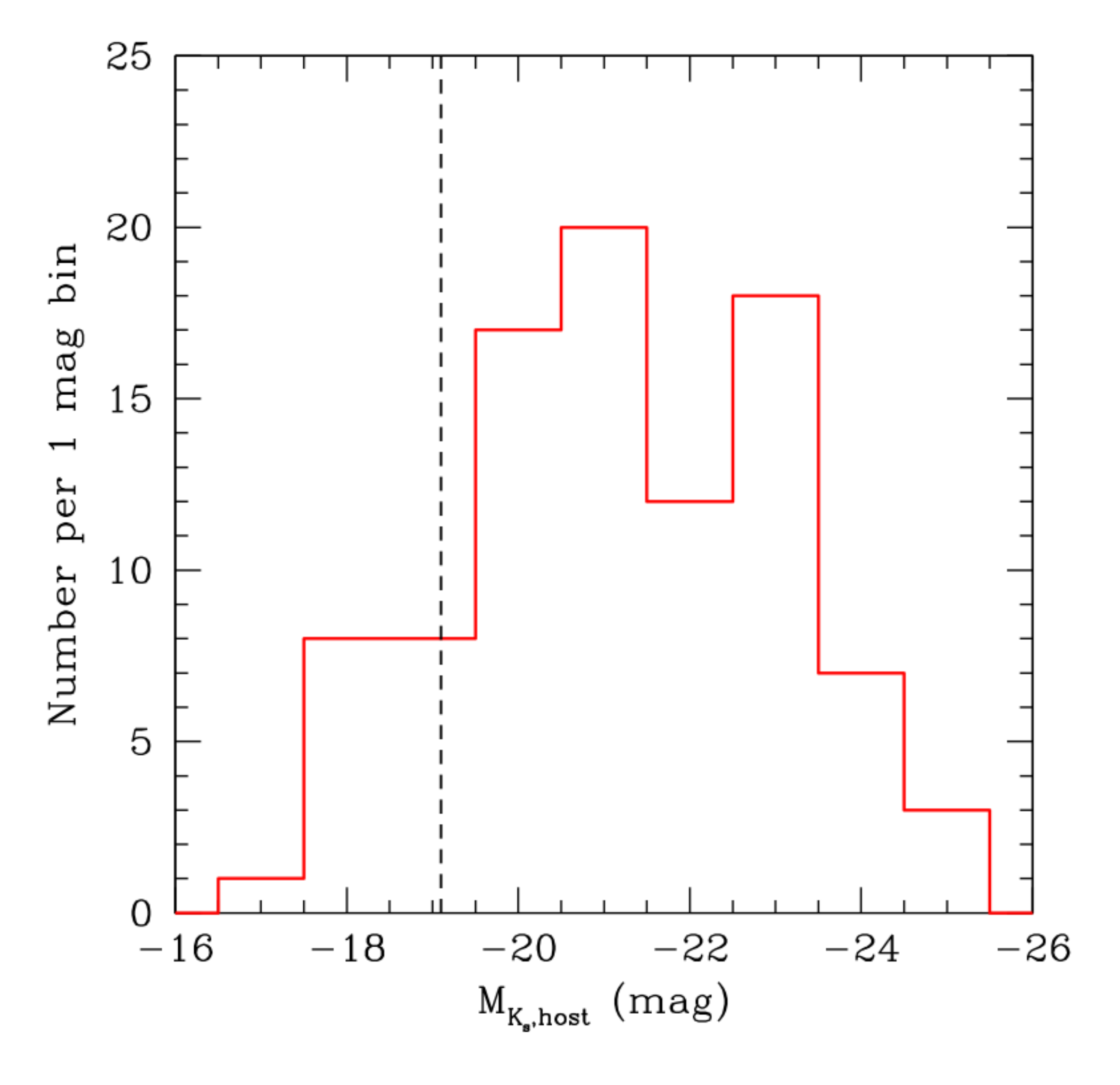}
\caption{Histogram with the distribution of absolute $K_s$-band magnitudes of the host galaxies of 94 Type~II SNe discovered by ASAS-SN in 2013-2017 (\citealt{holoien17,holoien17b,holoien17c,holoien18}). The dashed vertical line marks the absolute magnitude of the host galaxy of ASASSN-14jb, ESO 467-G051, indicating a relatively low stellar mass for a Type~II SN in this sample.}
\label{fig:histogram_host}
\end{figure}

Figure~\ref{fig:histogram_host} shows a histogram with the distribution of host galaxy $M_{K_s}$ magnitudes of Type~II SNe discovered by ASAS-SN in 2013-2017 from \cite{holoien17,holoien17b,holoien17c,holoien18}. The host galaxy of ASASSN-14jb is in the lower $K_s$ absolute magnitude (lower stellar mass) range of the Type~II SNe discovered by ASAS-SN.    

The stellar mass, age and SFR of ESO 467-G051 can be estimated from these data.
We fitted stellar population synthesis (SPS) models to the $FUV$, $NUV$, $BVgri$, $K_s$, 3.6~$\mu$m, and 4.5~$\mu$m magnitudes using FAST \citep{kriek09}. We assumed a Galactic extinction law, a Salpeter IMF, an exponentially declining star formation law ($\rm SFR \propto e^{-t/\tau}$, with $\tau = 1$~Gyr), and the \cite{bruzualcharlot03} SPS models. The best-fit model provides the following parameters: $M_{\star} \simeq 1 \times 10^{9}$~M$_{\odot}$, $\rm age \simeq 3.2$~Gyr, and $\rm SFR \simeq 0.07$~M$_{\odot}$/yr, where the uncertainties are $1\sigma$.
The SFR can be also estimated from the H$\alpha$ extinction-corrected fluxes \citep[see below,]{kennicutt98}, which provides 
 $\rm SFR(H_{\alpha})\sim 0.07$~M$_{\odot}$/yr. This is fully consistent with the SED SPS fit.  

Neutral hydrogen (HI 21~cm) observations suggest that the host galaxy of ASASSN-14jb is undergoing a direct encounter with  467-ESOG050 or NGC~7259, the host galaxy of SN~2009ip \citep[$\Delta V = 92.6 \pm 5.6 $km/s,][]{Nordgren97}. The total HI flux is $S_{HI} = 23.1 \pm 0.8$~Jy km/s and the derived HI mass is  $M_{HI} = 3.1 \pm 0.8 \ [10^9 \mathrm{M}_{\odot}]$ \footnote{$M_{HI} =  2.356 \times 10^5 \times D^2 \int{S_\nu d\nu}$~M$_{\odot}$, D = 23.5 Mpc}.
Using our distance estimate ($25 \pm 1$ Mpc) the HI mass scales to $M_{HI} = 3.5 \pm 0.9 \ [10^9 \mathrm{M}_{\odot}]$.
This gives a gas fraction of $M_{gas}/(M_{gas}+M_{*})\approx 0.77$.

The maximal HI rotational velocity is $60 \pm 1$ km/s (Hyperleda database\footnote{http://leda.univ-lyon1.fr/}). In our H$\alpha$ velocity map we observe this velocity limit more clearly on the East side of the galaxy. 

\subsection{H$\alpha$ and Oxygen Abundance Maps}

\begin{figure*}[ht!]
\includegraphics[width=\linewidth]{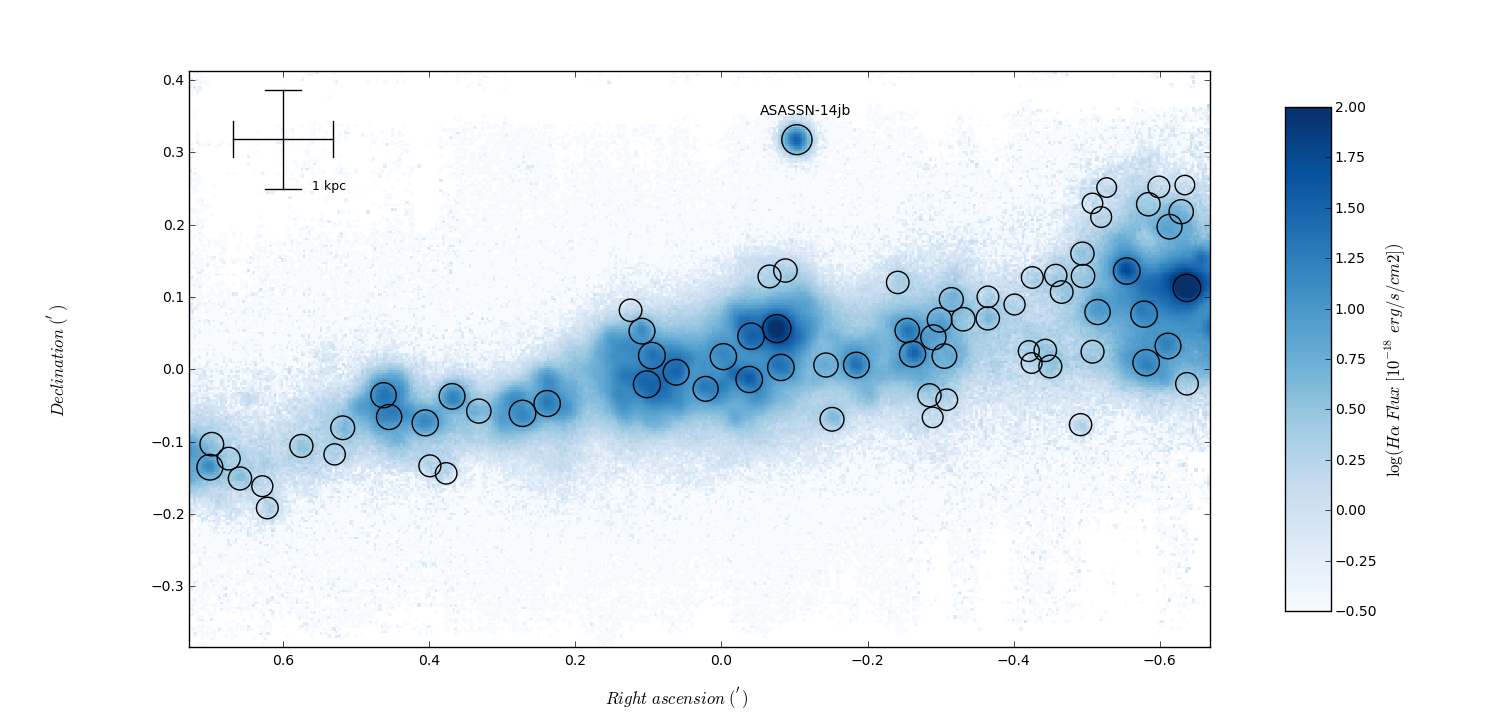}
\caption{H$\alpha$ MUSE map of the host galaxy and the supernova explosion site. Black circles marks the position of the H~II regions detected with H~II-Explorer. Coordinates are the right ascension and declination offsets from the center of ESO 467-G051, in arcminutes. The 1 kpc scale of the image is marked in the upper left.\label{fig:Halpha}}
\end{figure*}

Our map of the H$\alpha$ dereddened emission line fluxes is given in Figure~\ref{fig:Halpha}. It includes H~II regions, diffuse emission in the host, and the nebular emission from ASASSN-14jb.
The position of the H~II regions detected with the H~II-Explorer code, and that of the SN, are marked. ASASSN-14jb is at a vertical distance of $\approx 2.1$~kpc from the disk major axis. The H~II region nearest to the SN is at of $\approx 1.4$~kpc.  

\begin{figure}[ht!]
\includegraphics[width=9.5cm]{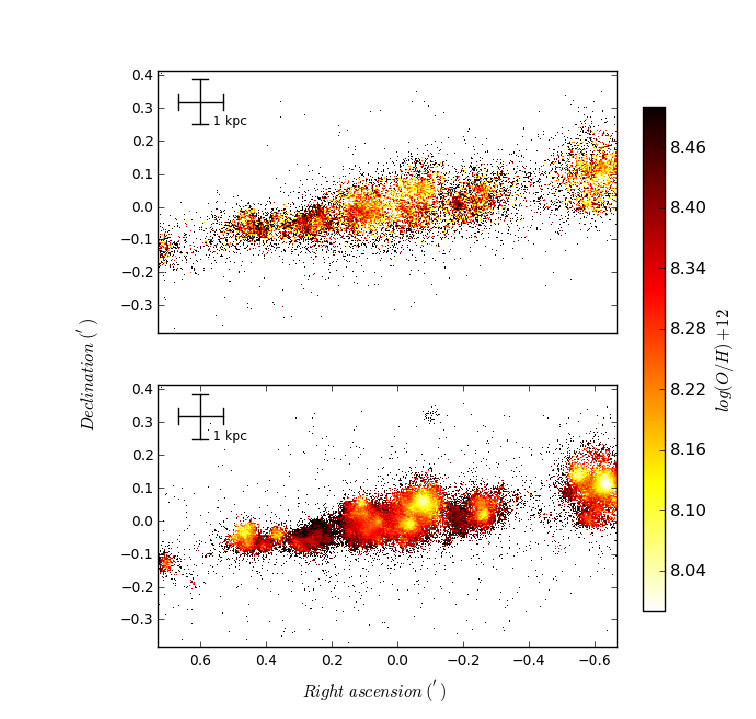}
\caption{D16 (top) and O3N2 (bottom) oxygen abundance maps of the host galaxy ESO467-G051 as seen by MUSE, shifting the D16 map by +0.4 dex to provide an easier comparison. Coordinates are the right ascension and declination offsets from the center of ESO 467-G051, in arc-minutes\label{fig:D12_O3N2}. The 1 kpc scale of the image is marked in the upper left.}
\end{figure}

The spectra of the H~II regions can be used to estimate the gas-phase oxygen abundance.
We did so using the O3N2 and N2 strong emission line methods \citep{marino13}. The median obtained, with $99\%$ confidence intervals, were $\rm 12 + log(O/H) = 8.27^{+0.16}_{-0.20}$ and $8.19^{+0.10}_{-0.24}$, respectively.
A new strong line calibration by \citealt{dopita16}, hereafter D16, which uses [S~II], [N~II], and H$\alpha$ lines, provides significantly lower values, with a median of $\rm 12 + log(O/H) = 7.77^{+0.10}_{-0.24}$.
The D16 and O3N2 maps are presented in Figure~\ref{fig:D12_O3N2}. Median values of the maps are $12+\log(O/H) = 7.77^{0.19}_{-0.1}$ and $12+\log(O/H) = 8.22^{0.11}_{-0.14}$, for the spaxels with S/N $\geq 10$.

\begin{figure}[ht!]
\includegraphics[width=9.5cm]{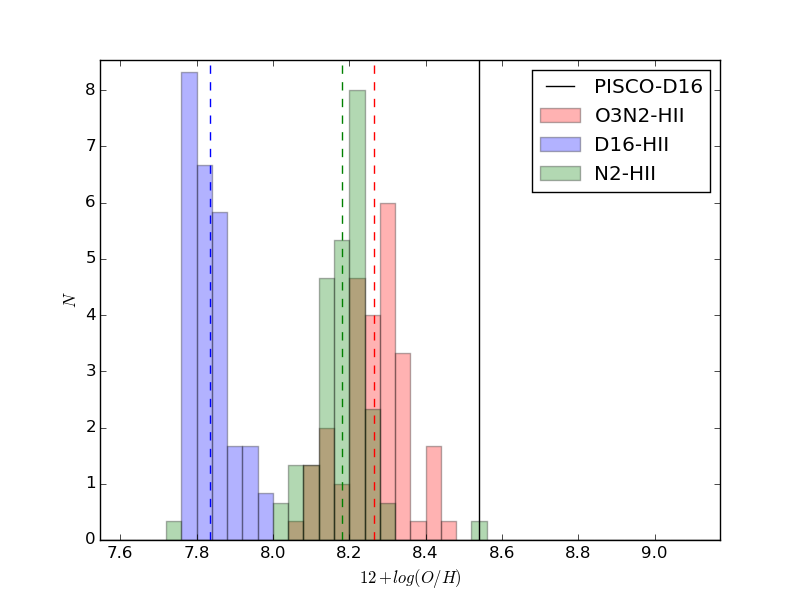}
\caption{Distribution of the gas-phase oxygen abundances used for ESO467-G051 H~II regions, based on the D16, O3N2 and N2 calibrations. The dashed lines represent the median for each distribution and the black vertical line indicates the average abundance for Type II SNe hosts from the PISCO compilation using the D16 calibration  \citep{galbany18}.\label{fig:Z_hist}}
\end{figure}

The distributions of oxygen abundances obtained with the N2, O3N2 and D16 nebular emission line diagnostics is given in Figure~\ref{fig:Z_hist}.
All the mean values are sub-Solar (using as reference $\rm 12+log(O/H)_{\odot} = 8.69$, from \citealt{Asplund09}), and significantly below $\rm 12+log(O/H)_{\odot} = 8.54$ \citep{galbany18}, the average for H~II regions near Type II SNe of the the PMAS/PPAK Integral-Field Supernova Host Compilation (PISCO).
Our host galaxy $B$-band absolute magnitude, of $M_B = -17.6$~mag, compares well with the mean $B$-band magnitude of $M^{host}_{B} = -17.66$~mag in the recent sample of Type~II SNe in low-metallicity, dwarf galaxies \citep{claudia18}. Also, as showed in \cite{claudia18}, lower luminosity galaxies hosts lower metallicity Type II SNe, filling the space between the 0.1 and 0.4 Z$_{\odot}$ metallicity models. This is expected given the strong correlation between stellar mass (or galaxy luminosity) and metallicity.
It is worthwhile to mention that the O3N2 map shows lower oxygen abundances in the cores of the H~II regions. This could be a consequence of the unaccounted internal gradient of the ionization parameter of each H~II region in the empirical strong line methods \citep{kruhler17}.

\subsection{D16 and the N/O ratio}

\begin{figure}[ht!]
\includegraphics[width=9.5cm]{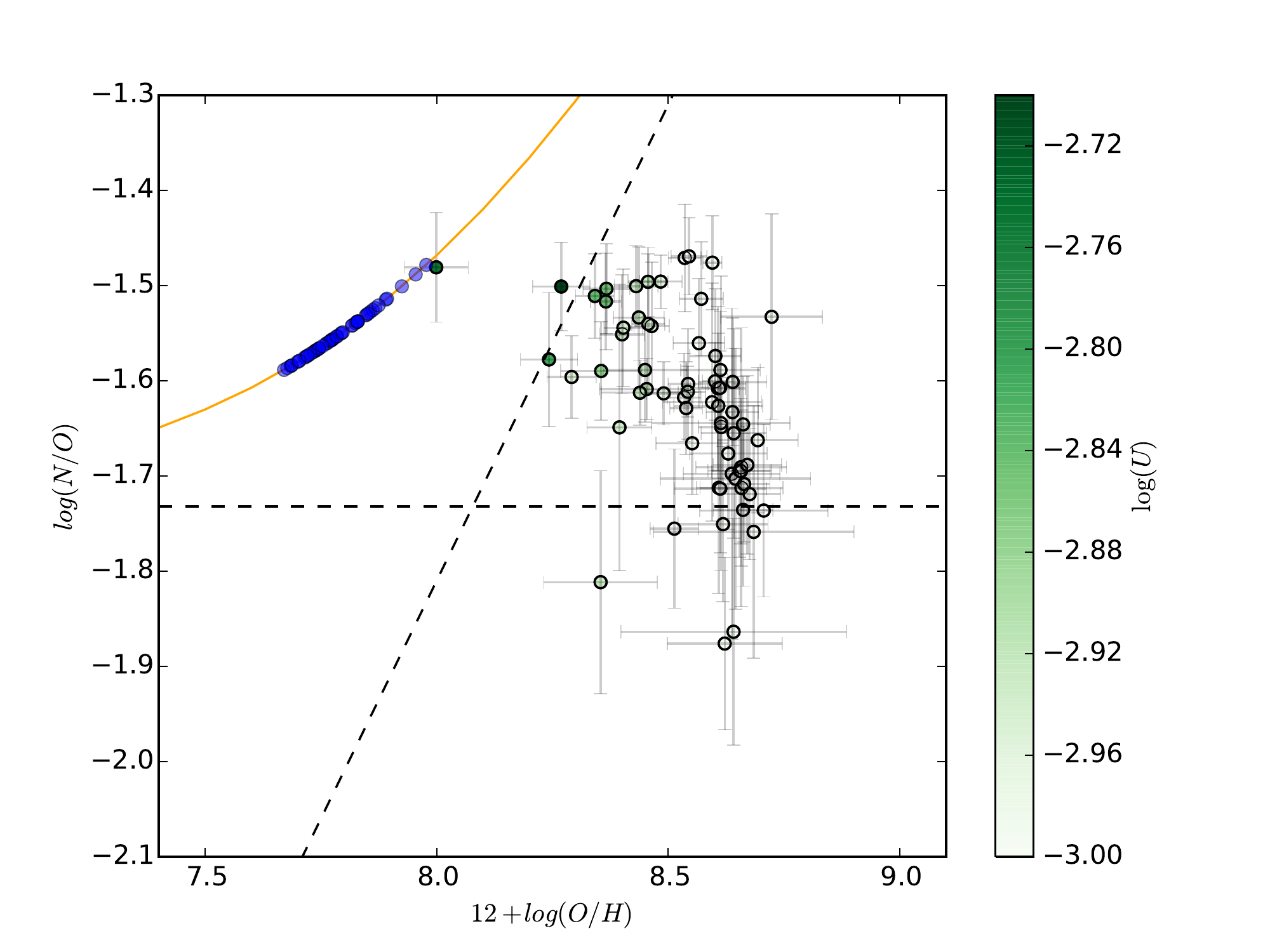}
\caption{Oxygen abundances versus N/O ratios obtained with the code developed in \citealt{Montero2014}. Each point has the errorbars delivered by the code and each point is colored according to the ionization parameter U delivered. The solid orange line indicates the relation as assumed in the D16 calibration and the blue dots correspond to the observed values for our H~II regions abundances using D16. The dashed lines indicate the asymptotes for orange line, at the very high and very low O/H ratios regimes.\label{fig:NOHU}}
\end{figure}

The significant discrepancy of the D16 with other strong line oxygen abundance calibrations calls our attention.%
This abundance scale relies on a specific relation between the N/O and O/H ratios \citep{dopita16} for which it is useful to consider a diagnostic independent on any a priori relation. \citet{Montero2014} developed a semi-empirical code, \textit{H~II-CHI-MISTRY}, that derives O/H, N/O and the ionization parameter U, using the reddening-corrected fluxes of [O II] $\lambda$3727 , [O III]  $\lambda$4363,  $\lambda$5007 , [N II]  $\lambda$6584  and [S II]  $\lambda\lambda$6717,6731,  relative-to-
H$\beta$.
When the O auroral lines are absent, as is our case, the code uses a limited grid of empirically constrained models to provide abundances that are consistent with the ``direct" electron temperature measurement method.
We applied \textit{H~II-CHI-MISTRY} v3.0 with the accessible lines  [O I] $\lambda$5007 , [N II]  $\lambda$6584  and [S II]  $\lambda\lambda$6717,6731. The output N/O ratio is shown as a function of the oxygen abundances in Figure \ref{fig:NOHU}. We observe a clear offset from the D16 calibration. The mean of the oxygen abundances is $8.53 \pm 0.13$ while the N/O ratio has a mean of $-1.63\pm 0.1$. There is a trend with the ionization parameter so that higher oxygen abundances show lower N/O and U values (same with H$\alpha$ luminosity instead of U) and, vice versa, the lower abundances correspond to higher N/O ratios and also they are closer to the D16 prescription. The negative correlation between N/O and O/H, together with the trend with H$\alpha$ luminosity, has been observed in blue compact dwarf galaxies \citep{Kumari2018}.

\section{Discussion}
\label{sec:discusion}
ASASSN-14jb is a normal Type~IIP SN located at a projected distance of $\sim 2.1$~kpc from the mid-plane and $\sim 1.4$~kpc from the closest, diffuse H~II region of ESO 467-G051.
Since Type~II SNe come from stars with short life times, the unusual location begs for an explanation. Several hypothesis have been advanced for SNe with no clear parent galaxy. One is that the host is an undetected low-surface brightness (LSB) galaxy, or even a small fragment of a disrupted satellite galaxy \citep{abadi09}. 
Other interesting possibility is that the progenitor of ASASSN-14jb traveled $\approx 1.4$~kpc from its birthplace driven by gravitational interaction.

\subsection{Type~II SNe from Runaway Stars}

Different scenarios have been proposed for SNe coming from runaway stars. One is the Binary Supernova Scenario (BSS) where they are pushed by the interaction with a massive binary companion that exploded as a supernova. Other is the Dynamical Ejection Scenario (DES), where the impulse comes from close encounters in clusters, including those with binaries, very massive stars, or intermediate or massive black-holes
 \citep{gvaramadze09,hansen03,gualandris07}.
Yet another is the hyper-velocity star (HVS) scenario, resulting from the so-called Hills Mechanism, where velocities higher than 100 km/s are provided by interaction with a supermassive black-hole at the center of the galaxy \citep[see ][for a review on HVS]{brown15}.

The approximate lifetime of a star with ZAMS mass of $\sim 10 M_{\odot}$ in single stellar evolution is $\sim 30$~Myr.
Covering the $\sim 1.4$~kpc from the nearest star forming region in this time would require a projected space velocity of $\sim 47$~km/s.
Considering the lifetime of the least massive stars that undergo core collapse in the models of \citet{zapartas17}, $\sim 48$~Myr for $\sim 8$~M$_{\odot}$ progenitors, reduces the requirement to $\sim 30$~km/s.
Both are plausible velocities given the observed population of runaway stars in the Galaxy and the Magellanic Clouds \citep[e.g.,][]{schnurr08}.
The velocity distribution of early type runaway stars in the Milky Way also supports these estimates \citep{silva11}. OB stars at high Galactic latitudes exceed the minimum velocity needed to reach 1~kpc off the Galactic disk. Scaling the result to ESO 467-G051 to account for its smaller mass, the maximum height above the disk that would be reached for a given velocity would be comparable to that of ASASSN-14jb. 
Let us now consider an illustrative case of the BSS scenario, with a primary star of 25 $M_{\odot}$ at ZAMS. The primary would explode $\approx 8$ Myr after formation \citep[e.g.,][assuming half solar metallicity in their grid]{Pols1998}, leaving $22$~Myr as the flight time for a secondary of 10 $M_{\odot}$ at ZAMS. Traveling the $\sim1.4$~kpc in this time requires a velocity of $v \simeq 61$~km/s.
Finally, the HVS scenario is also feasible. The projected distance from the position of ASASSN-14jb to the host galaxy center is $2.5$~kpc. Taking the lifetime of the longest living CCSN progenitors of \cite{zapartas17}(48$~Myr$) the lower limit for the space velocity is $v \geq 51$~km/s.
Comparing this with the case of the Type II SN~2006bx \citep{zinn11}, the progenitor of which would have been an HVS with $v\gtrsim 848$~km/s, an HVS progenitor for ASASSN-14jb seems unnecessary. 

On the other hand, theory is not as supportive of the runaway progenitor scenario. \cite{eldridge11} studied the distribution of spatial velocities and distances traveled for progenitors of CCSNe in the BSS scenario, ignoring the galaxy potential.
They obtained a total average of $190 \pm 380$~pc (non-Gaussian distribution) with a $17.5 \pm 18.7$~km/s space velocity for the secondary of the binary system exploding as a Type IIP like SN at Solar metallicity ($Z = 0.02$), while $Z = 0.004$ gives mildly higher velocities.
A more recent study by \cite{Renzo2018} confirms these results, showing that $2.5\%$ of binary systems have a disruption velocity $\gtrsim 30$~km/s after the explosion of the primary.
Hence, the bulk of the ejected populations does not reach the velocities needed by the progenitor of ASASSN-14jb and  taking the host galaxy potential into account would make the scenario less probable.
There are, however, tails in the ejected velocity distributions where the velocities are consistent with the requirements of the ASSASN-14bj progenitor.

\subsection{Disk Thickness and Extraplanar Star Formation}

\cite{hakobyan17c} studied the vertical distribution of both SNe~Ia and CCSNe in edge-on, late-type disk galaxies using a sample of 102 historical SNe. For Sb-Sc type galaxies the scale height, in units of the isophotal radius at a surface brightness in $g$-band of 25~mag/arcsec$^{2}$ or $R_{25}$, $0.090 \pm 0.016$ for CCSN.
ASASSN-14jb has $z_{SN}/R_{25} =  0.22$, more than $10 \sigma$ above this mean.
We estimated the height scale of $z_0/R_{25} =  0.082 \pm 0.01$ for ESO 467-G051 by fitting a a $sech^2 z/z_0$ profile to the H$\alpha$ profile along the perpendicular between the disk and the SN. 
Fits to the $V$ or the $R$-band images gave similar results.

The scale height of each supernova type is expected to match that of the stellar population supplying their progenitors \citep[e.g.,][]{kangas17}. For Type~II SNe we expect that the vertical distribution follows the distribution of recent star formation (OB stars). In particular, as the thick disk population is generally significantly older \citep{Dalcanton2002,Yoachim2008,Elmegreen2017}, we are less likely to observe a Type II SN in the thick disk. Nevertheless, \cite{Howk2018b} compiled a sample of 6 extraplanar H~II regions detected in nearby, edge-on disk galaxies. The projected vertical distances of these regions above their disk cover the range $0.9-3.0$~kpc, up to $0.11$ in units of $R_{25}$, which is within a factor of two of the height of ASASSN-14jb in the disk of ESO 467-G051. Given this, the scenario of in-situ star formation for the progenitor of ASASSN-14jb is possible (see \citealt{Howk2018b} for a discussion on the necessary conditions for extraplanar star formation). 

The MUSE nebular spectrum can be used to put constrains on the in-situ star formation at the position of ASASSN-14jb.
We fit the spectrum at the wavelength of the [O~III] $\lambda 5007$ line and obtain a 3$\sigma$ upper limit to the emission line luminosity of $\approx 10^{35}$~erg/s. We did not use H$\alpha$ because the strong and broad SN nebular emission line dominates at those wavelengths. Comparing the [O~III] $\lambda 5007$ luminosity upper limit with the ionization models of \cite{Lin18a}, we conclude that any underlying H~II region would have to be at least $\approx 10$~Myr old, implying M$_{ZAMS} < 15$~M$_{\odot}$ for a single star progenitor. All massive O stars would have been gone and exploded as CCSNe in such conditions. This indicates the need for a relatively low-mass progenitor of ASASSN-14jb in the in-situ star formation scenario, as well.

\subsection{Abundance Offset}

Another relevant piece of information is the expected abundance offset, $\Delta \epsilon$, of a former, or undetected, star forming region at the position of the SN \citep{Howk2018a,Howk2018b}.
To do so we need estimates of the oxygen abundance at the supernova position and the disk, given in the same scale.
\cite{Howk2018a} show that relative abundances in the O3N2 scale are reliable so we chose this one.

For the reference disk ($z=0$) we take the average of the H~II regions within 1$\sigma$ of their vertical distribution, which corresponds to a vertical distance of $\approx 0.25$ kpc. The reult is $\rm 12 + log(O/H) = 8.22 \pm 0.07$~dex.
For ASASSN-14jb we obtain the oxygen abundance from our estimate of the iron abundance derived in Section~\ref{subsec:Fe_abundance} of $Z = 0.3  \pm 0.1 Z_{\odot}$.
We adopt the linear relation between [O/H] and [Fe/H] given by \cite{stoll13}, and obtain $\rm 12 + log(O/H) = 8.4 \pm 0.1$~dex.
The oxygen abundance offset implied results $\Delta  \epsilon = 0.3 \pm 0.1 $~dex. 

Recent data from the Apache Point Observatory Galactic Evolution Experiment (APOGEE) survey (Majewski et al 2018), provides for an independent estimate.
Rojas-Arriagada et al. (2018, private comm., submitted) study abundance gradients using red giant branch stars, assumed to be likely bulge members by their spatial location. Adopting the same strategy on APOGEE data of our host we obtain relation between $[O/H]$ and $[Fe/H]$ similar to that of \cite{stoll13}.
A linear fit of the form $[Fe/H] = c_1 + c_2\times[O/H]$, using a re-sampling technique to estimate the errors, provides coefficients $c_1 = -0.2552 \pm 0.0014$ and $c_2 = 1.0452 \pm 0.0056$ was obtained.
This finally translates to an offset of $\Delta  \epsilon = 0.21 \pm 0.11 $~dex.

Another possible estimate of the the abundance at the supernova height is the extrapolation of the vertical gradient derived from the H~II regions in the disk. Doing a linear fit as a function $|z|$, we obtain a gradient of $-0.62 \pm 0.32$ dex per height normalized by $R_{25}$, which for ASASSN-14jb height corresponds to an oxygen abundance offset of $\Delta \epsilon = -0.14 \pm 0.07$~dex. 

\begin{figure}[ht!]
\includegraphics[width=9.4cm]{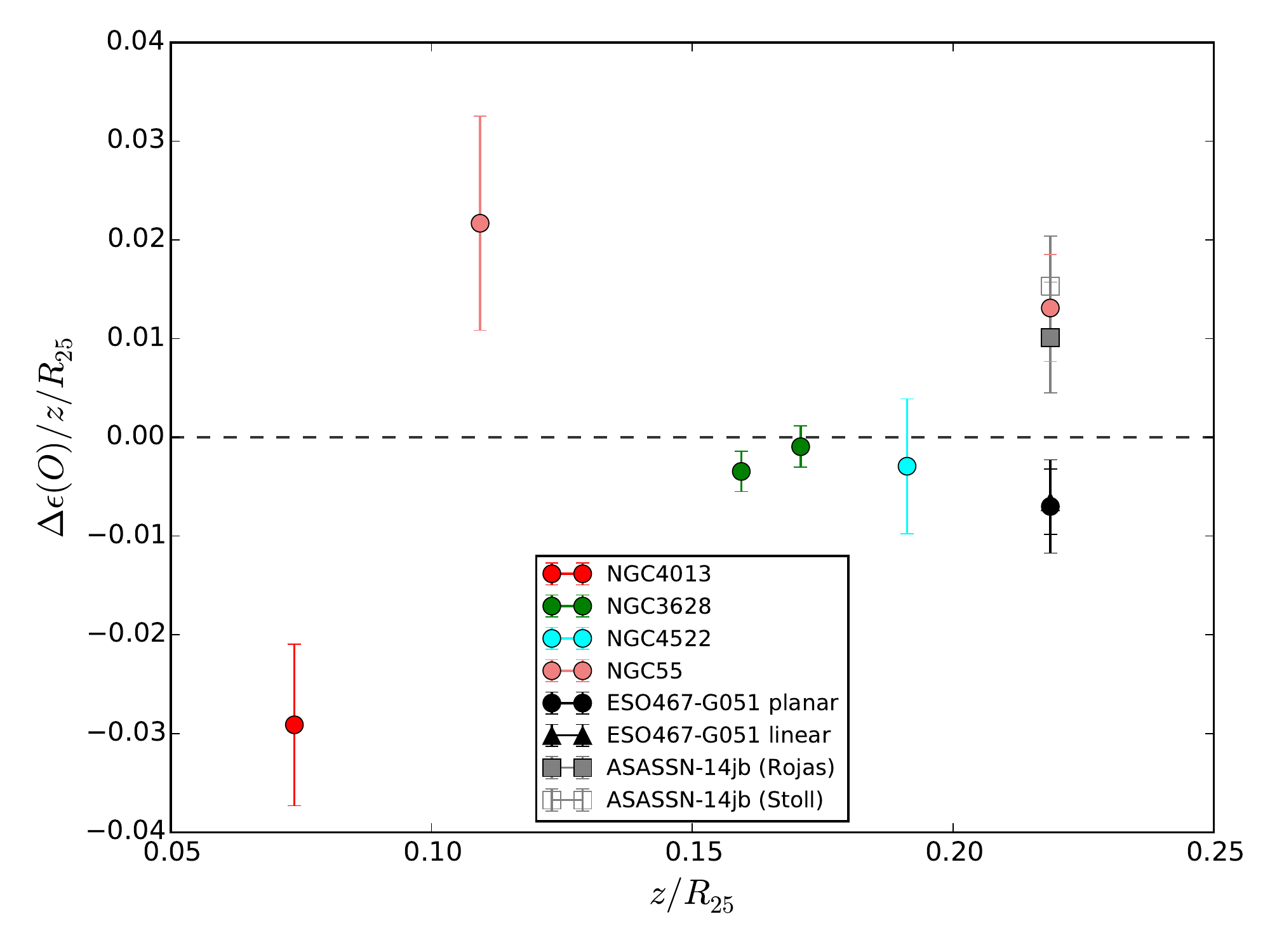}
\caption{Metalicity gradient obtained from the extraplanar H~II-regions from \cite{Howk2018b} and from ASASSN-14jb, as a function of the vertical distance to the disk  normalized by the isophotal radius $R_{25}$. ASASSN-14jb metallicity offset was obtained from the metallicity measurements from the FeII $\lambda5018$ line pEW and the calculated mean oxygen abundance of the host, ESO467-G051, from the O3N2 diagnostic.  A metallicity of $0.3 \pm 0.1 Z_\odot$ is used for ASASSN-14jb as obtained interpolating the models pEWs from \cite{luc13} and this was converted to an oxygen abundance using a linear scaling between [O/H] and [Fe/H] as done in \cite{stoll13} (marked with the unfilled square) while the calibration from Rojas et.al, (Submitted) was used in the value marked as a black and filled square. We also estimated the oxygen abundance at the supernova position extrapolating the vertical gradient obtained from our H~II-region abundances using the regions at any azimuthal position, dubbed ``linear", and from a two dimensional model,  dubbed ``planar", which are marked as a black triangle and square, respectively.\label{fig:z_grad_normed}}
\end{figure}

In Figure \ref{fig:z_grad_normed} we show the vertical oxygen abundance gradient $\Delta \epsilon /z$ derived for the extraplanar H~II region compilation of \cite{Howk2018a} and ASASSN-14jb, as a function of the height in kiloparsec in units of $R_{25}$. 
There are three interesting points to note from these plots: 1) There is a 2$\sigma$ offset between our abundance offset estimates (direct EW measurements from the SN versus vertical gradient of the disk abundance profile); 2) The metallicity estimate from ASASSN-14jb corresponds to a positive offset relative to the mean reference measurement of H~II regions in the galaxy disk; and 3) Including the $R_{25}$ normalization, the hypothetical H~II region where the progenitor of ASASSN-14jb could have formed (in an in-situ star-formation scenario) is farther than the \cite{Howk2018a} sample. 

An important caveat to consider in the previous analysis is that, because the galaxy is edge-on, we are probably observing mostly the outer H~II regions of the host galaxy that generally will not be representative of the whole disk. In general star-forming galaxies have higher metallicity gas in their inner regions, having a common radial gradient of $\approx -0.1$ dex/$R_{eff}$ (or $0.3 \pm 0.2$  dex$/R_{25}$) up to 2 $R_{eff}$ \citep[e.g.,][]{sanchez2012,Pilyugin2014,SanchezM2016}. As we do not actually know the radial coordinate of ASASSN-14jb and the observed H~II regions, there might be a systematic error on what we consider the reference abundance for the disk, being higher if the observed H~II regions are in average metal poor at the external disk. If we consider up to $\approx 0.2$  dex in the metallicity offset, given by the possible bias in our H~II regions, ASASSN-14jb metallicity is still consistent with the disk at the $1\sigma$ level. This is consistent with the possibility that the progenitor star actually came from the disk as a runaway or that the gas from which it formed at this height has significant contribution of enriched material from the disk.

\subsubsection{SN-Host connection}
We have constrained the properties of ASASSN-14jb both from the observations of the transient and from the observations of its host. Both sets of results provide a consistent picture.

From the early and late light curves we estimated an ejected mass ($\leq 10$ M$_\odot$), given a low explosion energy ($\leq 0.25$ foe) consistent with the transient energetics, and the low estimate of the $^{56}$Ni mass. Also, our analysis of the nebular spectra points to a low mass progenitor of $10-12$ ~M$_{\odot}$.

The analysis of the environment of ASASSN-14jb, led us to two possible explanations for the unusual location of the transient, which are the extraplanar in-situ star formation and the runaway scenarios.
In both cases, a low mass progenitor is required because their longer lifetime allows for a longer flight time in the runaway case and a longer time delay between the explosion of the more massive ionizing stars versus the less massive ones, given the strong upper limit of the possible underlying H~II region in the in-situ scenario.  

We have also obtained a constraint on the progenitor metallicity from a direct measurement of the Fe~II $\lambda5169$ pseudo equivalent width. The value of $0.3 \pm 0.1 Z_{\odot}$ is consistent, or mildly higher, than the values inferred for the gas-phase metallicity of the disk.
This, again we argue, is consistent with both scenarios as the progenitor metallicity could reflect that the star formed in the disk and was ejected or that the enriched gas itself was ejected and contributed to the abundance of the in-situ extraplanar region where the progenitor could have been born.
The efficiency of extraplanar enrichment depends on the depth of the galaxy potential well, which is a decreasing function of the stellar mass \citep{peeples2011,Brooks2014,Christensen2016,chisholm2018}.
Hence, the extraplanar relative enrichment is expected to be larger in low mass galaxies, as the ASASSN-14jb host, than, for example, Milky Way-like galaxies. The same reasoning applies for runaways stars as the maximum height that they achieve would be a decreasing function of the potential well depth, as well. 

\section{Conclusions}
\label{sec:conclusion}
ASASSN-14jb shows light curves and spectral evolution consistent with typical Type II SNe. The SN had a peak absolute $V$-band magnitude of  $M_{V,max} = -16.04 \pm 0.18$~mag and an estimated bolometric luminosity at 50 days of $\log{(L/L_{\odot})} = 8.4 \pm 0.18$, which put it in the low luminosity tail of the brightness distribution. Both the decline slope in the plateau phase ($s_{2,V} = -0.15 \pm 0.02$~mag per 100~days) and the expansion velocity at 50 days ($\rm v(FeII) = 2774 \pm 69$~km/s) follow the observed correlations with absolute magnitudes.

The color evolution of ASASSN-14jb points to a bluer continuum than other Type~IIP like SNe. This behavior may be due to a true difference in the temperature evolution or it may be an effect of lower line-blanketing in the blue part of the spectrum due to a lower metallicity of the progenitor star.

The spectral evolution presents a couple of interesting peculiarities. All our plateau observations show weaker lines (lower pseudo-EW) compared with a set of prototypical Type II SNe from the literature, which also points to a relatively bluer continuum or lower metallicity progenitor. Also, in photospheric spectra obtained at 64 and 80~days after explosion, we clearly detect high velocity absorption components at $\approx 7000$ km/s in the H$\alpha$ and H$\beta$ lines. These absorption components cannot be Si~II~$\lambda 6355$ or Ba absorption features, and signal a moderate interaction with circumstellar material. 

We used three empirical methods to measure Type II SNe distances and found internal consistency in the distance estimates, considering their uncertainties. The weighted average distance modulus of ASASSN-14jb was found to be $\mu = 32.00 \pm 0.18$~mag ($\rm D =25\pm 1$~Mpc).

From the late-time, nebular-phase photometry we estimate a $^{56}$Ni mass of $0.0210 \pm 0.0025 \ M_{\odot}$, slighly lower than the median for normal Type~II SNe. Based on the near-UV photometry from {\it Swift}/UVOT and analytic models from RW11 we estimate a progenitor radius of $R_* \simeq 579 \pm 28 R_{\odot}$, which is consistent with the color evolution of the mildly sub-Solar and compact RSG progenitor model (m15z8m3) presented in D13. Comparing our nebular-phase spectrum with models from D13 and J14 we conclude ASSASN14jb had a low to moderate mass progenitor ($10-12~M_{\odot}$).
The early light curve analysis yields $M_{ej} \approx 36.0 \times E_{exp}^{1.32}$~M$_{\odot}$, while at nebular times it implies $M_{ej} \approx 12.0 \times E_{exp}^{0.5}$~M$_{\odot}$. Both constraints intersect at the pair ($M_{ej}, E_{exp}$) $\approx$ (6 M$_{\odot}$, 0.25 foe). The estimated pseudo-equivalent width of the Fe~II $\lambda 5018$ line and its time evolution are more consistent with the theoretical expectations for a $(0.3 \pm 0.1) Z_{\odot}$ metallicity progenitor.
Given that the progenitor seems to be relatively compact ($R_* < 600$~$R_{\odot}$), the metallicity could be lower. More models would be needed to constrain this further.

We used the MUSE data to constrain the host galaxy gas-phase oxygen abundances. From the O3N2 and N2 strong line methods \citep{marino13} the H~II regions have a median abundance of $\rm 12 + log(O/H) = 8.27^{+0.16}_{-0.20}$ and $8.19^{+0.10}_{-0.24}$, respectively. Using the D16 \citep{dopita16} calibration we obtained a median of $\rm 12 + log(O/H) = 7.77^{+0.10}_{-0.24}$. These values are significantly below the average of the PISCO sample mean for Type II SNe nearby H~II regions, of $\rm 12 + log(O/H)= 8.54$~dex. This is independent of the diagnostics used. 

Finally, we discussed scenarios for the unusual explosion site of ASASSN-14jb. We conclude that, although the probablity of ejection from the disk is low from theory \citep{Renzo2018} and the specific mechanism to initiate the star formation at this heights is uncertain, both scenarios require a low mass progenitor with a ZAMS mass of $\approx 10 - 12 \ M_{\odot}$. This estimate is consistent with the physical parameters derived for the explosion.


\begin{acknowledgements}
NM acknowledges the Insitute of Astrophysics at PUC, the  Millennium Institute of Astrophysics (MAS), and the Astronomy Nucleus for University Diego Portales (UDP) for supporting this research. We thank Luc Dessart,  Enrique Perez-Montero, Ondrej Pejcha, Franz Bauer, Melina Bersten, Lin Xiao, and Antonia Bevan for valuable discussions. 

Support for NM and JLP was provided in part by FONDECYT through the grant 1151445. Support for NM, JLP, and AC was also provided by the Ministry of Economy, Development, and Tourism's Millennium Science Initiative
through grant IC120009, awarded to The Millennium Institute
of Astrophysics, MAS. The authors thank Las Cumbres Observatory and its staff for their continued support of ASAS-SN. ASAS-SN is supported by the Gordon and Betty Moore Foundation through grant GBMF5490 to the Ohio State University and NSF grant AST-1814440. Development of ASAS-SN has been supported by NSF grant AST-0908816, the Center for Cosmology and AstroParticle Physics at the Ohio State University, the Mt. Cuba Astronomical Foundation, the Chinese Academy of Sciences South America Center for Astronomy (CASSACA), and by George Skestos.

This research has made use of data from the Public ESO Spectroscopic Survey of Transient Objects \citep[PESSTO;][ESO program ID 191.D-0935]{smartt15} and {\it Spitzer} IRAC data from 2015-09-13 (program ID 11053), 2016-08-17 (program ID 12099) and 2014-09-05 (program ID 10139). This paper includes data gathered with the 6.5 meter Magellan Telescopes located at Las Campanas Observatory, Chile. This research has made use of the NASA/IPAC Extragalactic Database (NED) which is operated by the Jet Propulsion Laboratory, California Institute of Technology, under contract with the National Aeronautics. We acknowledge the use of the HyperLeda database (http://leda.univ-lyon1.fr). This work is based in part on observations made with the Spitzer Space Telescope, which is operated by the Jet Propulsion Laboratory, California Institute of Technology under a contract with NASA. Observations made with the NASA Galaxy Evolution Explorer (GALEX) were used in the analyses presented in this manuscript. Some of the data presented in this paper were obtained from the Mikulski Archive for Space Telescopes (MAST).

\end{acknowledgements}


\begin{appendix}
\section{Photometry}
Here we describe the photometric pipeline used to derive the magnitudes of ASASSN-14jb in the standard system. For each frame previously reduced we do the following steps to obtain the aperture photometry for the supernova:
\begin{itemize}
\item We run SExtractor \citep{sextractor} to obtain an initial estimate for the FWHM of the stars in the image, using bright stars far from the edges of the frame, and global background level.
\item We re-run SExtractor using the initial estimates and fixing the aperture to 1.6 times the median FWHM of the stars. After cross-matching this catalog with our local standard star catalog we reject the standard objects that have photometric uncertainties $\sigma_{\rm phot}^{i} > 0.1$. 
\item With the zero-point and uncertainty for each local standard star (no color term) $(\Delta m^{i},\sigma_{\rm phot}^{i})$ we calculate the mean zero-point of the image with its final error including both statistical photometric uncertainties and the rms of the zero-point. The typical value of the light curves deviations were $\lesssim 0.02$~mag. 
\begin{eqnarray}
\left<\Delta m \right> &=& \frac{\sum{\Delta m^{i}}}{N_{star}} \nonumber \\
Var(\left<\Delta m \right> ) = rms(\Delta m)^2 &+& \sum{\left(Var_{phot}^{i}+ Var_{catalog}^i\right)}
\end{eqnarray}
So with this the final photometry of the supernova and its variance are: 
\begin{eqnarray}
m_{SN}^i &=& m_{phot}^i + \left<\Delta m \right> \nonumber \\
Var (m_{SN}^i) &=& \sigma_{phot}^2 + Var(\left<\Delta m \right>)
\end{eqnarray}

\item Regarding color terms, as it is showed in Valenti et al. (in prep)\footnote{See \url{http://wiki.pessto.org/pessto-operation-groups/pessto-targets-alerts/lcogt-1m-telescope-time-for-pessto}.}, the only significant color term for LCOGT cameras in observations obtained in the  $BVgri$ filters are on the $g$-band. We apply this correction using the $g-r$ color but with the $r$-band magnitudes only corrected trough a mean zero-point term. That is, we follow the equations:\\
\begin{eqnarray}
g &=& g_{phot} + \left<\Delta g\right> + C(g-r) \nonumber \\
r &=& r_{phot} +  \left<\Delta r\right>
\end{eqnarray}
with the corresponding error propagation. The $g$-band color term correction obtained was important only at early times when the SN is bluest. 
\end{itemize}
To verify our results from aperture photometry, we ran PSF fitting photometry through our pipeline using the softwares DOPHOT and DAOPHOT in IRAF for selected frames, and in both cases we obtained consistent results. We chose the SExtractor aperture photometry because it was the most robust method. The PSF photometry of DOPHOT was limited because of the fixed gaussian profile shape and DAOPHOT required too much tuning of the parameters to obtain good results on frames that had significant sky variations or were noiser than average.

\section{P-Cygni profile fitting}

Expansion velocities were estimated using the velocity shift at maximum abpsorption in the, continuum normalized, P-Cygni profiles. For a given supernova we took each individual spectra $F_\lambda$ and, after masking the regions for the strong lines H$\alpha$ and H$\beta$ and NaI $\lambda\lambda5890,5896$ a black body is fitted. If the spectra is after 20 days past explosion the range $\lambda < 5100 $~\AA \ is also removed to avoid the contamination from line blanketing. Also a simple power law is fitted and if the residuals in this case are better this fit is adopted as the continuum. after estimating the continuum $F_C$ the spectra and wavelength is normalized as $ y = \frac{F_\lambda}{F_C}(x = \frac{\lambda -\lambda_0}{\lambda_0})$, where the 0 subscript refers to the line position (e.g. $\lambda_0 = 6563$~\AA \ for H$\alpha$).\\
To find a minima of a profile throughout a polinomial fitting, a function written in Python code was created. If no initial guess is given for the maximum abpsorption velocity $v_*$ (or equivalently $\lambda_* = \lambda_0 + \lambda_0 v_0/c$) first the data is smoothed with a high degree polinomial to find a guess. Then a new fitting window is defined, centered on the this guess, and a new low degree polinomial (from 2 up to 8) is fitted to find a new minima. In this final fit a weighting is applied to the data to avoid problems on the defined edges of the fitting window and the adopted guess. This weighting goes on the residuals as : 

\begin{eqnarray}
    \chi^2 = \sum_{i} w_i(y_i - y_{model}(x_i))^2 \nonumber \\ 
    w_i = \exp{(-|x-x_*|/s_x)}
\end{eqnarray}
The e-fold scale $s_x$ is used to only consider the points more closed to the assumed profile minima. For the sample of this work a systematic error was considered using different values for $s_x$ of 10,50 and 100 \AA, at least. This procedure effectively measures how biased is the chosen minima $x_*$.

We estimate a global statistical error, for each $s_x$, using a Jacknife procedure, where each data point is masked before fitting and a mean and standard deviation are obtained from the distribution.

\begin{figure}[ht!]
\centering
\includegraphics[scale = 0.5]{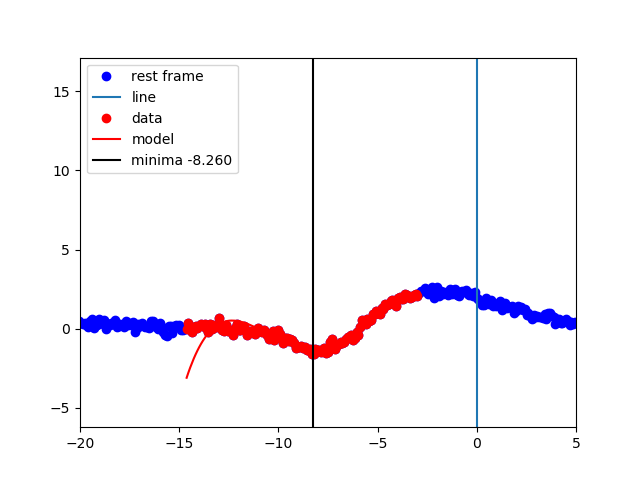}
\caption{Example of a P-Cygni profile fit. The x-axis is in units of 1000 km/s. The blue dots are the data points for the normalized spectra and the red points are the points actually used for the fitting procedure. The vertical line marks the position of the final velocity adopted and the cyan vertical indicates zero velocity at the line position. \label{fig:pcygni_fit}}
\end{figure}

\section{Distance Measurement}
\label{se:distances}
The methods for Type II SNe distance measurement can be categorized in theoretically based, such as the Expanding Photosphere Method \citep[EPM,][]{epm,epm2} or the spectra-fitting expanding atmosphere method \citep[SEAM,][]{seam}, and purely empirical such as the Standard Candle Method \citep[SCM ,][]{hamuy02}, the Photospheric Magnitude Method \citep[PMM][henceforth R14; R18]{Osmar14,Osmar18} or the Photometric Color Method \citep[DJ17][]{tomato17}.

We will compute distances to ASSASN-14bj using the three empirical ones.
We will not apply $K$-corrections since the low redshift of the source  makes them very small (e.g., less than 0.01~mag for the $V$-band).

\subsection{Photospheric Magnitude Method}

The PMM is a method to measure distances to Type II SNe that relies on a time-variable standardization of the photospheric magnitude evolution. This method can be viewed as a generalization for various epochs of the SCM (\citealt{hamuy02}) in the sense that both rely on the physical connection between the expansion velocity (a measure of the explosion energy) and the plateau luminosity, but the PMM has a different approach in the calibration. For a distance to be measured we need a good constraint on the shock-breakout epoch, magnitudes corrected from both host and local extinction and corrected by the $K$-correction, and expansion velocities inferred from the minima of the P-Cygni profiles (usually from the Fe~II lines) in the photospheric phase. Explicitly we have,\begin{eqnarray}
\mu_{\lambda} &=& m_{\lambda}^{corr} - \mathcal{M}_{\lambda} - \mathcal{R}  \nonumber \\
 \mathcal{R} &=& -5log\left(R_{ph}\right)+5 =  -5log\left(v_{ph}(t-t_0)\right)+5
\end{eqnarray}
 where $m^{corr}_{\lambda}$ is the rest-frame and extinction corrected magnitude at the wavelength $\lambda$, $\mathcal{M}_{\lambda}$ is the photospheric magnitude, $\mathcal{R}$ is the factor that considers the surface area of the photosphere and depends on the photospheric radius, which for homologous expansion is simply the product of photospheric velocity $v_{ph}$ and the time since the beginning of the expansion $t-t_0$ (see R14 for details). Although we have named $t_0$ the `explosion epoch`, this reference time is actually associated with the shock breakout. As our constraint on $t_0$ is of the order of the shock propagation time to reach a RSG surface after core-collapse \citep{arnett96} we do not make an explicit distinction. 
We have measured the velocities of ASASSN-14jb from the Fe II $\lambda5169$ line (see Figure \ref{fig:comp_vels}) and the extinction corrected magnitudes in the $V$-band without applying a $K$-correction. Averaging the distances obtained at the three epochs with Fe II $\lambda5169$ velocities, we obtain an average distance modulus of $\left<\mu\right>_{PMM} = 31.94 \pm 0.06 $ mag.

To make a direct comparison of the distance measurements of other methods we need consider that the PMM was calibrated using local distance ladder measurements, so the method carries an implicit value of the Hubble constant. As the value used by R14 is $H_0 = 68\, \rm km/s/Mpc$ for the $V$-band calibration, we will take this value as reference in the following analysis. We also use now the CMB frame redshift of $z_{CMB} = 0.00505$ for the ASASSN-14jb host. \citep{CMB_z}. 

\subsection{Standard Candle Method}

The most common method to measure Type IIP like distances is the SCM which relies on the tight correlation between brightness and the expansion velocity at a given epoch in the plateau phase. Similar to the Type Ia SNe calibrations, the SCM parameters are fitted directly in the Hubble diagram following the model \citep{hamuy02},
\begin{equation}
M_{\lambda_1} = -\alpha log\left(v_{ph}\right) + \beta (M_{\lambda_2}-M_{\lambda_1})
\end{equation}

where $M_{\lambda_k}$ is the absolute magnitude in the $k$-band, $v_{ph}$ is the photospheric velocity, usually at 50 days, and ($\alpha,\beta$) are the parameters to fit using a statistically significant sample. The distance modulus in the DJ17 prescription takes the form,
\begin{eqnarray}
\mu_{\lambda_1} &=& m_{\lambda_1} + \alpha log\left(\frac{v(H{\beta})}{<v(H{\beta})>}\right) - \beta (m_{\lambda_2}-m_{\lambda_1}) \nonumber \\
 &-& 5\log(H_0) - \mathcal{M}_{SCM} + 25
\end{eqnarray}. We use the parameters obtained it DJ17 for the SDSS $i,(r-i)$ color
combination\footnote{$\alpha_{SCM} = 3.18 \pm 0.41, \ \beta_{SCM} = 0.97 \pm 0.25 ,\ \mathcal{M}_{SCM} = -1.13 \pm 0.04$}. With the interpolated velocity in $H\beta$ at 40 days, using a power law model, of $v(H{\beta}) = 4181 \pm 146 \ km/s$ and the color at 45 days $(r-i)_0 = 0.03 \pm 0.13 $mag, the distance modulus is $\mu_{SCM} = 32.17 \pm 0.20$ mag. 

\subsection{Photometric Colour Method}

The PCM, presented in \citep{tomato15}, exploits the correlation between the light curve decline rate at the plateau, $s_2$, and the luminosity \citep{Joe14b}. In this method the distance modulus is obtained via,\begin{eqnarray}
\mu_{PCM}  &=& i - \alpha_{PCM} s_{2,i}+\beta_{PCM} (r-i)_0 + \mathcal{M}_{PCM} \nonumber \\
&+& 5log(H_0) - 25 
\end{eqnarray}
where the values of the parameters taken from DJ17 are $\alpha_{PCM} = 0.39\pm 0.08 , \ \beta_{PCM} =  0.8 \pm 0.48 , \ \mathcal{M}_{PCM} = -0.84 \pm 0.08$. Measuring the slope in the $i$-band light curve we find $s_{2,i} = -0.365 \pm 0.187$. With the corresponding color $(r-i)_0 = 0.027 \pm 0.16$ we obtain $\mu_{PCM} = 32.4 \pm 0.2$ mag.

\end{appendix}

\begin{table*}
\scriptsize 
\centering
\caption{$BVgri$ magnitudes of the local APASS standard star sequence used.\label{tab:std_table}}
\begin{tabular}{cccccccc}
\hline
{Star} & {$\alpha (deg)$} & {$\delta (deg)$} & {B}
& {V} & {g}  & {r} & {i} \\
\hline
1 & $335.744$ &  $-28.968$  &  $15.274$ ($0.020$) & $14.416$ ($0.010$) & $14.840$  ($0.015$) & $14.194$  ($0.012$) & 	 $13.947$ ($0.026$)\\ 
2 & $335.773$ &  $-28.917$  &  $15.742$ ($0.062$) & $15.154$ ($0.016$) & $15.437$  ($0.022$) & $15.024$  ($0.000$) & 	 $14.967$ ($0.111$)\\ 
3 & $335.805$ &  $-28.912$  &  $16.232$ ($0.104$) & $15.562$ ($0.082$) & $15.849$  ($0.037$) & $15.380$  ($0.016$) & 	 $15.309$ ($0.104$)\\ 
4 & $335.808$ &  $-29.040$  &  $17.149$ ($0.213$) & $16.056$ ($0.000$) & $16.523$  ($0.045$) & $15.998$  ($0.098$) & 	 $15.823$ ($0.026$)\\ 
5 & $335.829$ &  $-29.009$  &  $15.577$ ($0.037$) & $14.658$ ($0.024$) & $15.064$  ($0.014$) & $14.389$  ($0.014$) & 	 $14.146$ ($0.056$)\\ 
6 & $335.838$ &  $-28.927$  &  $15.838$ ($0.018$) & $14.829$ ($0.042$) & $15.325$  ($0.011$) & $14.527$  ($0.030$) & 	 $14.262$ ($0.036$)\\ 
7 & $335.844$ &  $-28.936$  &  $17.087$ ($0.043$) & $15.579$ ($0.067$) & $16.356$  ($0.006$) & $14.986$  ($0.039$) & 	 $14.083$ ($0.041$)\\ 
8 & $335.852$ &  $-29.013$  &  $14.603$ ($0.029$) & $13.906$ ($0.012$) & $14.229$  ($0.000$) & $13.712$  ($0.022$) & 	 $13.573$ ($0.046$)\\ 
9 & $335.863$ &  $-29.016$  &  $14.755$ ($0.017$) & $14.084$ ($0.041$) & $14.369$  ($0.010$) & $13.945$  ($0.025$) & 	 $13.824$ ($0.046$)\\ 
10 & $335.870$ &  $-28.995$  &  $15.693$ ($0.099$) & $14.964$ ($0.053$) & $15.272$  ($0.022$) & $14.794$  ($0.026$) & 	 $14.655$ ($0.136$)\\ 
\hline
\label{tab:std}
\end{tabular}
\end{table*}

\newpage
\begin{table*}{}
\caption{ASAS-SN $V$-band photometry of ASASSN-14jb, including pre-discovery and post-plateau upper limits. (*) : Days after the estimated explosion time, $t_0 = 56946.58$ (MJD).}
\scriptsize 
\centering
\begin{tabular}{lcccccl}
\hline
UT Date & MJD & $(t-t_0)^{(*)}$ & V ($1\sigma$ error) & Telescope\\
\hline
 & (days) & (days) & (mag) &\\ 
\hline
2014-10-13 & 56943.098 & -3.485 &$>18.343$ & ASAS-SN \\ 
2014-10-19 & 56949.093 & 2.510 &16.861 (0.129) & ASAS-SN \\ 
2014-10-20 & 56950.060 & 3.477 &16.284 (0.069) & ASAS-SN \\ 
2014-10-22 & 56952.058 & 5.475 &16.154 (0.065) & ASAS-SN \\ 
2014-10-23 & 56953.051 & 6.468 &16.072 (0.065) & ASAS-SN \\ 
2014-10-29 & 56959.084 & 12.501 &16.139 (0.069) & ASAS-SN \\ 
2014-11-04 & 56965.117 & 18.534 &16.040 (0.109) & ASAS-SN \\ 
2014-11-07 & 56968.064 & 21.481 &16.026 (0.112) & ASAS-SN \\ 
2014-11-11 & 56972.059 & 25.476 &16.071 (0.063) & ASAS-SN \\ 
2014-11-12 & 56973.144 & 26.561 &15.907 (0.063) & ASAS-SN \\ 
2014-11-16 & 56977.056 & 30.473 &16.118 (0.078) & ASAS-SN \\ 
2014-11-21 & 56982.058 & 35.475 &16.089 (0.068) & ASAS-SN \\ 
2014-11-24 & 56985.046 & 38.463 &16.106 (0.068) & ASAS-SN \\ 
2014-12-01 & 56992.053 & 45.470 &16.125 (0.088) & ASAS-SN \\ 
2014-12-04 & 56995.031 & 48.448 &16.230 (0.103) & ASAS-SN \\ 
2014-12-08 & 56999.035 & 52.452 &16.213 (0.084) & ASAS-SN \\ 
2014-12-09 & 57000.034 & 53.451 &15.995 (0.081) & ASAS-SN \\ 
2014-12-14 & 57005.026 & 58.443 &15.984 (0.108) & ASAS-SN \\ 
2014-12-19 & 57010.044 & 63.460 &16.030 (0.062) & ASAS-SN \\ 
2015-04-10 & 57122.421 & 175.838 &$>16.596$ & ASAS-SN \\ 
\hline
\end{tabular}

\label{tab:asassn}
\end{table*}


\begin{sidewaystable*}
\scriptsize 
\centering
\caption{LCOGT $BVgri$ and NTT/EFOSC $BV$ photometry of ASASSN-14jb. (*) : Days after the estimated explosion time, $t_0 = 56946.58$.}

\begin{tabular}{ ccccccccc } 
\hline
UT Date & MJD & $(t-t_0) (*)$ & $B$  & $V$  & $g$ & $r$ & $i$ &  Telescope \\
\hline
2014-10-19 & 56949.785 &    3.202 & \ldots & \ldots &16.077 (0.023) & 16.402 (0.019) &  \ldots &LCOGT~1m/SAAO \\ 
2014-10-19 & 56949.901 &    3.318 & \ldots & \ldots &16.072 (0.020) & 16.376 (0.014) &  \ldots &LCOGT~1m/SAAO \\ 
2014-10-20 & 56950.393 &    3.810 & \ldots & \ldots &15.986 (0.022) & 16.237 (0.015) &  \ldots &LCOGT~1m/SSO \\ 
2014-10-20 & 56950.802 &    4.219 & \ldots & \ldots &15.960 (0.021) & 16.159 (0.014) &  \ldots &LCOGT~1m/SAAO \\ 
2014-10-21 & 56951.536 &    4.953 &16.084 (0.036) & 16.016 (0.020) & 15.973 (0.021) & 16.080 (0.021) & 16.231 (0.033) & LCOGT~1m/SSO \\ 
2014-10-22 & 56952.532 &    5.949 &16.055 (0.025) & 16.022 (0.016) & 15.948 (0.021) & 16.055 (0.013) & 16.217 (0.021) & LCOGT~1m/SSO \\ 
2014-10-23 & 56953.527 &    6.944 &15.992 (0.025) & 15.970 (0.017) & 15.944 (0.018) & 15.984 (0.012) & 16.116 (0.023) & LCOGT~1m/SSO \\ 
2014-10-24 & 56954.141 &    7.558 &15.995 (0.027) & 15.941 (0.016) & 15.913 (0.022) & 15.982 (0.014) & 16.195 (0.027) & LCOGT~1m/MDO \\ 
2014-10-24 & 56954.543 &    7.960 &16.021 (0.033) & 15.960 (0.016) & 15.927 (0.020) & 15.945 (0.011) & 16.091 (0.022) & LCOGT~1m/SSO \\ 
2014-10-26 & 56956.830 &   10.247 &16.026 (0.023) & 15.974 (0.016) &  \ldots &15.948 (0.012) & 16.128 (0.020) & LCOGT~1m/SAAO \\ 
2014-10-29 & 56959.546 &   12.963 &16.097 (0.023) & 15.980 (0.015) & 15.967 (0.021) & 15.911 (0.011) & 16.084 (0.020) & LCOGT~1m/SSO \\ 
2014-11-01 & 56962.518 &   15.935 &16.166 (0.036) & 16.020 (0.026) & 16.064 (0.026) & 15.910 (0.017) & 16.096 (0.042) & LCOGT~1m/SSO \\ 
2014-11-04 & 56965.576 &   18.993 &16.211 (0.028) & 15.955 (0.019) & 16.023 (0.030) & 15.861 (0.014) & 15.991 (0.023) & LCOGT~1m/SSO \\ 
2014-11-07 & 56968.505 &   21.921 &16.309 (0.032) & 15.984 (0.024) & 16.046 (0.029) & 15.860 (0.023) & 15.934 (0.037) & LCOGT~1m/SSO \\ 
2014-11-10 & 56971.505 &   24.921 &16.348 (0.029) & 15.949 (0.020) & 16.127 (0.024) & 15.833 (0.016) & 15.948 (0.027) & LCOGT~1m/SSO \\ 
2014-11-13 & 56974.763 &   28.180 &16.440 (0.028) & 16.018 (0.017) & 16.193 (0.022) & 15.866 (0.012) & 15.980 (0.021) & LCOGT~1m/SAAO \\ 
2014-11-17 & 56978.766 &   32.183 &16.510 (0.032) & 16.023 (0.019) & 16.212 (0.023) & 15.867 (0.014) & 15.939 (0.023) & LCOGT~1m/SAAO \\ 
2014-11-21 & 56982.777 &   36.194 &16.566 (0.034) & 15.998 (0.019) & 16.223 (0.024) & 15.853 (0.014) & 15.861 (0.029) & LCOGT~1m/SAAO \\ 
2014-11-25 & 56986.781 &   40.198 &16.582 (0.026) & 15.980 (0.017) & 16.274 (0.022) & 15.848 (0.013) & 15.902 (0.021) & LCOGT~1m/SAAO \\ 
2014-11-28 & 56989.854 &   43.271 &16.633 (0.031) & 15.972 (0.021) & 16.259 (0.024) & 15.839 (0.014) & 15.830 (0.022) & LCOGT~1m/SAAO \\ 
2014-12-01 & 56992.776 &   46.193 &16.679 (0.050) & 15.979 (0.025) & 16.255 (0.025) & 15.847 (0.023) & 15.805 (0.033) & LCOGT~1m/SAAO \\ 
2014-12-04 & 56995.836 &   49.253 &16.660 (0.059) & 15.961 (0.026) & 16.259 (0.030) & 15.784 (0.021) & 15.751 (0.028) & LCOGT~1m/SAAO \\ 
2014-12-08 & 56999.102 &   52.519 &16.681 (0.029) & 15.962 (0.016) & 16.262 (0.021) & 15.802 (0.012) & 15.797 (0.019) & LCOGT~1m/CTIO \\ 
2014-12-13 & 57004.784 &   58.201 &16.669 (0.027) & 15.955 (0.015) & 16.239 (0.020) & 15.785 (0.011) & 15.788 (0.019) & LCOGT~1m/SAAO \\ 
2014-12-16 & 57007.794 &   61.211 &16.678 (0.026) & 15.954 (0.016) & 16.273 (0.020) & 15.815 (0.011) & 15.802 (0.020) & LCOGT~1m/SAAO \\ 
2014-12-26 & 57017.040 &   70.457 &16.750 (0.032) & 15.974 (0.017) & 16.304 (0.020) & 15.825 (0.012) & 15.807 (0.020) & LCOGT~1m/CTIO \\ 
2014-12-29 & 57020.430 &   73.847 &16.765 (0.035) & 15.997 (0.016) & 16.320 (0.023) & 15.843 (0.012) & 15.822 (0.021) & LCOGT~1m/SSO \\ 
2015-04-11 & 57123.408 &  176.824 &20.600 (0.109) & 19.114 (0.034) &  \ldots & \ldots & \ldots &ESO-NTT \\ 
2015-10-11 & 57306.204 &  359.621 &21.938 (0.042) & 21.108 (0.021) &  \ldots & \ldots & \ldots &ESO-NTT \\ 
2015-10-11 & 57306.237 &  359.654 &22.035 (0.032) &  \ldots & \ldots & \ldots & \ldots &ESO-NTT \\ 
2015-12-11 & 57367.058 &  420.475 &22.752 (0.069) &  \ldots & \ldots & \ldots & \ldots &ESO-NTT \\ 
2015-12-11 & 57367.091 &  420.508 & \ldots &21.827 (0.063) &  \ldots & \ldots & \ldots &ESO-NTT \\
\hline
\label{tab:allphot}
\end{tabular}

\end{sidewaystable*}

\begin{table*}
\scriptsize 
\centering
\caption{Log of spectroscopic observations of ASASSN-14jb. (*): Days after the estimated explosion time, $t_0 = 56946.58$.}
\begin{tabular}{ccccc}
\hline
UT Date & MJD &  $(t-t_0)^{(*)}$  & Airmass & Telescope\\
\hline
2014-10-20 & 56950.15 & 3.57 & 2.03 & FLWO~1.5m \\ 
2014-10-21 & 56951.19 & 4.61 & 2.11 & FLWO~1.5m \\ 
2014-10-22 & 56952.16 & 5.58 & 2.04 & FLWO 1.5m \\ 
2014-10-23 & 56953.17 & 6.59 & 2.05 & FLWO 1.5m \\ 
2014-10-24 & 56954.17 & 7.59 & 2.08 & FLWO 1.5m \\ 
2014-10-25 & 56955.18 & 8.59 & 2.11 & FLWO 1.5m \\ 
2014-10-27 & 56957.13 & 10.55 & 2.03 & FLWO 1.5m \\ 
2014-11-20 & 56981.09 & 34.51 & 2.05 & FLWO 1.5m \\ 
2014-12-20 & 57011.08 & 64.50 & 2.66 & FLWO 1.5m \\ 
2015-01-04 & 57026.55 & 79.96 & 2.20 & Baade 6.5m \\ 
2015-11-14 & 57340.10 & 393.52 & 1.28 & VLT-UT4 8.4m \\ 
\hline
\label{tab:spectra}
\end{tabular}
\end{table*}

\begin{table*}
\scriptsize 
\centering

\caption{Pcygni abpsorption velocities for ASASSN-14jb. All velocities quoted assuming the heliocentric redshift of z = 0.006031.\label{tab:vel}}

\begin{tabular}{ccccccc}
\hline
{UT Date} & {MJD} & {$t-t_0$} & {$H\beta$}  & {Fe II $\lambda$5018} & {Fe II $\lambda$5169}  & {
H$\alpha$} \\
\hline

2014-10-20 & 56950.15 & 3.57 &7900 (148) &\ldots &\ldots &10784 ( 386)\\ 
2014-10-21 & 56951.19 & 4.61 &8624 (54) &\ldots &\ldots &\ldots \\ 
2014-10-22 & 56952.16 & 5.58 &8631 (17) &\ldots &\ldots &9324 (310)\\ 
2014-10-23 & 56953.17 & 6.59 &8313 (43) &\ldots &\ldots &9146 (256)\\ 
2014-10-24 & 56954.17 & 7.59 &8038 (16) &\ldots &\ldots &8667 (316)\\ 
2014-10-25 & 56955.18 & 8.59 &7721 (127) &\ldots &\ldots &8573 (231)\\ 
2014-10-27 & 56957.13 & 10.55 &7432 (41) &\ldots &\ldots &8205 (189)\\ 
2014-11-20 & 56981.09 & 34.51 &4826 (21) &3052 (112) &3221 (88) &6354 (110)\\ 
2014-12-20 & 57011.08 & 64.50 &3300 (65) &2296 (47) &2263 (110) &4605 (115)\\ 
2015-01-04 & 57026.05 & 79.46 &2803 (72) &2024 (20) &1976 (32) &4211 (24)\\ 
\hline
\end{tabular}
\end{table*}

\begin{table*}
\centering
\scriptsize 
\caption{Relevant information of the SN~II in the literature used for comparison with ASASSN-14jb.\label{tab:comp_Table}.}

\begin{tabular}{ccccccc}
\hline
{SN Name} & {$t_0$} & {$E(B-V)$} & {Host}
& {$z(*)$} & {$\mu$}  & {References$(+)$} \\
\hline
SN1999em & $2451476.44$ & $0.100$ & NGC 1637 & $0.00239$ & $30.360$ & 1,2,3,12 \\
SN1999gi & $2451518.32$ & $0.210$ & NGC 3184 & $0.00198$ & $30.830$ & 4,12 \\ 
SN2005cs & $2453548.92$ & $0.050$ & M51 & $0.00200$ & $29.630$ & 5,12 \\ 
SN2007pk & $2454412.32$ & $0.100$ & NGC 3953 & $0.01665$ & $34.230$ & 6 \\ 
SN2009N & $2454845.46$ & $0.132$ & NGC 4487 & $0.00450$ & $31.840$ & 7 \\ 
SN2012aw & $2456002.50$ & $0.086$ & M95 & $0.00260$ & $30.150$ & 8,9,12 \\ 
SNLSQ13fn & $2456293.98$ & $0.054$ & GALEXASCJ115117.64-293638.0 & $0.06300$ & $37.100$ &  10\\ 
SN2013ab & $2456340.00$ & $0.044$ & NGC 5669 & $0.00456$ & $31.920$ &  11\\ 

\hline
\end{tabular}\\
(*): Heliocentric redshift from NED. (+): (1) \cite{faran14a},(2) \cite{hamuy01},(3) \cite{leonard02a},(4) \cite{leonard02b},(5) \cite{pastorello09}, (6) \cite{Inserra13},(7) \cite{katy14},(8) \cite{bose13},(9) \cite{dallora14},(10) \cite{polshaw16},(11) \cite{bose15}, (12) Distance from \cite{Osmar14}
\end{table*}

\begin{table*}
\centering
\scriptsize 
\caption{Integrated magnitudes of the host galaxy of ASASSN-14jb, ESO\label{tab:hostmags}}

\begin{tabular}{cccc}
\hline 
Filter/band & Magnitude & System & Reference\\
\hline 
Galex $FUV$ & $17.27 \pm 0.05$ & AB & 1 \\
Galex $NUV$ & $15.24 \pm 0.05$ & AB & 1 \\
$B$   & $14.45 \pm 0.05$ & Vega & this work \\
$V$   & $14.10 \pm 0.05$ & Vega & this work \\
$g$   & $14.23 \pm 0.05$ & AB & this work \\
$r$   & $13.92 \pm 0.05$ & AB & this work \\ 
$i$  & $13.50 \pm 0.05$ & AB & this work \\
$K_s$ & $12.89\pm 0.10$ & Vega & this work \\
Spitzer $3.6$~$\mu$m & $14.90 \pm 0.05$  &  AB & 2 \\
Spitzer $4.5$~$\mu$m & $15.33 \pm 0.05$  &  AB & 2 \\
\hline 
(1) \cite{galex}, (2) \cite{sheth10}
\end{tabular}
\end{table*}


\begin{thebibliography}{}

\bibitem[Abadi et al.(2009)]{abadi09} Abadi, M.~G., Navarro, J.~F., \& Steinmetz, M.\ 2009, \apjl, 691, L63
\bibitem[Alard \& Lupton(1998)]{alardlupton98} Alard, C., \& Lupton, R.~H.\ 1998, \apj, 503, 325
\bibitem[Alard(2000)]{alard2000} Alard, C.\ 2000, \aaps, 144, 363
\bibitem[Alatalo et al.(2016)]{BPTshocks} Alatalo, K., Cales, S.~L., Rich, J.~A., et al.\ 2016, \apjs, 224, 38 
\bibitem[Anderson \& James(2009)]{Joe09} Anderson, J.~P., \& James, P.~A.\ 2009, \mnras, 399, 559 
\bibitem[Anderson et al.(2010)]{Joe10} Anderson, J.~P., Covarrubias, R.~A., James, P.~A., Hamuy, M., \& Habergham, S.~M.\ 2010, \mnras, 407, 2660
\bibitem[Anderson et al.(2012)]{Joe12} Anderson, J.~P., Habergham, S.~M., James, P.~A., \& Hamuy, M.\ 2012, \mnras, 424, 1372
\bibitem[Anderson et al.(2014)]{Joe14} Anderson, J.~P., Dessart, L., Gutierrez, C.~P., et al.\ 2014, \mnras, 441, 671
\bibitem[Anderson et al.(2014)]{Joe14b} Anderson, J.~P., Gonz{\'a}lez-Gait{\'a}n, S., Hamuy, M., et al.\ 2014, \apj, 786, 67
\bibitem[Anderson et al.(2015)]{Joe15} Anderson, J.P., James, P.A., Habergham, S.M., Galbany, L., \& Kuncarayakti, H.\ 2015, \pasa, 32, e019
\bibitem[Anderson et al.(2016)]{Joe16} Anderson, J.~P., Guti{\'e}rrez, C.~P., Dessart, L., et al.\ 2016, \aap, 589, A110 
\bibitem[Arcavi et al.(2010)]{Arcavi10} Arcavi, I., Gal-Yam, A., Kasliwal, M.~M., et al.\ 2010, \apj, 721, 777 
\bibitem[Arnett(1996)]{arnett96} Arnett, D.\ 1996, Supernovae and Nucleosynthesis: An Investigation of the History of Matter, from the Big Bang to the Present, by D.~Arnett.~Princeton: Princeton University Press, 1996.,  
\bibitem[Asplund et al.(2009)]{Asplund09} Asplund, M., Grevesse, N., Sauval, A.~J., \& Scott, P.\ 2009, \araa, 47, 481 
\bibitem[Bacon et al.(2010)]{bacon10} Bacon, R., Accardo, M., Adjali, L., et al.\ 2010, \procspie, 7735, 773508
\bibitem[Baldwin et al.(1981)]{BPT81} Baldwin, J.~A., Phillips, M.~M., \& Terlevich, R.\ 1981, \pasp, 93, 5 
\bibitem[Barbon et al.(1979)]{Barbon79} Barbon, R., Ciatti, F., \& Rosino, L.\ 1979, \aap, 72, 287
\bibitem[Baron et al.(2004)]{seam} Baron, E., Nugent, P.~E., Branch, D., \& Hauschildt, P.~H.\ 2004, \apjl, 616, L91
\bibitem[Baron et al.(2005)]{PHOENIX} Baron, E., Nugent, P.~E., Branch, D., \& Hauschildt, P.~H.\ 2005, 1604-2004: Supernovae as Cosmological Lighthouses, 342, 351 
\bibitem[Bartunov et al.(1994)]{Bartunov94} Bartunov, O.~S., Tsvetkov, D.~Y., \&
\bibitem[Becker(2015)]{becker15} Becker, A.\ 2015, Astrophysics Source Code Library, ascl:1504.004 Filimonova, I.~V.\ 1994, \pasp, 106, 1276 
\bibitem[Bersten \& Hamuy(2009)]{melina09} Bersten, M.~C., \& Hamuy, M.\ 2009, \apj, 701, 200 
\bibitem[Bersten et al.(2011)]{melina11} Bersten, M.~C., Benvenuto, O., \& Hamuy, M.\ 2011, \apj, 729, 61
\bibitem[Bertin \& Arnouts(1996)]{sextractor} Bertin, E., \& Arnouts, S.\ 1996, \aaps, 117, 393
\bibitem[Blaauw(1993)]{blaauw93} Blaauw, A.\ 1993, Massive Stars:  Their Lives in the Interstellar Medium, 35, 207
\bibitem[Blinnikov \& Bartunov(1993)]{Blinnikov1993} Blinnikov, S.~I., \& Bartunov, O.~S.\ 1993, \aap, 273, 106 
\bibitem[Bose et al.(2013)]{bose13} Bose, S., Kumar, B., Sutaria, F., et al.\ 2013, \mnras, 433, 1871
\bibitem[Bose et al.(2015)]{bose15} Bose, S., Valenti, S., Misra, K., et al.\ 2015, \mnras, 450, 2373
\bibitem[Bose et al.(2016)]{bose16} Bose, S., Kumar, B., Misra, K., et al.\ 2016, \mnras, 455, 2712 
\bibitem[Bose et al.(2018)]{bose18} Bose, S., Dong, S., Kochanek, C.~S., et al.\ 2018, arXiv:1804.00025 
\bibitem[Brimacombe et al.(2014)]{discovery} Brimacombe, J., Kiyota, S., Holoien, T.~W.-S., et al.\ 2014, The Astronomer's Telegram, 6592, 
\bibitem[Brook et al.(2014)]{Brooks2014} Brook, C.~B., Stinson, G., Gibson, B.~K., et al.\ 2014, \mnras, 443, 3809 
\bibitem[Brown et al.(2013)]{brown13} Brown, T.~M., Baliber, N., Bianco, F.~B., et al.\ 2013, \pasp, 125, 1031
\bibitem[Branch \& Wheeler(2017)]{BandW2017} Branch, D., \& Wheeler, J.~C.\ 2017, Supernova Explosions: Astronomy and Astrophysics Library, ISBN 978-3-662-55052-6.~Springer-Verlag GmbH Germany, Cap. 1.3
\bibitem[Brown et al.(2014)]{sousa} Brown, P.~J., Breeveld, A.~A., Holland, S., Kuin, P., \& Pritchard, T.\ 2014, \apss, 354, 89
\bibitem[Brown(2015)]{brown15} Brown, W.~R.\ 2015, \araa, 53, 15
\bibitem[Bruzual \& Charlot(2003)]{bruzualcharlot03} Bruzual, G., \& Charlot, S.\ 2003, \mnras, 344, 1000 
\bibitem[Cardelli et al.(1989)]{cardelli} Cardelli, J.~A., Clayton, G.~C., \& Mathis, J.~S.\ 1989, \apj, 345, 245
\bibitem[Challis(2014)]{classification1} Challis, P.\ 2014, The Astronomer's Telegram, 6600,  
\bibitem[Christensen et al.(2016)]{Christensen2016} Christensen, C.~R., Dav{\'e}, R., Governato, F., et al.\ 2016, \apj, 824, 57 
\bibitem[Chevalier \& Soderberg(2010)]{Chevalier2010} Chevalier, R.~A., \& Soderberg, A.~M.\ 2010, \apjl, 711, L40 
\bibitem[Chisholm et al.(2018)]{chisholm2018} Chisholm, J., Tremonti, C., \& Leitherer, C.\ 2018, \mnras,  
\bibitem[Chugai(1994)]{Chugai94} Chugai, N.~N.\ 1994, Circumstellar Media in Late Stages of Stellar Evolution, 148
\bibitem[Chugai et al.(2005)]{Chugai2005} Chugai, N.~N., Fabrika, S.~N., Sholukhova, O.~N., et al.\ 2005, Astronomy Letters, 31, 792 
\bibitem[Cid Fernandes et al.(2009)]{cidfernandes09} Cid Fernandes, R., Schoenell, W., Gomes, J.~M., et al.\ 2009, Revista Mexicana de Astronomia y Astrofisica Conference Series, 35, 127 
\bibitem[Clocchiatti \& Wheeler(1997)]{clocchi97} Clocchiatti, A., \& Wheeler, J.~C.\ 1997, \apj, 491, 375
\bibitem[Cutri et al.(2003)]{2mass} Cutri, R.~M., Skrutskie, M.~F., van Dyk, S., et al.\ 2003, VizieR Online Data Catalog, 2246 
\bibitem[Dalcanton \& Bernstein(2002)]{Dalcanton2002} Dalcanton, J.~J., \& Bernstein, R.~A.\ 2002, \aj, 124, 1328 
\bibitem[Dall'Ora et al.(2014)]{dallora14} Dall'Ora, M., Botticella, M.~T., Pumo, M.~L., et al.\ 2014, \apj, 787, 139
\bibitem[Dessart \& Hillier(2005)]{Luc05b} Dessart, L., \& Hillier, D.~J.\ 2005, \aap, 439, 671
\bibitem[Dessart \& Hillier(2005)]{Luc05a} Dessart, L., \& Hillier, D.~J.\ 2005, \aap, 437, 667 
\bibitem[Dessart \& Hillier(2005)]{CMFGEN} Dessart, L., \& Hillier, D.~J.\ 2005, The Fate of the Most Massive Stars, 332, 427 
\bibitem[Dessart \& Hillier(2010)]{luc10} Dessart, L., \& Hillier, D.~J.\ 2010, \mnras, 405, 2141
\bibitem[Dessart et al.(2013)]{luc13} Dessart, L., Hillier, D.~J., Waldman, R., \& Livne, E.\ 2013, \mnras, 433, 1745 (D13)
\bibitem[de Vaucouleurs et al.(1991)]{devac91} de Vaucouleurs, G., de Vaucouleurs, A., Corwin, H.~G., Jr., et al.\ 1991, Third Reference Catalogue of Bright Galaxies (New York: Springer)
\bibitem[Dopita et al.(1984)]{Dopita1984} Dopita, M.~A., Evans, R., Cohen, M., \& Schwartz, R.~D.\ 1984, \apjl, 287, L69 
\bibitem[Dopita et al.(2016)]{dopita16} Dopita, M.~A., Kewley, L.~J., Sutherland, R.~S., \& Nicholls, D.~C.\ 2016, \apss, 361, 61 
\bibitem[Dressler et al.(2011)]{dressler11} Dressler, A., Bigelow, B., Hare, T., et al.\ 2011, \pasp, 123, 288
\bibitem[Dwek et al.(1983)]{Dwek1983} Dwek, E., A'Hearn, M.~F., Becklin, E.~E., et al.\ 1983, \apj, 274, 168 
\bibitem[Eastman et al.(1996)]{eastman96} Eastman, R.~G., Schmidt, B.~P., \& Kirshner, R.\ 1996, \apj, 466, 911 
\bibitem[Ekstr{\"o}m et al.(2012)]{ekstrom12} Ekstr{\"o}m, S., Georgy, C., Eggenberger, P., et al.\ 2012, \aap, 537, A146 
\bibitem[Eldridge et al.(2011)]{eldridge11} Eldridge, J.~J., Langer, N., \& Tout, C.~A.\ 2011, \mnras, 414, 3501
\bibitem[Eldridge et al.(2013)]{Eldridge13} Eldridge, J.~J., Fraser, M., Smartt, S.~J., Maund, J.~R., \& Crockett, R.~M.\ 2013, \mnras, 436, 774 
\bibitem[Eldridge et al.(2017)]{eldridge17} Eldridge, J.~J., Stanway, E.~R., Xiao, L., et al.\ 2017, \pasa, 34, e058 
\bibitem[Elmegreen et al.(2017)]{Elmegreen2017} Elmegreen, B.~G., Elmegreen, D.~M., Tompkins, B., \& Jenks, L.~G.\ 2017, \apj, 847, 14 
\bibitem[Elmhamdi et al.(2003)]{elmhamdi03_99em} Elmhamdi, A., Danziger, I.~J., Chugai, N., et al.\ 2003, \mnras, 338, 939
\bibitem[Fabricant et al.(1998)]{fabricant98} Fabricant, D., Cheimets, P., Caldwell, N., \& Geary, J.\ 1998, \pasp, 110, 79
\bibitem[Faran et al.(2014)]{faran14a} Faran, T., Poznanski, D., Filippenko, A.~V., et al.\ 2014, \mnras, 442, 844
\bibitem[Fazio et al.(2004)]{fazio04} Fazio, G.~G., Hora, J.~L., Allen, L.~E., et al.\ 2004, \apjs, 154, 10 
\bibitem[Filippenko(1988)]{Filipenko1988} Filippenko, A.~V.\ 1988, \aj, 96, 1941 
\bibitem[Filippenko et al.(1993)]{Filipenko1993} Filippenko, A.~V., Matheson, T., \& Ho, L.~C.\ 1993, \apjl, 415, L103 
\bibitem[Filippenko(1997)]{Filipenko1997} Filippenko, A.~V.\ 1997, \araa, 35, 309 
\bibitem[Folatelli et al.(2015)]{Folatelli15} Folatelli, G., Bersten, M.~C., Kuncarayakti, H., et al.\ 2015, \apj, 811, 147 
\bibitem[Folatelli et al.(2016)]{Folatelli16} Folatelli, G., Van Dyk, S.~D., Kuncarayakti, H., et al.\ 2016, \apjl, 825, L22 
\bibitem[Fransson \& Chevalier(1989)]{fransson89} Fransson, C., \& Chevalier, R.~A.\ 1989, \apj, 343, 323
\bibitem[Fraser et al.(2013)]{Fraser13} Fraser, M., Inserra, C., Jerkstrand, A., et al.\ 2013, \mnras, 433, 1312 
\bibitem[Galbany et al.(2014)]{galbany14} Galbany, L., Stanishev, V., Mour{\~a}o, A.~M., et al.\ 2014, \aap, 572, A38
\bibitem[Galbany et al.(2016c)]{galbany16d} Galbany, L., Stanishev, V., Mour{\~a}o, A.~M., et al.\ 2016, \aap, 591, A48 
\bibitem[Galbany et al.(2016b)]{galbany16b} Galbany, L., Anderson, J.~P., Rosales-Ortega, F.~F., et al.\ 2016, \mnras, 455, 4087 
\bibitem[Galbany et al.(2016a)]{galbany16a} Galbany, L., Hamuy, M., Phillips, M.~M., et al.\ 2016, \aj, 151, 33 
\bibitem[Galbany et al.(2018)]{galbany18} Galbany, L., Anderson, J.~P., S{\'a}nchez, S.~F., et al.\ 2018, \apj, 855, 107 
\bibitem[Gall et al.(2015)]{gall15} Gall, E.~E.~E., Polshaw, J., Kotak, R., et al.\ 2015, \aap, 582, A3 
\bibitem[Gezari et al.(2010)]{gezari10} Gezari, S., Rest, A., Huber, M.~E., et al.\ 2010, \apjl, 720, L77 
\bibitem[Gonz{\'a}lez-Gait{\'a}n et al.(2015)]{santiago15} Gonz{\'a}lez-Gait{\'a}n, S., Tominaga, N., Molina, J., et al.\ 2015, \mnras, 451, 2212 
\bibitem[Gualandris \& Portegies Zwart(2007)]{gualandris07} Gualandris, A., \& Portegies Zwart, S.\ 2007, \mnras, 376, L29
\bibitem[Guti{\'e}rrez et al.(2014)]{claudia14} Guti{\'e}rrez, C.~P., Anderson, J.~P., Hamuy, M., et al.\ 2014, \apjl, 786, L15
\bibitem[Guti{\'e}rrez et al.(2017)]{claudia17a} Guti{\'e}rrez, C.~P., Anderson, J.~P., Hamuy, M., et al.\ 2017, \apj, 850, 89 
\bibitem[Guti{\'e}rrez et al.(2017)]{claudia17b} Guti{\'e}rrez, C.~P., Anderson, J.~P., Hamuy, M., et al.\ 2017, \apj, 850, 90 
\bibitem[Guti{\'e}rrez et al.(2018)]{claudia18} Guti{\'e}rrez, C.~P., Anderson, J.~P., Sullivan, M., et al.\ 2018, \mnras, 479, 3232 
\bibitem[Green et al.(2014)]{Green2014} Green, A.~W., Glazebrook, K., McGregor, P.~J., et al.\ 2014, \mnras, 437, 1070 
\bibitem[Gvaramadze et al.(2009)]{gvaramadze09} Gvaramadze, V.~V., Gualandris, A., \& Portegies Zwart, S.\ 2009, \mnras, 396, 570
\bibitem[Hakobyan et al.(2008)]{hakobyan08b} Hakobyan, A.~A., Petrosian, A.~R., McLean, B., et al.\ 2008, \aap, 488, 523
\bibitem[Hakobyan et al.(2009)]{hakobyan09} Hakobyan, A.~A., Mamon, G.~A., Petrosian, A.~R., Kunth, D., \& Turatto, M.\ 2009, \aap, 508, 1259
\bibitem[Hakobyan et al.(2014)]{hakobyan14} Hakobyan, A.~A., Nazaryan, T.~A., Adibekyan, V.~Z., et al.\ 2014, Multiwavelength AGN Surveys and Studies, 304, 339
\bibitem[Hakobyan et al.(2014)]{hakobyan14b} Hakobyan, A.~A., Nazaryan, T.~A., Adibekyan, V.~Z., et al.\ 2014, \mnras, 444, 2428
\bibitem[Hakobyan et al.(2016)]{hakobyan16} Hakobyan, A.~A., Karapetyan, A.~G., Barkhudaryan, L.~V., et al.\ 2016, \mnras, 456, 2848
\bibitem[Hakobyan et al.(2017)]{hakobyan17c} Hakobyan, A.~A., Barkhudaryan, L.~V., Karapetyan, A.~G., et al.\ 2017, \mnras, 471, 1390
\bibitem[Hamuy \& Pinto(2002)]{hamuy02} Hamuy, M., \& Pinto, P.~A.\ 2002, \apjl, 566, L63
\bibitem[Hamuy et al.(2001)]{hamuy01} Hamuy, M., Pinto, P.~A., Maza, J., et al.\ 2001, \apj, 558, 615
\bibitem[Hamuy(2003)]{hamuy03} Hamuy, M.\ 2003, \apj, 582, 905
\bibitem[Hansen \& Milosavljevi{\'c}(2003)]{hansen03} Hansen, B.~M.~S., \& Milosavljevi{\'c}, M.\ 2003, \apjl, 593, L77
\bibitem[Heger et al.(2003)]{heger03} Heger, A., Fryer, C.~L., Woosley, S.~E., Langer, N., \& Hartmann, D.~H.\ 2003, \apj, 591, 288
\bibitem[Henden et al.(2012)]{henden12} Henden, A.~A., Levine, S.~E., Terrell, D., Smith, T.~C., \& Welch, D.\ 2012, Journal of the American Association of Variable Star Observers (JAAVSO), 40, 430
\bibitem[Henry \& Worthey(1999)]{Henry99} Henry, R.~B.~C., \& Worthey, G.\ 1999, \pasp, 111, 919 
\bibitem[Hillebrandt \& Niemeyer(2000)]{Hillebrandt2000} Hillebrandt, W., \& Niemeyer, J.~C.\ 2000, \araa, 38, 191 
\bibitem[Holoien et al.(2017a)]{holoien17} Holoien, T.~W.-S., Stanek, K.~Z., Kochanek, C.~S., et al.\ 2017a, \mnras, 464, 2672 
\bibitem[Holoien et al.(2017b)]{holoien17b} Holoien, T.~W.-S., Brown, J.~S., Stanek, K.~Z., et al.\ 2017b, \mnras, 467, 1098
\bibitem[Holoien et al.(2017c)]{holoien17c} Holoien, T.~W.-S., Brown, J.~S., Stanek, K.~Z., et al.\ 2017c, \mnras, 471, 4966 
\bibitem[Holoien et al.(2018)]{holoien18} Holoien, T.~W.-S., Brown, J.~S., Vallely, P.~J., et al.\ 2018, MNRAS, submitted, arXiv:1811.08904 
\bibitem[Hoogerwerf et al.(2001)]{hoogerwerf01} Hoogerwerf, R., de Bruijne, J.~H.~J., \& de Zeeuw, P.~T.\ 2001, \aap, 365, 49
\bibitem[Howk et al.(2018b)]{Howk2018b} Howk, J.~C., Rueff, K.~M., Lehner, N., et al.\ 2018, \apj, 856, 167 
\bibitem[Howk et al.(2018a)]{Howk2018a} Howk, J.~C., Rueff, K.~M., Lehner, N., et al.\ 2018, \apj, 856, 166 
\bibitem[Huang et al.(2016)]{huang16} Huang, F., Wang, X., Zampieri, L., et al.\ 2016, \apj, 832, 139 
\bibitem[Huang et al.(2018)]{huang18} Huang, F., Wang, X.-F., Hosseinzadeh, G., et al.\ 2018, \mnras, 475, 3959
\bibitem[Inserra et al.(2013)]{Inserra13} Inserra, C., Pastorello, A., Turatto, M., et al.\ 2013, \aap, 555, A142
\bibitem[Jerkstrand et al.(2012)]{jerkstrand12} Jerkstrand, A., Fransson, C., Maguire, K., et al.\ 2012, \aap, 546, A28
\bibitem[Jerkstrand et al.(2014)]{jerkstrand14} Jerkstrand, A., Smartt, S.~J., Fraser, M., et al.\ 2014, \mnras, 439, 3694 (J14)
\bibitem[Jerkstrand(2017)]{jerkstrand17} Jerkstrand, A.\ 2017, arXiv:1702.06702
\bibitem[Jerkstrand et al.(2018)]{Jerkstrand18} Jerkstrand, A., Ertl, T., Janka, H.-T., et al.\ 2018, \mnras, 475, 277 (J18)
\bibitem[Jos{\'e}(2016)]{Jose2016} Jos{\'e}, J.\ 2016, Stellar Explosions: Hydrodynamics and Nucleosynthesis, CRC/Taylor and Francis
\bibitem[Kangas et al.(2017)]{kangas17} Kangas, T., Portinari, L., Mattila, S., et al.\ 2017, \aap, 597, A92 
\bibitem[Kasen \& Woosley(2009)]{kasen09} Kasen, D., \& Woosley, S.~E.\ 2009, \apj, 703, 2205
\bibitem[Kennicutt(1998)]{kennicutt98} Kennicutt, R.~C., Jr.\ 1998, \araa, 36, 189 
\bibitem[Kewley et al.(2001)]{Kewley2001} Kewley, L.~J., Dopita, M.~A., Sutherland, R.~S., Heisler, C.~A., \& Trevena, J.\ 2001, \apj, 556, 121 
\bibitem[Khazov et al.(2016)]{khazov16} Khazov, D., Yaron, O., Gal-Yam, A., et al.\ 2016, \apj, 818, 3
\bibitem[Kirshner \& Kwan(1974)]{epm} Kirshner, R.~P., \& Kwan, J.\ 1974, \apj, 193, 27
\bibitem[Kriek et al.(2009)]{kriek09} Kriek, M., van Dokkum, P.~G., Labb{\'e}, I., et al.\ 2009, \apj, 700, 221 
\bibitem[Kr{\"u}hler et al.(2017)]{kruhler17} Kr{\"u}hler, T., Kuncarayakti, H., Schady, P., et al.\ 2017, \aap, 602, A85
\bibitem[Kuncarayakti et al.(2013)]{hanin13} Kuncarayakti, H., Doi, M., Aldering, G., et al.\ 2013, \aj, 146, 31 
\bibitem[Kuncarayakti et al.(2015)]{hanin15} Kuncarayakti, H., Maeda, K., Bersten, M.~C., et al.\ 2015, \aap, 579, A95 
\bibitem[Kuncarayakti et al.(2018)]{hanin17} 
Kuncarayakti, H., Anderson, J.~P., Galbany, L., et al.\ 2018, \aap, 613, A35 
\bibitem[Kumari et al.(2018)]{Kumari2018} Kumari, N., James, B.~L., Irwin, M.~J., Amor{\'{\i}}n, R., \& P{\'e}rez-Montero, E.\ 2018, \mnras, 476, 3793 
\bibitem[Lang et al.(2010)]{lang10} Lang, D., Hogg, D.~W., Mierle, K., Blanton, M., \& Roweis, S.\ 2010, \aj, 139, 1782
\bibitem[Leonard et al.(2002)]{leonard02b} Leonard, D.~C., Filippenko, A.~V., Li, W., et al.\ 2002, \aj, 124, 2490
\bibitem[Leonard et al.(2002)]{leonard02a} Leonard, D.~C., Filippenko, A.~V., Gates, E.~L., et al.\ 2002, \pasp, 114, 35
\bibitem[Litvinova \& Nadezhin(1985)]{Litvinova1985} Litvinova, I.~Y., \& Nadezhin, D.~K.\ 1985, Soviet Astronomy Letters, 11, 145 
\bibitem[McMillan \& Ciardullo(1996)]{Macmillan96} McMillan, R.~J., \& Ciardullo, R.\ 1996, \apj, 473, 707 
\bibitem[Maguire et al.(2012)]{maguire12} Maguire, K., Jerkstrand, A., Smartt, S.~J., et al.\ 2012, \mnras, 420, 3451
\bibitem[Marino et al.(2013)]{marino13} Marino, R.~A., Rosales-Ortega, F.~F., S{\'a}nchez, S.~F., et al.\ 2013, \aap, 559, A114 
\bibitem[Mauerhan et al.(2013)]{mauerhan13} Mauerhan, J.~C., Smith, N., Filippenko, A.~V., et al.\ 2013, \mnras, 430, 1801
\bibitem[McEvoy et al.(2017)]{mcevoy17} McEvoy, C.~M., Dufton, P.~L., Smoker, J.~V., et al.\ 2017, \apj, 842, 32
\bibitem[McSwain et al.(2007)]{mcswain07} McSwain, M.~V., Boyajian, T.~S., Grundstrom, E.~D., \& Gies, D.~R.\ 2007, \apj, 655, 473
\bibitem[Mikhailova et al.(2007)]{Mikhailova2007} Mikhailova, G.~A., Bartunov, O.~S., \& Tsvetkov, D.~Y.\ 2007, Astronomy Letters, 33, 715 
\bibitem[Minkowski(1941)]{minkowski41} Minkowski, R.\ 1941, \pasp, 53, 224
\bibitem[Modjaz et al.(2008)]{Modjaz08} Modjaz, M., Kewley, L., Kirshner, R.~P., et al.\ 2008, \aj, 135, 1136 
\bibitem[Moriya et al.(2016)]{Moriya2016} Moriya, T.~J., Pruzhinskaya, M.~V., Ergon, M., \& Blinnikov, S.~I.\ 2016, \mnras, 455, 423 
\bibitem[Morrissey et al.(2007)]{galex} Morrissey, P., Conrow, T., Barlow, T.~A., et al.\ 2007, \apjs, 173, 682 
\bibitem[Mould et al.(2000)]{CMB_z} Mould, J.~R., Huchra, J.~P., Freedman, W.~L., et al.\ 2000, \apj, 529, 786 
\bibitem[M{\"u}ller et al.(2017)]{muller17} M{\"u}ller, T., Prieto, J.~L., Pejcha, O., \& Clocchiatti, A.\ 2017, \apj, 841, 127 
\bibitem[Nakar \& Sari(2010)]{nakar10} Nakar, E., \& Sari, R.\ 2010, \apj, 725, 904 
\bibitem[Nomoto et al.(1993)]{Nomoto1993} Nomoto, K., Suzuki, T., Shigeyama, T., et al.\ 1993, \nat, 364, 507 
\bibitem[Nordgren et al.(1997)]{Nordgren97} Nordgren, T.~E., Chengalur, J.~N., Salpeter, E.~E., \& Terzian, Y.\ 1997, \aj, 114, 913 
\bibitem[Olivares E.~et al.(2010)]{olivares10} Olivares E., F., Hamuy, M., Pignata, G., et al.\ 2010, \apj, 715, 833
\bibitem[Pastorello et al.(2009)]{pastorello09} Pastorello, A., Valenti, S., Zampieri, L., et al.\ 2009, \mnras, 394, 2266
\bibitem[Pastorello et al.(2013)]{pastorello13} Pastorello, A., Cappellaro, E., Inserra, C., et al.\ 2013, \apj, 767, 1
\bibitem[Peeples \& Shankar(2011)]{peeples2011} Peeples, M.~S., \& Shankar, F.\ 2011, \mnras, 417, 2962 
\bibitem[Pejcha \& Prieto(2015)]{pejcha15} Pejcha, O., \& Prieto, J.~L.\ 2015, \apj, 806, 225
\bibitem[Pejcha \& Prieto(2015)]{pejcha15b} Pejcha, O., \& Prieto, J.~L.\ 2015, \apj, 799, 215
\bibitem[P{\'e}rez-Montero(2014)]{Montero2014} P{\'e}rez-Montero, E.\ 2014, \mnras, 441, 2663 
\bibitem[Petrosian et al.(2005)]{Petrosian2005} Petrosian, A., Navasardyan, H., Cappellaro, E., et al.\ 2005, \aj, 129, 1369 
\bibitem[Phillips et al.(2013)]{phillips13} Phillips, M.~M., Simon, J.~D., Morrell, N., et al.\ 2013, \apj, 779, 38 
\bibitem[Pilyugin et al.(2014)]{Pilyugin2014} Pilyugin, L.~S., Grebel, E.~K., \& Kniazev, A.~Y.\ 2014, \aj, 147, 131 
\bibitem[Pols et al.(1998)]{Pols1998} Pols, O.~R., Schr{\"o}der, K.-P., Hurley, J.~R., Tout, C.~A., \& Eggleton, P.~P.\ 1998, \mnras, 298, 525 
\bibitem[Polshaw et al.(2016)]{polshaw16} Polshaw, J., Kotak, R., Dessart, L., et al.\ 2016, \aap, 588, A1
\bibitem[Popov(1993)]{Popov1993} Popov, D.~V.\ 1993, \apj, 414, 712 
\bibitem[Potashov et al.(2013)]{potashov} Potashov, M., Blinnikov, S., Baklanov, P., \& Dolgov, A.\ 2013, \mnras, 431, L98 
\bibitem[Prieto et al.(2008)]{prieto08} Prieto, J.~L., Stanek, K.~Z., \& Beacom, J.~F.\ 2008, \apj, 673, 999 
\bibitem[Prieto et al.(2009)]{prieto09} Prieto, J.~L., Sellgren, K., Thompson, T.~A., \& Kochanek, C.~S.\ 2009, \apj, 705, 1425
\bibitem[Prieto et al.(2012)]{prieto12} Prieto, J.~L., Lee, J.~C., Drake, A.~J., et al.\ 2012, \apj, 745, 70 
\bibitem[Prieto et al.(2013)]{prieto13} Prieto, J.~L., Brimacombe, J., Drake, A.~J., \& Howerton, S.\ 2013, \apjl, 763, L27
\bibitem[Prieto et al.(2016)]{prieto16} Prieto, J.~L., Kr{\"u}hler, T., Anderson, J.~P., et al.\ 2016, \apjl, 830, L32
\bibitem[Rabinak \& Waxman(2011)]{rabinak11} Rabinak, I., \& Waxman, E.\ 2011, \apj, 728, 63
\bibitem[Renzo et al.(2018)]{Renzo2018} Renzo, M., Zapartas, E., de Mink, S.~E., et al.\ 2018, arXiv:1804.09164
\bibitem[Rodr{\'{\i}}guez et al.(2014)]{Osmar14} Rodr{\'{\i}}guez, {\'O}., Clocchiatti, A., \& Hamuy, M.\ 2014, \aj, 148, 107 
\bibitem[Rodr{\'{\i}}guez et al.(2014)]{Osmar18} Rodr{\'{\i}}guez, {\'O}., Pignata, A., et al., in preparation (R18)
\bibitem[Rubin et al.(2016)]{rubin16} Rubin, A., Gal-Yam, A., De Cia, A., et al.\ 2016, \apj, 820, 33 
\bibitem[Sana et al.(2012)]{Sana2012} Sana, H., de Mink, S.~E., de Koter, A., et al.\ 2012, Science, 337, 444 
\bibitem[Sana(2017)]{Sana2017} Sana, H.\ 2017, The Lives and Death-Throes of Massive Stars, 329, 110 
\bibitem[S{\'a}nchez et al.(2012)]{sanchez2012} S{\'a}nchez, S.~F., Rosales-Ortega, F.~F., Marino, R.~A., et al.\ 2012, \aap, 546, A2
\bibitem[S{\'a}nchez et al.(2014)]{sanchez2014} S{\'a}nchez, S.~F., Rosales-Ortega, F.~F., Iglesias-P{\'a}ramo, J., et al.\ 2014, \aap, 563, A49 
\bibitem[S{\'a}nchez-Menguiano et al.(2016)]{SanchezM2016} S{\'a}nchez-Menguiano, L., S{\'a}nchez, S.~F., P{\'e}rez, I., et al.\ 2016, \aap, 587, A70 
\bibitem[Sanders et al.(2015)]{Sanders15} Sanders, N.~E., Soderberg, A.~M., Gezari, S., et al.\ 2015, \apj, 799, 208
\bibitem[Sapir \& Waxman(2017)]{sapir2017} Sapir, N., \& Waxman, E.\ 2017, \apj, 838, 130 
\bibitem[Sarangi \& Cherchneff(2015)]{Sarangi2015} Sarangi, A., \& Cherchneff, I.\ 2015, \aap, 575, A95 
\bibitem[Schlafly \& Finkbeiner(2011)]{schlafly11} Schlafly, E.~F., \& Finkbeiner, D.~P.\ 2011, \apj, 737, 103
\bibitem[Schlegel(1990)]{Schlegel1990} Schlegel, E.~M.\ 1990, \mnras, 244, 269 
\bibitem[Schmidt et al.(1992)]{epm2} Schmidt, B.~P., Kirshner, R.~P., \& Eastman, R.~G.\ 1992, \apj, 395, 366 
\bibitem[Schnurr et al.(2008)]{schnurr08} Schnurr, O., Moffat, A.~F.~J., St-Louis, N., Morrell, N.~I., \& Guerrero, M.~A.\ 2008, \mnras, 389, 806
\bibitem[Shappee et al.(2014)]{shappee14} Shappee, B.~J., Prieto, J.~L., Grupe, D., et al.\ 2014, \apj, 788, 48
\bibitem[Sheth et al.(2010)]{sheth10} Sheth, K., Regan, M., Hinz, J.~L., et al.\ 2010, \pasp, 122, 1397 
\bibitem[Silva \& Napiwotzki(2011)]{silva11} Silva, M.~D.~V., \& Napiwotzki, R.\ 2011, \mnras, 411, 2596
\bibitem[Silverman et al.(2017)]{silverman17} Silverman, J.~M., Pickett, S., Wheeler, J.~C., et al.\ 2017, \mnras, 467, 369
\bibitem[Smartt et al.(2009)]{Smartt09} Smartt, S.~J., Eldridge, J.~J., Crockett, R.~M., \& Maund, J.~R.\ 2009, \mnras, 395, 1409
\bibitem[Smartt(2015)]{Smartt15} Smartt, S.~J.\ 2015, \pasa, 32, e016 
\bibitem[Smartt et al.(2015)]{smartt15} Smartt, S.~J., Valenti, S., Fraser, M., et al.\ 2015, \aap, 579, A40
\bibitem[Smith et al.(2014)]{smith14} Smith, N., Mauerhan, J.~C., \& Prieto, J.~L.\ 2014, \mnras, 438, 1191 
\bibitem[Smith et al.(2016)]{smith16} Smith, N., Andrews, J.~E., \& Mauerhan, J.~C.\ 2016, \mnras, 463, 2904
\bibitem[Stathakis \& Sadler(1991)]{Stathakis1991} Stathakis, R.~A., \& Sadler, E.~M.\ 1991, \mnras, 250, 786 
\bibitem[Stoll et al.(2013)]{stoll13} Stoll, R., Prieto, J.~L., Stanek, K.~Z., \& Pogge, R.~W.\ 2013, \apj, 773, 12 
\bibitem[Sukhbold et al.(2016)]{Sukhbold2016} Sukhbold, T., Ertl, T., Woosley, S.~E., Brown, J.~M., \& Janka, H.-T.\ 2016, \apj, 821, 38
\bibitem[Szalai et al.(2018)]{szalai18} Szalai, T., Zs{\'{\i}}ros, S., Fox, O.~D., Pejcha, O., \& M{\"u}ller, T.\ 2018, arXiv:1803.02571
\bibitem[Tak{\'a}ts \& Vink{\'o}(2012)]{katy12} Tak{\'a}ts, K., \& Vink{\'o}, J.\ 2012, \mnras, 419, 2783
\bibitem[Tak{\'a}ts et al.(2014)]{katy14} Tak{\'a}ts, K., Pumo, M.~L., Elias-Rosa, N., et al.\ 2014, \mnras, 438, 368
\bibitem[Tartaglia et al.(2018)]{tartaglia18} Tartaglia, L., Sand, D.~J., Valenti, S., et al.\ 2018, \apj, 853, 62 
\bibitem[Terreran et al.(2016)]{terreran16} Terreran, G., Jerkstrand, A., Benetti, S., et al.\ 2016, \mnras, 462, 137
\bibitem[Turatto et al.(2007)]{Turatto2007} Turatto, M., Benetti, S., \& Pastorello, A.\ 2007, Supernova 1987A: 20 Years After: Supernovae and Gamma-Ray Bursters, 937, 187 
\bibitem[Uomoto(1986)]{uomoto86} Uomoto, A.\ 1986, \apjl, 310, L35
\bibitem[Utrobin \& Chugai(2009)]{Utrobin2009} Utrobin, V.~P., \& Chugai, N.~N.\ 2009, \aap, 506, 829 
\bibitem[Valenti et al.(2014)]{valenti14} Valenti, S., Sand, D., Pastorello, A., et al.\ 2014, \mnras, 438, L101 
\bibitem[Valenti et al.(2015)]{valenti15} Valenti, S., Sand, D., Stritzinger, M., et al.\ 2015, \mnras, 448, 2608 
\bibitem[Valenti et al.(2016)]{valenti16} Valenti, S., Howell, D.~A., Stritzinger, M.~D., et al.\ 2016, \mnras, 459, 3939 
\bibitem[Van Dyk et al.(1999)]{VanDyk1999} Van Dyk, S.~D., Peng, C.~Y., Barth, A.~J., \& Filippenko, A.~V.\ 1999, \aj, 118, 2331 
\bibitem[van Zee \& Haynes(2006)]{VanZee06} van Zee, L., \& Haynes, M.~P.\ 2006, \apj, 636, 214
\bibitem[Weilbacher et al.(2014)]{weilbacher14} Weilbacher, P.~M.,  Streicher, O., Urrutia, T., et al.\ 2014, Astronomical Data Analysis Software and Systems XXIII, 485, 451
\bibitem[Xiao et al.(2018)]{Lin18a} Xiao, L., Stanway, E.~R., \& Eldridge, J.~J.\ 2018, \mnras, 477, 904 
\bibitem[Yaron et al.(2017)]{Yaron2017} Yaron, O., Perley, D.~A., Gal-Yam, A., et al.\ 2017, Nature Physics, 13, 510 
\bibitem[Yoachim \& Dalcanton(2008)]{Yoachim2008} Yoachim, P., \& Dalcanton, J.~J.\ 2008, \apj, 683, 707 
\bibitem[Zapartas et al.(2017)]{zapartas17} Zapartas, E., de Mink, S.~E., Izzard, R.~G., et al.\ 2017, \aap, 601, A29 
\bibitem[Zhang \& Wang(2014)]{classification2} Zhang, J., \& Wang, X.\ 2014, The Astronomer's Telegram, 6601,  
\bibitem[Zinn et al.(2011)]{zinn11} Zinn, P.-C., Grunden, P., \& Bomans, D.~J.\ 2011, \aap, 536, A103
\bibitem[de Jaeger et al.(2015)]{tomato15} de Jaeger, T., Gonz{\'a}lez-Gait{\'a}n, S., Anderson, J.~P., et al.\ 2015, \apj, 815, 121
\bibitem[de Jaeger et al.(2017)]{tomato17} de Jaeger, T., Gonz{\'a}lez-Gait{\'a}n, S., Hamuy, M., et al.\ 2017, \apj, 835, 166

\end{thebibliography}
\end{document}